\documentclass[prd,nofootinbib,showpacs,amsmath,superscriptaddress]{revtex4}

\usepackage{times}
\usepackage{hyperref}

\usepackage{graphicx}
\usepackage{dcolumn}
\usepackage{bm}

\begin{document}

\title{Renormalization of the baryon axial vector current in large-$N_c$ chiral perturbation theory: Effects of the decuplet-octet mass difference and flavor symmetry breaking}

\author{Rub\'en Flores-Mendieta}
\affiliation{Instituto de F{\'\i}sica, Universidad Aut\'onoma de San Luis Potos{\'\i}, \'Alvaro Obreg\'on 64, Zona Centro, San Luis Potos{\'\i}, S.L.P.\ 78000, M\'exico}

\author{Mar{\'\i}a A.\ Hern\'andez-Ruiz}
\affiliation{Instituto de F{\'\i}sica, Universidad Aut\'onoma de San Luis Potos{\'\i}, \'Alvaro Obreg\'on 64, Zona Centro, San Luis Potos{\'\i}, S.L.P.\ 78000, M\'exico}
\affiliation{Facultad de Ciencias Qu{\'\i}micas, Universidad Aut\'onoma de Zacatecas, Apartado Postal 585, Zacatecas, Zacatecas 98060, M\'exico}

\author{Christoph P.\ Hofmann}
\affiliation{Facultad de Ciencias, Universidad de Colima, Bernal D{\'\i}az del Castillo 340, Colima, Colima 28045, M\'exico}

\date{\today}

\begin{abstract}
The baryon axial vector current is computed at one-loop order in large-$N_c$ baryon chiral perturbation theory, where $N_c$ is the number of colors. Loop graphs with octet and decuplet intermediate states are systematically incorporated into the analysis and the effects of the decuplet-octet mass difference and SU(3) flavor symmetry breaking are accounted for. As expected, large-$N_c$ cancellations between different one-loop graphs are observed as a consequence of the large-$N_c$ spin-flavor symmetry of QCD baryons. Fitting our analytical formulas against experimental data on baryon semileptonic decays and the strong decays of decuplet baryons, a detailed numerical analysis regarding the determination of the basic parameters of large-$N_c$ baryon chiral perturbation theory as well as the extraction of the baryon axial vector couplings is performed. The large-$N_c$ baryon chiral perturbation theory predictions are in very good agreement both with the expectations from the $1/N_c$ expansion and with the experimental data.
\end{abstract}


\maketitle

\section{Introduction}

From the theoretical point of view, the analysis of baryon semileptonic decays, $B_i\to B_je^-\overline{\nu}_e$, is rather involved due to the participation of both vector and axial-vector currents. In the past, the understanding of the consequences of the weak hadronic currents relied on the Cabibbo model \cite{cab63}, i.e., on an exact flavor SU(3) symmetry. However, with the advent of high-statistics experiments \cite{part}, the departure from the limit of exact symmetry has now become more evident and one is prompted to compute the effects of SU(3) symmetry breaking in the form factors.

The leading vector form factors in baryon semileptonic decays are protected by the Ademollo-Gatto theorem \cite{ag64} against SU(3) breaking corrections to lowest order in $\epsilon = m_s-\hat{m}$. For this reason, the theoretical framework to compute them is under reasonable control within the limits of experimental precision. However, in the case of the axial-vector form factors one faces larger theoretical uncertainties because of the appearance of first-order SU(3) breaking effects.

Although one can assert that in recent years lattice QCD calculations have demonstrated remarkable progress in computing hadron properties from first principles with high accuracy, analytical calculations of these properties are not possible because QCD is strongly coupled at low energies. Thus, a number of different methods have been developed to understand the low-energy QCD hadron dynamics. In particular, SU(3) breaking corrections to baryon semileptonic decay form factors have been analyzed within quark and soliton models \cite{dhk87,sch95,kpg00,fghl08,fghikl08,lskg08,sdcg09,cpw10}, lattice QCD \cite{edwEtAl,glps07,yamEtAl,lo09,sy09,goeEtAl}, the 1/$N_c$ expansion \cite{djm94,djm95,ddjm96,fjm98}, (heavy) baryon chiral perturbation theory \cite{kra90,al93,sw97,dhb99,bor99,zpr01,kai01,zsr02,bhlo02,glmpsv05,vil06,lkm07,jt08}, as well as within the combined framework of large-$N_c$ baryon chiral perturbation theory \cite{fh06,aco06,acu07}.

Large-$N_c$ QCD is the generalization of QCD from three colors $N_c=3$ to $N_c \gg 3$. The baryon sector in the large-$N_c$ limit of QCD exhibits an exact SU($2N_f$) contracted spin-flavor symmetry, where $N_f$ is the number of light quark flavors. In the large-$N_c$ limit, decuplet and octet baryon states become degenerate, the difference $\Delta$ between the SU(3) invariant masses of the decuplet and octet baryons given by $\Delta\equiv M_T-M_B \propto 1/N_c$. The spin-flavor symmetry allows one to classify large-$N_c$ baryon states and matrix elements and to compute static properties of large-$N_c$ baryons in a systematic expansion in 1/$N_c$ \cite{djm94,djm95}. This formalism has been applied to a variety of physical quantities, including the baryon axial vector current \cite{djm94,djm95,ddjm96,fjm98,fhj00,fhjm00}.

Another systematic and model-independent method is chiral perturbation theory which is based on the spontaneously broken chiral symmetry $\mathrm{SU(3)}_L \times \mathrm{SU(3)}_R$ of the QCD Lagrangian. It is the low-energy effective field theory of QCD, formulating the dynamics in terms of the pseudoscalar octet of Goldstone bosons. Physical observables can be expanded systematically, order by order, in powers of $p^2/\Lambda_\chi$ and $m^2_{\Pi}/\Lambda_\chi$, where $p$ is the meson momentum, $m_{\Pi}$ is the Goldstone boson mass and $\Lambda_\chi$ is the scale of chiral symmetry breaking \cite{wei79,gl85,leu94}. The inclusion of heavy particles such as the proton or the neutron, whose masses do not vanish in the chiral limit $m_q \to 0$, can be performed within the framework of (Lorentz invariant) baryon chiral perturbation theory \cite{gss88,bkkm92,bl99,fgjs03} or heavy baryon chiral perturbation theory \cite{jm91a,dobogoko}. In this work we will use the latter approach which involves velocity-dependent baryon fields and where the expansion of the baryon chiral Lagrangian in
powers of $m_q$ and 1/$M_B$ (where $M_B$ is the baryon mass) is manifest.

In particular, chiral logarithmic corrections due to meson loops were considered in Refs.~\cite{jm91a,dobogoko,jm91b,jlms93,ms97,dh98,pr00}. A crucial observation was that while these corrections are large when only octet baryon intermediate states are kept, the inclusion of decuplet baryon intermediate states yields sizable cancellations between one-loop corrections. This observation, as we will illustrate in the present work, can be rigorously explained in the context of the 1/$N_c$ expansion.

A very powerful method -- the one we will use in the present work -- is constituted by the combined use of the $1/N_c$ expansion and chiral perturbation theory \cite{jen96}. It describes the interactions between the spin-$\frac{1}{2}$ baryon octet and the spin-$\frac{3}{2}$ baryon decuplet with the pseudoscalar Goldstone boson octet augmented by the $\eta^\prime$. Observables that have been calculated within this combined framework include baryon masses \cite{jen96,bl96,ow99,CG12}, baryon magnetic moments \cite{lmw95,flo09,ahu10,afh10,jen12} and the baryon axial vector current \cite{fhj00,fhjm00,fh06,aco06,acu07,CG12}. When computing the renormalization of the baryon axial vector current at one-loop order, large-$N_c$ cancellations between various Feynman diagrams occur, provided that both octet and decuplet intermediate states are considered. While the general structure of these cancellations was analyzed in Ref.~\cite{fhjm00}, the explicit evaluation of the corresponding operator expressions, which involve complicated structures of commutators and/or anticommutators of the SU(6) spin-flavor operators, was discussed in Ref.~\cite{fh06}.

In the present work we go beyond the analysis of Ref.~\cite{fh06} in various ways. As we will discuss in detail in Sec.~\ref{renormalization}, we extend the operator analysis by including all effects which are suppressed by $1/N_c^2$ relative to the tree-level value, which includes to take into account the non-vanishing decuplet-octet mass difference $\Delta$. Moreover, effects related to SU(3) flavor symmetry breaking are included as follows: On the one hand, at tree level, we include all relevant operators which explicitly break SU(3) at leading order. On the other hand, in the one-loop corrections, SU(3) symmetry breaking is accounted for implicitly, since the loop integrals depend on the pion, kaon and $\eta$ masses. As an application of our formalism, given the precision of our analytic expressions, we are then able to perform various fits in order to determine the basic parameters of large-$N_c$ baryon chiral perturbation theory as well as to extract baryon axial vector couplings from baryon semileptonic decays and from the pion decays of decuplet baryons.

The rest of the paper is organized as follows. In Sec.~\ref{overview} we provide an outline of the basic ingredients of large-$N_c$ baryon chiral perturbation theory needed in the present work. In Sec.~\ref{renormalization}, the renormalization of the baryon axial vector current is presented and the large-$N_c$ cancellations are illustrated. The incorporation of SU(3) symmetry breaking effects into the operator analysis is performed in Sec.~\ref{FlavorSymmetryBreaking}, while Sec.~\ref{fits} contains a detailed numerical analysis regarding the determination of the basic parameters of large-$N_c$ baryon chiral perturbation theory and the extraction of $g_A$ from baryon semileptonic decays and the strong decays of decuplet baryons, our conclusions are presented in Sec.~\ref{conclusions}. Finally, technical details regarding loop integrals, commutator/anticommutator operator structures, flavor $\mathbf{8}$ and $\mathbf{27}$ contributions to the baryon axial vector couplings, as well as matrix elements of baryon operators are relegated to four different appendices.

\section{Overview of large-$N_c$ chiral perturbation theory}
\label{overview}

In order to introduce our notation and conventions, in this section we provide an overview of the chiral Lagrangian for baryons in the $1/N_c$ expansion introduced first in Ref.~\cite{jen96}. This Lagrangian, which incorporates nonet symmetry and the contracted spin-flavor symmetry for baryons in the large-$N_c$ limit, can be written as
\begin{equation}
\mathcal{L}_{\text{baryon}} = i \mathcal{D}^0 - \mathcal{M}_{\text{hyperfine}} + \text{Tr} \left(\mathcal{A}^k \lambda^c \right) A^{kc} + \frac{1}{N_c} \text{Tr} \left(\mathcal{A}^k \frac{2I}{\sqrt 6}\right) A^k + \ldots, \label{eq:ncch}
\end{equation}
with
\begin{equation}
\mathcal{D}^0 = \partial^0 \openone + \text{Tr} \left(\mathcal{V}^0 \lambda^c\right) T^c. \label{eq:kin}
\end{equation}

The ellipses in Eq.~(\ref{eq:ncch}) denote higher partial wave pion couplings which occur at subleading orders in the $1/N_c$ expansions for $N_c > 3$. In the large $N_c$ limit, all of these higher partial waves vanish, and the pion coupling to baryons is purely $p$ wave. Therefore, the only terms relevant for our analysis are those displayed in Eq.~(\ref{eq:ncch}).

Meson fields enter the Lagrangian (\ref{eq:ncch}) through the vector and axial-vector combinations
\begin{equation}
\mathcal{V}^0 = \frac12 \left(\xi \partial^0 \xi^\dagger + \xi^\dagger \partial^0 \xi\right), \qquad
\mathcal{A}^k = \frac{i}{2} \left(\xi \nabla^k \xi^\dagger - \xi^\dagger \nabla^k \xi\right), \qquad \qquad
\xi(x)=\exp[i\Pi(x)/f],
\end{equation}
where $\Pi(x)$ represents the nonet of Goldstone boson fields and $f \approx 93$ $\mathrm{MeV}/c^2$ is the pion decay constant.

Each of the different terms in the Lagrangian (\ref{eq:ncch}) contains a baryon operator. While the baryon kinetic energy term involves the spin-flavor identity, $\mathcal{M}_{\text{hyperfine}}$ represents the hyperfine baryon mass operator which incorporates the spin splittings of the tower of baryon states with spins $1/2,\ldots, N_c/2$ in the flavor representations. The quantities $A^k$ and $A^{kc}$ stand for the flavor singlet and flavor octet baryon axial vector currents, respectively. All these baryon operators can be written as polynomials in the SU(6) spin-flavor generators \cite{djm95}
\begin{equation}
J^k = q^\dagger \frac{\sigma^k}{2} q, \qquad T^c = q^\dagger \frac{\lambda^c}{2} q, \qquad G^{kc} = q^\dagger
\frac{\sigma^k}{2}\frac{\lambda^c}{2} q. \label{eq:su6gen}
\end{equation}
Here $q^\dagger$ and $q$ are SU(6) operators that create and annihilate states in the fundamental representation of SU(6), and
$\sigma^k$ and $\lambda^c$ are the Pauli spin and Gell-Mann flavor matrices, respectively. The SU(6) spin-flavor generators satisfy the commutation relations listed in Table \ref{tab:surel}.\begingroup
\begin{table}
\caption{\label{tab:surel}$\mathrm{SU}(2 N_f)$ commutation relations.}
\bigskip
\label{tab:su2fcomm}
\centerline{\vbox{ \tabskip=0pt \offinterlineskip
\halign{
\strut\quad $ # $\quad\hfil&\strut\quad $ # $\quad \hfil\cr
\multispan2\hfil $\left[J^i,T^a\right]=0,$ \hfil \cr
\noalign{\medskip}
\left[J^i,J^j\right]=i\epsilon^{ijk} J^k,
&\left[T^a,T^b\right]=i f^{abc} T^c,\cr
\noalign{\medskip}
\left[J^i,G^{ja}\right]=i\epsilon^{ijk} G^{ka},
&\left[T^a,G^{ib}\right]=i f^{abc} G^{ic},\cr
\noalign{\medskip}
\multispan2\hfil$\displaystyle [G^{ia},G^{jb}] = \frac{i}{4}\delta^{ij}
f^{abc} T^c + \frac{i}{2N_f} \delta^{ab} \epsilon^{ijk} J^k + \frac{i}{2} \epsilon^{ijk} d^{abc} G^{kc}.$ \hfill\cr
}}}
\end{table}
\endgroup

The $1/N_c$ expansions of the baryon flavor singlet and octet axial vector currents were derived in Ref.~\cite{djm95}. Taking into account that $A^k$ is a spin-1 object and a singlet under SU(3), its $1/N_c$ expansion amounts to
\begin{equation}
A^k = \sum_{n=1,3}^{N_c} b_n^{1,1} \frac{1}{N_c^{n-1}} \mathcal{D}_n^k, \label{eq:asin}
\end{equation}
where $\mathcal{D}_1^k = J^k$ and $\mathcal{D}_{2m+1}^k = \{J^2,\mathcal{D}_{2m-1}^k\}$ for $m\geq 1$. The superscript on the operator coefficients of $A^k$ denotes that they refer to the baryon singlet current. At the physical value $N_c=3$, Eq.~(\ref{eq:asin}) reduces to
\begin{equation}
A^k = b_1^{1,1} J^k + b_3^{1,1} \frac{1}{N_c^2} \{J^2,J^k\}.
\end{equation}

The flavor octet current $A^{kc}$, on the other hand, is a spin-1 object, an octet under SU(3) and odd under time reversal. Its $1/N_c$ expansion reads \cite{djm94,djm95}
\begin{equation}
A^{kc} = a_1 G^{kc} + \sum_{n=2,3}^{N_c} b_n \frac{1}{N_c^{n-1}} \mathcal{D}_n^{kc} + \sum_{n=3,5}^{N_c} c_n
\frac{1}{N_c^{n-1}} \mathcal{O}_n^{kc}, \label{eq:akcfull}
\end{equation}
where the unknown coefficients $a_1$, $b_n$, and $c_n$ have expansions in powers of $1/N_c$ and are order unity at leading order in the $1/N_c$ expansion. The first few operators in expansion (\ref{eq:akcfull}) are
\begin{eqnarray}
\mathcal{D}_2^{kc} & = & J^kT^c, \label{eq:d2kc} \\
\mathcal{D}_3^{kc} & = & \{J^k,\{J^r,G^{rc}\}\}, \label{eq:d3kc} \\
\mathcal{O}_3^{kc} & = & \{J^2,G^{kc}\} - \frac12 \{J^k,\{J^r,G^{rc}\}\}, \label{eq:o3kc}
\end{eqnarray}
while higher order terms can be obtained as $\mathcal{D}_n^{kc}=\{J^2,\mathcal{D}_{n-2}^{kc}\}$ and $\mathcal{O}_n^{kc}=\{J^2,\mathcal{O}_{n-2}^{kc}\}$ for $n\geq 4$. Notice that $\mathcal{D}_n^{kc}$ are diagonal operators with non-zero matrix elements only between states with the same spin, and the $\mathcal{O}_n^{kc}$ are purely off-diagonal operators with non-zero matrix elements only between states with different spin. At $N_c = 3$ the series (\ref{eq:akcfull}) can be truncated as
\begin{equation}
A^{kc} = a_1 G^{kc} + b_2 \frac{1}{N_c} \mathcal{D}_2^{kc} + b_3 \frac{1}{N_c^2} \mathcal{D}_3^{kc} + c_3 \frac{1}{N_c^2} \mathcal{O}_3^{kc}. \label{eq:akc}
\end{equation}
Let us stress the fact that at the physical value $N_c=3$, the $1/N_c$ expansion only extends up to 3-body operators, such that there are only four terms in the expansion for the baryon octet axial vector current Eq.~(\ref{eq:akc}) in the flavor SU(3) limit. Note again that the unknown coefficients $a_1, b_2, b_3$ and $c_3$ are all of order unity for large $N_c$ (see Ref.~\cite{ddjm96}).

The matrix elements of the space components of $A^{kc}$ between SU(6) symmetric states yield the values of the axial vector couplings. For the octet baryons, the axial vector couplings are $g_A$, as defined in experiments in baryon semileptonic decays, normalized in such a way that $g_A\approx 1.27$ for neutron $\beta$ decay. For decuplet baryons, the axial vector couplings are $g$, which are extracted from the widths of the strong decays of the decuplet baryons into octet baryons and pions.

Finally, the $1/N_c$ expansion of the baryon mass operator $\mathcal{M}$ takes the form \cite{djm94,djm95}
\begin{eqnarray}
\mathcal{M} = m_0 N_c \openone + \sum_{n=2,4}^{N_c-1} m_{n} \frac{1}{N_c^{n-1}} J^n, \label{eq:mop}
\end{eqnarray}
where $m_n$ are unknown coefficients. The first term on the right-hand side is the overall spin-independent mass of the baryon multiplet and is removed from the chiral Lagrangian by the heavy baryon field redefinition~\cite{jm91a}. The other terms are spin-dependent and represent $\mathcal{M}_{\text{hyperfine}}$ introduced in the chiral Lagrangian (\ref{eq:ncch}). For $N_c=3$ the hyperfine mass expansion reduces to a single operator
\begin{eqnarray}
\mathcal{M} _{\text{hyperfine}} = \frac{m_2}{N_c} J^2 . \label{eq:smop}
\end{eqnarray}
Again we emphasize that at the physical value $N_c=3$, the $1/N_c$ expansion for the baryon mass operator $\mathcal{M}$ only extends up to 3-body operators, such that there are only two terms in Eq.~(\ref{eq:mop}).

While the above expression for the flavor octet axial current $A^{kc}$ refers to the SU(3) symmetry limit, we also want to include effects into the operator analysis which result from explicit SU(3) flavor symmetry breaking. Indeed, as we discuss in detail in Sec.~\ref{FlavorSymmetryBreaking}, the flavor octet axial current $A^{kc}$ will receive additional terms which account for SU(3) breaking.

\section{Renormalization of the baryon axial vector current}
\label{renormalization}

The $1/N_c$ baryon chiral Lagrangian displayed in Eq.~(\ref{eq:ncch}) has been applied to the calculation of non-analytic meson-loop corrections to various static properties of baryons. Among them, the chiral corrections to the axial vector coupling $g_A$ have been tackled in Refs.~\cite{fhjm00,fh06}.

\begin{figure}[ht]
\scalebox{0.9}{\includegraphics{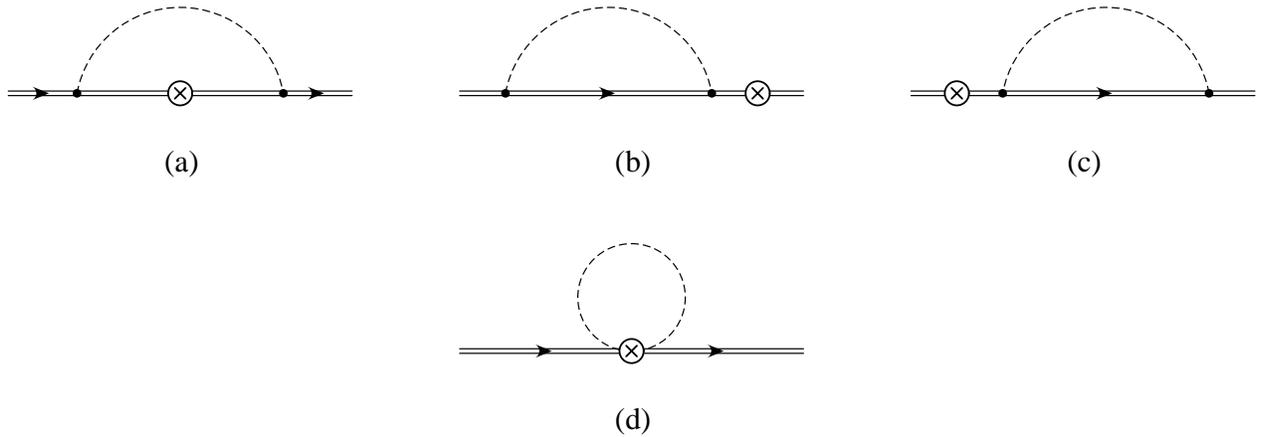}}
\caption{\label{fig:eins}One-loop corrections to the baryon axial vector current.}
\end{figure}

The one-loop diagrams that renormalize the baryon axial vector current $A^{kc}$ are displayed in Fig.~\ref{fig:eins}.
Previous analyses \cite{djm94,djm95,ddjm96,fhjm00,fh06} have shown that these loop graphs have a calculable dependence on the ratio $m_\Pi/\Delta$, where $m_\Pi$ denotes the meson mass and $\Delta \equiv M_T - M_B$ is the decuplet-octet mass difference. However, in order for the theory to be valid, the conditions $m_\Pi \ll \Lambda_\chi$ and $\Delta \ll \Lambda_\chi$ must be met, while the ratio $m_\Pi/\Delta$ is not constrained and can take any value. Also, the meson-baryon vertex is proportional to $g_A/f$. In the large-$N_c$ limit, $g_A\propto N_c$ and $f\propto\sqrt{N_c}$ so the pion-baryon vertex is of order $\mathcal{O}(\sqrt{N_c})$. The meson and baryon propagators are independent of $N_c$ and so are the loop integrals because in the $\overline{\mathrm{MS}}$ scheme they are given by the pole structure of the propagators.

The loop graphs of Fig.~\ref{fig:eins} possess a rather different $N_c$-dependence. Due to the fact that Figs.~\ref{fig:eins}(a,b,c) can be combined into a single structure \cite{fhjm00,fh06}, we first deal with the correction arising from these diagrams and postpone the discussion of diagram \ref{fig:eins}(d). The contribution from Fig.~\ref{fig:eins}(a,b,c) contains the full dependence on the ratio $\Delta/m_\Pi$ and can be written as \cite{fhjm00}
\begin{eqnarray}
\delta A^{kc} & = & \frac12 \left[A^{ja},\left[A^{jb},A^{kc}\right]\right] \Pi_{(1)}^{ab} - \frac12 \left\{ A^{ja}, \left[A^{kc},\left[\mathcal{M},A^{jb}\right] \right] \right\} \Pi_{(2)}^{ab} \nonumber \\
& & \mbox{} + \frac16 \left(\left[A^{ja}, \left[\left[\mathcal{M}, \left[ \mathcal{M},A^{jb}\right]\right],A^{kc}\right] \right] - \frac12 \left[\left[\mathcal{M},A^{ja}\right], \left[\left[\mathcal{M},A^{jb}\right],A^{kc}\right]\right]\right) \Pi_{(3)}^{ab} + \ldots \, . \label{eq:dakc}
\end{eqnarray}
Here $A^{kc}$ is the baryon axial vector current operator given in Eq.~(\ref{eq:akc}), $\mathcal{M}$ is the baryon mass operator given in Eq.~(\ref{eq:mop}) and $\Pi_{(n)}^{ab}$ represents a symmetric tensor which contains meson loop integrals with the exchange of a single meson: A meson of flavor $a$ is emitted and a meson of flavor $b$ is reabsorbed. This tensor decomposes into flavor singlet, flavor $\mathbf{8}$, and flavor $\mathbf{27}$ representations as \cite{jen96}
\begin{eqnarray}
\Pi_{(n)}^{ab} = F_\mathbf{1}^{(n)} \delta^{ab} + F_\mathbf{8}^{(n)} d^{ab8} + F_\mathbf{27}^{(n)} \left[ \delta^{a8} \delta^{b8} - \frac18 \delta^{ab} - \frac35 d^{ab8} d^{888}\right], \label{eq:pisym}
\end{eqnarray}
where
\begin{eqnarray}
F_\mathbf{1}^{(n)} & = & \frac18 \left[3F^{(n)}(m_\pi,0,\mu) + 4F^{(n)}(m_K,0,\mu) + F^{(n)}(m_\eta,0,\mu) \right], \label{eq:F1} \\
F_\mathbf{8}^{(n)} & = & \frac{2\sqrt 3}{5} \left[\frac32 F^{(n)}(m_\pi,0,\mu) - F^{(n)}(m_K,0,\mu) - \frac12 F^{(n)}(m_\eta,0,\mu) \right], \label{eq:F8} \\
F_\mathbf{27}^{(n)} & = & \frac13 F^{(n)}(m_\pi,0,\mu) - \frac43 F^{(n)}(m_K,0,\mu) + F^{(n)}(m_\eta,0,\mu). \label{eq:F27}
\end{eqnarray}
Equations (\ref{eq:F1})-(\ref{eq:F27}) are linear combinations of $F^{(n)}(m_\pi,0,\mu)$, $F^{(n)}(m_K,0,\mu)$, and $F^{(n)}(m_\eta,0,\mu)$, which account for the loop integrals. Indeed, $F^{(n)}(m_\Pi,0,\mu)$ represents the degeneracy limit $\Delta/m_\Pi \to 0$ of the general function $F^{(n)}(m_\Pi,\Delta,\mu)$, defined as
\begin{equation}
F^{(n)}(m_\Pi,\Delta,\mu) \equiv \frac{\partial^n F(m_\Pi,\Delta,\mu)}{\partial \Delta^n},
\end{equation}
where $\mu$ is the scale parameter of dimensional regularization. The function $F(m_\Pi,\Delta,\mu)$ along with its derivatives are given explicitly in Appendix \ref{app:loopint}. In the degeneracy limit one finds
\begin{subequations}
\label{eq:fprimes}
\begin{eqnarray}
F^{(1)} (m_\Pi, 0, \mu) & = & - \frac{m_\Pi^2}{16\pi^2f^2} \ln{\frac{m_\Pi^2}{\mu^2}}, \label{eq:fprime} \\
F^{(2)}(m_\Pi, 0, \mu) & = & - \frac{1}{8 \pi f^2}m_\Pi, \label{eq:fprime2} \\
F^{(3)} (m_\Pi, 0, \mu) & = & \frac{1}{4\pi^2 f^2}\ln{\frac{m_\Pi^2}{\mu^2}}. \label{eq:fprime3}
\end{eqnarray}
\end{subequations}
Notice that in Eq.~(\ref{eq:fprimes}) we have kept non-analytic terms in the quark mass explicitly. Analytic terms are scheme dependent and have the same form as higher dimension terms in the chiral Lagrangian so they have been omitted. It is important to note that the one-loop correction to the axial current Eq.~(\ref{eq:dakc}) takes into account SU(3) flavor symmetry breaking. These leading non-analytic corrections to the SU(3) symmetry limit are contained in the loop integrals which depend on the meson masses and thus break SU(3) symmetry implicitly through the terms $m_\Pi^2 \ln{\frac{m_\Pi^2}{\mu^2}},m_\Pi$ and $\ln{\frac{m_\Pi^2}{\mu^2}}$.

The computation of the group theoretic structure involved in the loop graphs of Fig.~\ref{fig:eins} is quite subtle. Here we are interested in computing corrections of relative order $\mathcal{O}(1/N_c^2)$ to $A^{kc}$, which is order $\mathcal{O}(N_c)$. In other words, we need to retain terms up to order $\mathcal{O}(1/N_c^3)$ in $\delta A^{kc}$ in Eq.~(\ref{eq:dakc}). To facilitate the computation, we keep in mind two things. First, we can make use of the $1/N_c$ power-counting scheme introduced in previous works \cite{fhjm00,fh06}, which states that, for baryons with spins of order one,
\begin{equation}
T^a \sim N_c, \qquad G^{ia} \sim N_c, \qquad J^i \sim 1. \label{eq:crules}
\end{equation}
This is equivalent to state that factors of $J^i/N_c$ are $1/N_c$ suppressed relative to factors of $T^a/N_c$ and $G^{ia}/N_c$. We can safely implement this $N_c$ counting rule provided that we restrict ourselves to the lowest-lying baryon states, namely, those that constitute the $\mathbf{56}$ dimensional representation of SU(6).

Second, we should also take into account that an odd or an even number of insertions of the baryon mass operator in Eq.~(\ref{eq:dakc}) yields structures with a rather different order in $N_c$. This $N_c$-dependence was determined in Ref.~\cite{fhjm00} throughout a detailed analysis. The basic idea is quite simple: one needs to count powers of $J$ because of the $1/N_c$ suppression the factor $J/N_c$ introduces. For instance, in $A^{kc}$ and $\mathcal{M}$ the spin operator $J$ appears a minimum of 0 and 2 times, respectively. Let $r$ be the number of $J$'s from $A^{kc}$ and $\mathcal{M}$ beyond these minimum values in a given structure. Thus, contributions with no mass insertion in Eq.~(\ref{eq:dakc}) are order $\mathcal{O}(N_c^0)$ for $r=0,1$ and $\mathcal{O}(N_c^{2-r})$ for $r\geq 2$. For one mass insertion, they are order $\mathcal{O}(N_c^0)$ for $r=0,1$ and $\mathcal{O}(N_c^{1-r})$ for $r\geq 2$. For two mass insertions, they are order $\mathcal{O}(N_c^{-r})$ \cite{fhjm00}. Let us remark that this power counting already includes a $1/N_c$ suppression due to the overall factor $1/f^2$ which comes along with the loop integral.

Let us analyze the implications this power counting scheme has on the different summands in Eq.~(\ref{eq:dakc}). The first one is
\begin{equation}
\frac12 \left[A^{ja},\left[A^{jb},A^{kc}\right]\right] \Pi_{(1)}^{ab}. \label{eq:fsm}
\end{equation}
It corresponds to the degeneracy limit $\Delta/m_\Pi\to 0$ and has been analyzed in Ref.~\cite{fh06}. Naively, one would expect the double commutator alone in (\ref{eq:fsm}) to be $\mathcal{O}(N_c^3)$: one factor of $N_c$ from each $A^{kc}$. However, there are large-$N_c$ cancellations between the Feynman diagrams of Figs.~\ref{fig:eins}(a,b,c), provided that all baryon states in a complete multiplet of the large-$N_c$ SU(6) spin-flavor symmetry are included in the sum over intermediate states and that the axial coupling ratios predicted by this spin-flavor symmetry are used \cite{fhjm00}. By explicit computation, it has been shown that this double commutator is of order $\mathcal{O}(N_c)$ at most \cite{fh06}. Therefore, expression (\ref{eq:fsm}) yields an overall correction of order $\mathcal{O}(1/N_c)$ to the tree-level value if one takes into account that $f$ is of order $\mathcal{O}(\sqrt{N_c})$. By using the counting rules discussed above, the terms with $r=0,1,2$ in the product $AAA$, namely, $GGG$, $GG\mathcal{D}_2$, $G\mathcal{D}_2\mathcal{D}_2$, $GG\mathcal{D}_3$ and $GG\mathcal{O}_3$, are found to contribute to the same order to the double commutator, namely, order $\mathcal{O}(N_c)$. Thus, these terms, along with the factor $1/f^2$ from the loop integral, make up the corrections of order $\mathcal{O}(1/N_c)$ to the tree-level value --which is order $\mathcal{O}(N_c)$-- discussed above. At next subleading order, the terms with $r=3$, i.e. $\mathcal{D}_2\mathcal{D}_2\mathcal{D}_2$, $G\mathcal{D}_2\mathcal{D}_3$, and $G\mathcal{D}_2\mathcal{O}_3$, will make up corrections of order $\mathcal{O}(1/N_c^2)$ to the tree-level value, including again the factor $1/f^2$.

The second summand in Eq.~(\ref{eq:dakc}), with one mass insertion, is
\begin{equation}
- \frac12 \left\{ A^{ja}, \left[A^{kc}, \left[\mathcal{M},A^{jb}\right] \right] \right\} \Pi_{(2)}^{ab}.
\end{equation}
This is one of the corrections we are concerned with in the present work. Although the baryon mass operator $\mathcal{M}$ enters explicitly in the above expression, one is left with the hyperfine mass splitting operator $\mathcal{M}_{\textrm{hyperfine}}$ instead, because the spin-independent term in $\mathcal{M}$ in Eq.~(\ref{eq:mop}) is proportional to the identity operator and hence drops out of the commutator. According to the $N_c$ power counting rules, for $r=0,1$ the terms in the product $AAA\mathcal{M}$, namely, $GGGJ^2$ and $GG\mathcal{D}_2J^2$, produce corrections of order $\mathcal{O}(1/N_c)$ to the tree-level value, whereas at next subleading order, for $r=2$, the contributions $G\mathcal{D}_2\mathcal{D}_2J^2$, $GG\mathcal{D}_3J^2$ and $GG\mathcal{O}_3J^2$ yield corrections of order $\mathcal{O}(1/N_c^2)$ relative to the tree-level value.

Finally, for two mass insertions the expression reads
\begin{equation}
\frac16 \left(\left[A^{ja}, \left[\left[\mathcal{M}, \left[ \mathcal{M},A^{jb}\right]\right],A^{kc}\right] \right]
- \frac12 \left[\left[\mathcal{M},A^{ja}\right], \left[\left[\mathcal{M},A^{jb}\right],A^{kc}\right]\right]\right) \Pi_{(3)}^{ab}.
\end{equation}
By using the $N_c$ power counting rules, we infer that in the product $AAA\mathcal{M}\mathcal{M}$ terms with $r=0$, i.e. $GGGJ^2J^2$, and $r=1$, i.e. $GG\mathcal{D}_2J^2J^2$, will yield corrections of orders $\mathcal{O}(1/N_c)$ and $\mathcal{O}(1/N_c^2)$ to the tree-level value, respectively. Also, an interesting piece of information we can extract is that the dominant $1/N_c$ corrections from the baryon mass splittings are due to multiple insertions of the $J^2$ operator rather than contributions of powers of $J^2$. For instance, two insertions of $J^2$ --like in $GGGJ^2J^2$-- are larger (by one power of $N_c$) than one insertion of $J^4$ --like in $GGGJ^4$.

Now that we have identified the various contributions required to the order of approximation implemented in the present study, we proceed to evaluate them along the lines discussed in Ref.~\cite{fh06}. The structures involved contain $n$-body operators,\footnote{An $n$-body operator is one with $n$ $q$'s and $n$ $q^\dagger$'s. It can be written as a polynomial of order $n$ in $J^i$, $T^a$, and $G^{ia}$ \cite{djm95}.} with $n>N_c$, which are complicated commutators and/or anticommutators of the one-body operators $J^k$, $T^c$, and $G^{kc}$. All these complicated operator structures should be reduced and rewritten as linear combinations of the operator basis, with $n \leq N_c$. The reduction, although lengthy and tedious in view of the considerable amount of group theory involved, is nevertheless doable because the operator basis is complete and independent \cite{djm94,djm95}. All the necessary reductions are listed in Appendix \ref{app:reduc} for the sake of completeness.

Without further ado, the one-loop correction to $A^{kc}$ to relative order $\mathcal{O}(1/N_c^2)$ can be written as
\begin{equation}
\delta A^{kc} = \delta A_{\mathbf{1}}^{kc} + \delta A_{\mathbf{8}}^{kc} + \delta A_{\mathbf{27}}^{kc}, \label{eq:dasplit}
\end{equation}
where
\begin{equation}
\delta A_{\mathbf{1}}^{kc} = \sum_{i=1}^{7} s_i S_i^{kc}, \label{eq:das}
\end{equation}
\begin{equation}
\delta A_{\mathbf{8}}^{kc} = \sum_{i=1}^{30} o_i O_i^{kc}, \label{eq:dao}
\end{equation}
and
\begin{equation}
\delta A_{\mathbf{27}}^{kc} = \sum_{i=1}^{61} t_i T_i^{kc}. \label{eq:dat}
\end{equation}
The subscript in each summand in Eq.~(\ref{eq:dasplit}) denotes the SU(3) flavor representation it comes from.

For the singlet contribution the operator basis is
\begin{eqnarray}
\begin{array}{lll}
S_{1}^{kc} = G^{kc}, &
S_{2}^{kc} = \mathcal{D}_2^{kc}, &
S_{3}^{kc} = \mathcal{D}_3^{kc}, \\
S_{4}^{kc} = \mathcal{O}_3^{kc}, &
S_{5}^{kc} = \mathcal{D}_4^{kc}, &
S_{6}^{kc} = \mathcal{D}_5^{kc}, \\
S_{7}^{kc} = \mathcal{O}_5^{kc}, & &
\end{array}
\end{eqnarray}
and the various coefficients that enter Eq.~(\ref{eq:das}) read
\begin{eqnarray}
s_{1} & = & \left[\frac{23}{24} a_1^3 - \frac{N_c+3}{3N_c} a_1^2b_2 + \frac{N_c^2+6N_c-54}{12N_c^2} a_1b_2^2 - \frac{N_c^2+6N_c+2}{2N_c^2} a_1^2b_3 - \frac{N_c^2+6N_c-3}{2N_c^2} a_1^2c_3 - \frac{6(N_c+3)}{N_c^3} a_1b_2b_3 \right] F_{\mathbf{1}}^{(1)} \nonumber \\
& & \mbox{} + \left[\frac14 a_1^3 - \frac{N_c+3}{N_c} a_1^2b_2 - \frac{N_c^2+6N_c+6}{N_c^2} a_1^2b_3 \right] \frac{\Delta}{N_c} F_{\mathbf{1}}^{(2)} + \frac{1}{12} (N_c^2+6N_c-3) a_1^3 \frac{\Delta^2}{N_c^2} F_{\mathbf{1}}^{(3)}, \label{eq:s1}
\end{eqnarray}
\begin{eqnarray}
s_{2} & = & \left[ \frac{101}{24N_c} a_1^2b_2 + \frac{2(N_c+3)}{3N_c^2} a_1b_2^2 + \frac{N_c^2+6N_c-18}{12N_c^3} b_2^3 - \frac{3(N_c+3)}{2N_c^2} a_1^2b_3 - \frac{N_c+3}{4N_c^2} a_1^2c_3 + \frac{N_c^2+6N_c+2}{2N_c^3} a_1b_2b_3 \right. \nonumber \\
& & \mbox{} \left. - \frac{3(N_c^2+6N_c-24)}{4N_c^3} a_1b_2c_3 \right] F_{\mathbf{1}}^{(1)} + \left[ - \frac14 (N_c+3) a_1^3 - \frac{N_c^2+6N_c-29}{4N_c} a_1^2b_2 - \frac{5(N_c+3)}{N_c^2} a_1^2b_3 \right. \\
& & \mbox{} \left. - \frac{3(N_c+3)}{2N_c^2}a_1^2c_3 \right] \frac{\Delta}{N_c} F_{\mathbf{1}}^{(2)} + \left[ - \frac{11}{24}(N_c+3) a_1^3 - \frac{3(N_c^2+6N_c-16)}{8N_c} a_1^2b_2 \right] \frac{\Delta^2}{N_c^2} F_{\mathbf{1}}^{(3)},
\end{eqnarray}
\begin{eqnarray}
s_{3} & = & \left[ \frac{11}{8N_c^2} a_1b_2^2 + \frac{51}{8N_c^2} a_1^2b_3 + \frac{1} {N_c^2}a_1^2c_3 + \frac{17(N_c+3)}{6N_c^3} a_1b_2b_3 - \frac{9(N_c+3)}{4N_c^3} a_1b_2c_3 \right] F_{\mathbf{1}}^{(1)} \nonumber \\
& & \mbox{} + \left[ \frac14 a_1^3 - \frac{N_c+3}{4N_c} a_1^2b_2 - \frac{2N_c^2+12N_c-53}{4N_c^2} a_1^2b_3 + \frac{9}{4N_c^2} a_1^2c_3 \right] \frac{\Delta}{N_c} F_{\mathbf{1}}^{(2)} + \left[ \frac12 a_1^3 - \frac{19(N_c+3)}{24N_c} a_1^2b_2 \right] \frac{\Delta^2}{N_c^2} F_{\mathbf{1}}^{(3)},
\end{eqnarray}
\begin{eqnarray}
s_{4} & = & \left[\frac{3}{4N_c^2} a_1b_2^2 + \frac{7}{6N_c^2} a_1^2b_3 + \frac{167}{24N_c^2} a_1^2c_3 + \frac{5(N_c+3)}{3N_c^3} a_1b_2b_3 - \frac{N_c+3}{3N_c^3} a_1b_2c_3 \right] F_{\mathbf{1}}^{(1)} \nonumber \\
& & \mbox{} +\left[ \frac12 a_1^3 - \frac{5}{2N_c^2} a_1b_2^2 + \frac{1}{N_c^2} a_1^2b_3 - \frac{2N_c^2+12N_c-37}{4N_c^2} a_1^2c_3 \right] \frac{\Delta}{N_c} F_{\mathbf{1}}^{(2)} + \left[\frac{2}{3} a_1^3 - \frac{N_c+3}{3N_c} a_1^2b_2 \right] \frac{\Delta^2}{N_c^2} F_{\mathbf{1}}^{(3)},
\end{eqnarray}
\begin{equation}
s_{5} = \left[ \frac{5}{4N_c^3} b_2^3 + \frac{11}{6N_c^3} a_1b_2b_3 + \frac{19}{2N_c^3} a_1b_2c_3\right] F_{\mathbf{1}}^{(1)}
+ \left[\frac{1}{N_c} a_1^2b_2 - \frac{N_c+3}{2N_c^2} a_1^2b_3 - \frac{N_c+3}{2N_c^2} a_1^2c_3 \right] \frac{\Delta}{N_c} F_{\mathbf{1}}^{(2)} + \frac{49}{12N_c} a_1^2b_2 \frac{\Delta^2}{N_c^2} F_{\mathbf{1}}^{(3)},
\end{equation}
\begin{equation}
s_{6} = \left[ \frac{3}{2N_c^2} a_1^2b_3 + \frac{1}{2N_c^2} a_1^2c_3 \right] \frac{\Delta}{N_c} F_{\mathbf{1}}^{(2)},
\end{equation}
\begin{equation}
s_{7} = \frac{5}{2N_c^2} a_1^2c_3 \frac{\Delta}{N_c} F_{\mathbf{1}}^{(2)}. \label{eq:s7}
\end{equation}
The $\mathbf{8}$ and $\mathbf{27}$ contributions can be found in Appendix \ref{sec:827} for the sake of completeness.

Equations (\ref{eq:s1})-(\ref{eq:s7}) and their analogues for the $\mathbf{8}$ and $\mathbf{27}$ contributions listed in Appendix \ref{sec:827} have been rearranged to display leading and subleading terms in $1/N_c$ explicitly. Although the resultant expressions are rather breathtaking, they are indeed illustrative. It is now evident that large-$N_c$ cancellations occur in the evaluation of the structures appearing in Eq.~(\ref{eq:dakc}), both for $\Delta=0$ and $\Delta \neq 0$, such that $\delta A^{kc}$ is at most of order $\mathcal{O}(1)$, or equivalently, $\mathcal{O}(1/N_c)$ times the tree-level value. This is consistent with being a quantum correction. Also, in the definitions of the coefficients $s_i$, $o_i$ and $t_i$ we have set $m_2=\Delta$, which is a consequence of the one-to-one correspondence between the parameters of the octet and decuplet chiral Lagrangian and the coefficients of the $1/N_c$ chiral Lagrangian for $N_c=3$ \cite{jen96}, namely,
\begin{equation}
M_B = 3m_0 + \frac14 m_2, \qquad M_T = 3m_0 + \frac54 m_2,
\end{equation}
so $\Delta$ is trivially given in terms of $m_2$ for $N_c=3$.

Finally, as far as the one-loop graph of Fig.~\ref{fig:eins}(d) is concerned, it does not depend on the ratio $\Delta/m_\Pi$. Its analysis has been discussed in full in Ref.~\cite{fh06} and will not be repeated here. At any rate, this contribution is taken into account in the present analysis.

\section{The axial-vector current with perturbative SU(3) breaking}
\label{FlavorSymmetryBreaking}

One important piece of information that should be accounted for in the present analysis is the issue of perturbative SU(3) symmetry breaking for the baryon axial vector current operator $A^{kc}$. Let us recall that $A^{kc}$ is a spin-1 object and transforms as a flavor octet under SU(3). Flavor symmetry breaking also transforms as an octet under SU(3).

If we neglect isospin breaking and include first order SU(3) symmetry breaking, then $A^{kc}$ has pieces transforming according to all SU(3) representations contained in the tensor product $(1,\mathbf{8}\otimes \mathbf{8})=(1,\mathbf{1}) \oplus (1,\mathbf{8}_S) \oplus (1,\mathbf{8}_A) \oplus (1,\mathbf{10+\overline{10}}) \oplus (1,\mathbf{27})$, namely,
\begin{equation}
\delta A_{\mathrm{SB}}^{kc} = \delta A_{\mathrm{SB},\mathbf{\mathbf{1}}}^{kc} + \delta A_{\mathrm{SB},\mathbf{\mathbf{8}}}^{kc} + \delta A_{\mathrm{SB},\mathbf{\mathbf{10+\overline{10}}}}^{kc} + \delta A_{\mathrm{SB},\mathbf{\mathbf{27}}}^{kc}. \label{eq:akcsb}
\end{equation}
In principle, $\delta A_{\mathrm{SB}}^{kc}$ is of order $\mathcal{O}(\epsilon N_c)$ and can not be neglected compared to the terms retained in $A^{kc}$, Eq.~(\ref{eq:akc}). We follow the lines of Refs.~\cite{djm95,ddjm96,jen12} in order to construct the operators that occur in (\ref{eq:akcsb}) to relative order $1/N_c^2$.

\subsection{$(1,\mathbf{1})$}

The $1/N_c$ expansion for the $(1,\mathbf{1})$ operator, to relative order $1/N_c^2$, contains two terms, namely,
\begin{equation}
\delta A_{\mathrm{SB},\mathbf{1}}^{kc} = c_1^{1,\mathbf{1}} \delta^{c8}J^k + c_3^{1,\mathbf{1}} \frac{1}{N_c^2} \delta^{c8} \{J^2,J^k\}, \label{eq:ex1}
\end{equation}
where the superscripts attached to the coefficients $c_i^{1,\mathbf{1}}$ indicate the representation. Higher order terms can be obtained by anticommuting the operators retained with $J^2/N_c^2$. The above contribution is only relevant to the baryon magnetic moment operator \cite{flo09,jen12}.

\subsection{$(1,\mathbf{8})$}

The $1/N_c$ expansion for the $(1,\mathbf{8})$ operator has a similar structure as $A^{kc}$ in Eq.~(\ref{eq:akc}). Thus, the $(1,\mathbf{8})$ breaking correction reads,
\begin{equation}
\delta A_{\mathrm{SB},\mathbf{8}}^{kc} = c_1^{1,\mathbf{8}} d^{ce8} G^{ke} + b_2^{1,\mathbf{8}} \frac{1}{N_c} d^{ce8} \mathcal{D}_2^{ke} +
b_3^{1,\mathbf{8}} \frac{1}{N_c^2} d^{ce8} \mathcal{D}_3^{ke} + c_3^{1,\mathbf{8}} \frac{1}{N_c^2} d^{ce8} \mathcal{O}_3^{ke}. \label{eq:ex8}
\end{equation}
Time reversal rules out a similar series with the $d$ symbol replaced by the $f$ symbol. There is another series for the $(1,\mathbf{8})$ operator; it starts with the term
\begin{equation}
c_2^{1,\mathbf{8}} \frac{1}{N_c} f^{ce8}\epsilon^{ijk}\{J^i,G^{je}\}, \label{eq:oth8}
\end{equation}
and higher order terms can be constructed by anticommuting the leading operator with $J^2/N_c^2$. Let us notice that
\begin{equation}
f^{ce8}\epsilon^{ijk}\{J^i,G^{je}\} = [J^2,[T^8,G^{kc}]]. \label{eq:iden8}
\end{equation}
The right-hand side of Eq.~(\ref{eq:iden8}) shows that the operator only contributes to processes where both spin and strangeness are changed. These processes have not been observed, so the series (\ref{eq:oth8}) will be excluded.

\subsection{$(1,\mathbf{10+\overline{10}})$}

To relative order $1/N_c^2$, the series for the $(1,\mathbf{10+\overline{10}})$ operator contains a two- and a three-body operator, namely,
\begin{subequations}
\begin{eqnarray}
& & \{G^{kc},T^8\}-\{G^{k8},T^c\}, \\
& & \{G^{kc},\{J^r,G^{r8}\}\}-\{G^{k8},\{J^r,G^{rc}\}\},
\end{eqnarray}
\end{subequations}
which require subtractions of the flavor-octet operators \cite{djm95}. The series for the $(1,\mathbf{10+\overline{10}})$ symmetry breaking term can thus be written as
\begin{eqnarray}
\delta A_{\mathrm{SB},\mathbf{\mathbf{10+\overline{10}}}}^{kc} & = & c_2^{1,\mathbf{10+\overline{10}}} \frac{1}{N_c} \left(\{G^{kc},T^8\}-\{G^{k8},T^c\} - \frac13 f^{ce8}f^{egh}(\{G^{kg},T^h\}-\{G^{kh},T^g\})\right) \nonumber \\
&  & \mbox{} + c_3^{1,\mathbf{10+\overline{10}}} \frac{1}{N_c^2} \Bigg(\{G^{kc},\{J^r,G^{r8}\}\}-\{G^{k8},\{J^r,G^{rc}\}\} \nonumber \\
&  & \mbox{\hglue2.8truecm} - \frac13 f^{ce8}f^{egh}(\{G^{kg},\{J^r,G^{rh}\}\}-\{G^{kh},\{J^r,G^{rg}\}\}) \Bigg). \label{eq:sb8}
\end{eqnarray}
Further reductions imply that
\begin{equation}
\frac13 f^{ce8}f^{egh}(\{G^{kg},T^h\}-\{G^{kh},T^g\}) = \frac23[J^2,[T^8,G^{kc}]],
\end{equation}
and
\begin{equation}
\frac13 f^{ce8}f^{egh}(\{G^{kg},\{J^r,G^{rh}\}\}-\{G^{kh},\{J^r,G^{rg}\}\}) = \frac13 (N_c+N_f)[J^2,[T^8,G^{kc}]],
\end{equation}
so the subtracted terms in (\ref{eq:sb8}) are irrelevant as they correspond to processes where both spin and strangeness are changed.

\subsection{$(1,\mathbf{27})$}

Finally, to relative order $1/N_c^2$, the series for the $(1,\mathbf{27})$ operator contains three terms: one two-body operator and two three-body operators, which read
\begin{subequations}
\label{eq:op27sb}
\begin{eqnarray}
& & \{G^{kc},T^8\}+\{G^{k8},T^c\}, \\
& & \{J^k,\{T^c,T^8\}\}, \\
& & \{G^{kc},\{J^r,G^{r8}\}\}+\{G^{k8},\{J^r,G^{rc}\}\}.
\end{eqnarray}
\end{subequations}
These operators require subtractions of the flavor-singlet and flavor-octet pieces \cite{djm95}. The $(1,\mathbf{27})$ symmetry breaking series thus reads,
\begin{eqnarray}
\delta A_{\mathrm{SB},\mathbf{27}}^{kc} & = & c_2^{1,\mathbf{27}} \frac{1}{N_c} \left(\{G^{kc},T^8\}+\{G^{k8},T^c\} - \frac{2}{N_f^2-1} \delta^{c8} \{G^{ke},T^e\} - \frac{2N_f}{N_f^2-4} d^{ce8}d^{egh} \{G^{kg},T^h\}\right) \nonumber \\
&  & \mbox{} + c_3^{1,\mathbf{27}} \frac{1}{N_c^2} \left(\{J^k,\{T^c,T^8\}\} - \frac{1}{N_f^2-1} \delta^{c8} \{J^{k},\{T^e,T^e\}\} - \frac{N_f}{N_f^2-4} d^{ce8}d^{egh} \{J^k,\{T^g,T^h\}\} \right) \nonumber \\
&  & \mbox{} + \bar{c}_3^{1,\mathbf{27}} \frac{1}{N_c^2} \Bigg( \{G^{kc},\{J^r,G^{r8}\}\}+\{G^{k8},\{J^r,G^{rc}\}\} - \frac{2}{N_f^2-1} \delta^{c8} \{G^{ke},\{J^r,G^{re}\}\} \nonumber \\
&  & \mbox{\hglue2.5truecm} - \frac{2N_f}{N_f^2-4} d^{ce8}d^{egh} \{G^{kg},\{J^r,G^{rh}\}\} \Bigg). \label{eq:ex27}
\end{eqnarray}
Again, further reductions yield
\begin{equation}
\frac{2}{N_f^2-1} \delta^{c8} \{G^{ke},T^e\} + \frac{2N_f}{N_f^2-4} d^{ce8}d^{egh} \{G^{kg},T^h\} = \frac{2(N_c+N_f)}{N_f+2}d^{c8e}G^{ke} + \frac{2(N_c+N_f)}{N_f(N_f+1)}\delta^{c8}J^k+\frac{2}{N_f+2}d^{c8e}\mathcal{D}_2^{ke},
\end{equation}
\begin{eqnarray}
&  & \frac{1}{N_f^2-1} \delta^{c8} \{J^{k},\{T^e,T^e\}\} + \frac{N_f}{N_f^2-4} d^{ce8}d^{egh} \{J^k,\{T^g,T^h\}\} = \frac{N_c(N_c+2N_f)(N_f-2)}{N_f(N_f^2-1)}\delta^{c8}J^k \nonumber \\
&  & \mbox{} + \frac{2(N_c+N_f)(N_f-4)}{N_f^2-4}d^{c8e}\mathcal{D}_2^{ke} + \frac{2N_f}{N_f^2-4}d^{c8e}\mathcal{D}_3^{ke} +  \frac{2}{N_f^2-1}\delta^{c8}\{J^2,J^k\},
\end{eqnarray}
and
\begin{eqnarray}
&  & \frac{2}{N_f^2-1} \delta^{c8} \{G^{ke},\{J^r,G^{re}\}\} + \frac{2N_f}{N_f^2-4} d^{ce8}d^{egh} \{G^{kg},\{J^r,G^{rh}\}\} = \frac{2N_f}{N_f+2} d^{c8e}G^{ke} + \frac{(N_c+2)(N_c+2N_f-2)}{2(N_f^2-1)}\delta^{c8}J^k \nonumber \\
&  & \mbox{} + \frac{N_f(N_c+N_f)}{N_f^2-4}d^{c8e}\mathcal{D}_2^{ke} + \frac{N_f-4}{N_f^2-4}d^{c8e}\mathcal{D}_3^{ke} + \frac{2}{N_f+2}d^{c8e}\mathcal{O}_3^{ke} + \frac{N_f-2}{N_f(N_f^2-1)}\delta^{c8}\{J^2,J^k\},
\end{eqnarray}
As expected, the subtraction of flavor-singlet and flavor-octet pieces in the $1/N_c$ expansion (\ref{eq:ex27}) contains operators already defined in the series (\ref{eq:ex1}) and (\ref{eq:ex8}), such that in the $1/N_c$ expansion (\ref{eq:ex27}) we only have to keep the terms displayed in Eq.~(\ref{eq:op27sb}).

\subsection{Total correction to the baryon axial-vector current}

The baryon axial-vector current $A^{kc}$, Eq.~(\ref{eq:akc}), gets corrections due to one-loop and perturbative SU(3) symmetry breaking contributions alike. The one-loop correction, $\delta A_{\mathrm{1L}}^{kc}$, arises from Figs.~\ref{fig:eins}(a,b,c), Eq.~(\ref{eq:dasplit}), and Fig.~\ref{fig:eins}(d), discussed in Ref.~\cite{fh06}. The perturbative SU(3) breaking corrections come from Eq.~(\ref{eq:akcsb}). The overall correction to the baryon axial-vector current thus amounts to
\begin{equation}
A^{kc} + \delta A^{kc} = A^{kc} + \delta A_{\mathrm{1L}}^{kc} + \delta A_{\mathrm{SB}}^{kc}. \label{eq:totalakc}
\end{equation}

The matrix elements of the space components of $A^{kc}+\delta A^{kc}$ between SU(6) symmetric states yield the values of the axial vector couplings. Again, for the octet baryons, the axial vector couplings are $g_A$, as defined in baryon semileptonic decays, normalized such that $g_A\approx 1.27$ for neutron $\beta$ decay. For decuplet baryons, the axial vector couplings correspond to the quantities $g$, which are extracted from the widths of the strong decays of decuplet baryons into octet baryons and pions. In the next section we provide various numerical analyses in order to compare our expressions with the experimental measurements.

\section{Fitting the data\label{fits}}

In this section we perform a detailed comparison of the cumbersome expression (\ref{eq:totalakc}) with the available experimental data through some least-squares fits, in order to get information about the free parameters of the theory. The numerical analysis can be performed in several ways. We first choose to study the effects of the one-loop corrections only by comparing the theoretical expressions with the available data on baryon semileptonic decays. Then we proceed to incorporate the effects of both one-loop and perturbative SU(3) breaking corrections into the analysis, using the experimental data on baryon semileptonic decays and the strong decays of the decuplet baryons.

\subsection{Fits to the data on baryon semileptonic decays: Effects of one-loop corrections}

The available experimental data on baryon semileptonic decays is listed in Table \ref{tab:tab1} in the form of the total decay rate $R$, angular correlation coefficients $\alpha_{e\nu}$ and spin-asymmetry coefficients $\alpha_e$, $\alpha_\nu$, $\alpha_B$, $A$ and $B$, along with the measured $g_A/g_V$ ratios. A word of caution is in order here. Most data on angular correlation and asymmetry coefficients are rather old, dating back from the 80's of the past century. We have borrowed the world averages reported in Ref.~\cite{gk} for hyperon semileptonic decays. The decay rates and $g_A/g_V$ ratios, on the other hand, are found in Ref.~\cite{part}, except for the ratio $g_A/g_V$ of the $\Xi^-\to \Sigma^0$ process, which is also given in Ref.~\cite{gk}. For the $n\to p$ process, however, from present experimental results \cite{part} for the order-zero angular coefficients $B$, $A$ and $a$, we have obtained the corresponding angular coefficients $\alpha_\nu$, $\alpha_{e\nu}$ and $\alpha_e$ listed in Table \ref{tab:tab1}.

\begingroup
\squeezetable
\begin{table}
\caption{\label{tab:tab1} Experimental data on eight observed baryon semileptonic decays. The units of $R$ are $10^{-3}\, \textrm{s}^{-1}$ for neutron decay and $10^6 \, \textrm{s}^{-1}$ for the others.}
\begin{center}
\begin{tabular}{
l
r@{.}l@{\,$\pm$\,}r@{.}l r@{.}l@{\,$\pm$\,}r@{.}l
r@{.}l@{\,$\pm$\,}r@{.}l r@{.}l@{\,$\pm$\,}r@{.}l
r@{.}l@{\,$\pm$\,}r@{.}l r@{.}l@{\,$\pm$\,}r@{.}l
r@{.}l@{\,$\pm$\,}r@{.}l r@{.}l@{\,$\pm$\,}r@{.}l
} \hline \hline
&
\multicolumn{4}{c}{$n \to p e^- \overline \nu_e$} &
\multicolumn{4}{c}{$\Sigma^+ \to \Lambda e^+ \nu_e$} &
\multicolumn{4}{c}{$\Sigma^- \to \Lambda e^- \overline \nu_e$} &
\multicolumn{4}{c}{$\Lambda \to p e^- \overline \nu_e$} &
\multicolumn{4}{c}{$\Sigma^- \to n e^-\overline \nu_e$} &
\multicolumn{4}{c}{$\Xi^- \to \Lambda e^- \overline \nu_e$} &
\multicolumn{4}{c}{$\Xi^-\to \Sigma^0 e^- \overline \nu_e$} &
\multicolumn{4}{c}{$\Xi^0\to \Sigma^+ e^- \overline \nu_e$} \\
\hline
$R$ &
1 & 1362 & 0 & 0014 & 0 & 249 & 0 & 062 & 0 & 387 & 0 & 018 & 3 & 161 & 0 & 058 & 6 & 876 & 0 & 235 & 3 & 44 & 0 & 19 & 0 & 53 & 0 & 10 & 0 & 872 & 0 & 039 \\
$\alpha_{e\nu}$ &
$-$0 & 0788 & 0 & 0008 & $-$0 & 35 & 0 & 15 & $-$0 & 404 & 0 & 044 & $-$0 & 019 & 0 & 013 & 0 & 347 & 0 & 024 & 0 & 53 & 0 & 10 & \multicolumn{4}{c}{} & \multicolumn{4}{c}{} \\
$\alpha_e$ &
$-$0 & 0871 & 0 & 0010 & \multicolumn{4}{c}{} & \multicolumn{4}{c}{} & 0 & 125 & 0 & 066 & $-$0 & 519 & 0 & 104 & \multicolumn{4}{c}{} & \multicolumn{4}{c}{} & \multicolumn{4}{c}{} \\
$\alpha_\nu$ &
0 & 9875 & 0 & 0044 & \multicolumn{4}{c}{} & \multicolumn{4}{c}{} & 0 & 821 & 0 & 060 & $-$0 & 230 & 0 & 061 & \multicolumn{4}{c}{} & \multicolumn{4}{c}{} & \multicolumn{4}{c}{} \\
$\alpha_B$ &
\multicolumn{4}{c}{} & \multicolumn{4}{c}{} & \multicolumn{4}{c}{} & $-$0 & 508 & 0 & 065 & 0 & 509 & 0 & 102 & \multicolumn{4}{c}{} & \multicolumn{4}{c}{} & \multicolumn{4}{c}{} \\
$A$ &
\multicolumn{4}{c}{} & \multicolumn{4}{c}{} & 0 & 07 & 0 & 07 & \multicolumn{4}{c}{} & \multicolumn{4}{c}{} & 0 & 62 & 0 & 10 & \multicolumn{4}{c}{} & \multicolumn{4}{c}{} \\
$B$ &
\multicolumn{4}{c}{} & \multicolumn{4}{c}{} & 0 & 85 & 0 & 07 & \multicolumn{4}{c}{} & \multicolumn{4}{c}{} & \multicolumn{4}{c}{} &\multicolumn{4}{c}{} & \multicolumn{4}{c}{} \\
$g_A/g_V$ &
1 & 2701 & 0 & 0025 & \multicolumn{4}{c}{} & \multicolumn{4}{c}{} & 0 & 718 & 0 & 015 & $-$0 & 340 & 0 & 017 & 0 & 25 & 0 & 05 & 1 & 287 & 0 & 158 & 1 & 21 & 0 & 05 \\ \hline \hline
\end{tabular}
\end{center}
\end{table}
\endgroup

The theoretical expressions for the integrated observables in baryon semileptonic decays can be found in Refs.~\cite{gk,flo04,aco06}. These expressions require several inputs. First, the hadronic matrix element is written in terms of $f_1(q^2)$ and $g_1(q^2)$, the vector and axial-vector form factors, $f_2(q^2)$ and $g_2(q^2)$, the weak magnetism and electricity form factors, and $f_3(q^2)$ and $g_3(q^2)$, the induced scalar and pseudoscalar form factors, respectively, where $q^2$ is the momentum transfer squared. Time reversal invariance requires the form factors to be real. In the limit of zero momentum transfer, $f_1(0)$ and $g_1(0)$ reduce to the vector and axial-vector coupling constants $g_V$ and $g_A$, respectively.

In the limit of exact SU(3) flavor symmetry, the hadron weak vector and axial vector currents belong to SU(3) octets, so the form factors of different baryon semileptonic decays are related by SU(3) flavor symmetry and given in terms of some reduced forms factors and Clebsch-Gordan coefficients. The weak currents and the electromagnetic current are members of the same SU(3) octet, so all the vector form factors for baryon semileptonic decays are related at $q^2=0$ to the electric charges and the anomalous magnetic moments of the nucleons. Furthermore, $f_3(q^2)$ vanishes in the SU(3) symmetry limit. In turn, the leading axial-vector form factor is given in terms of two reduced form factors $F$ and $D$. Also, in the SU(3) symmetry limit $g_2=0$. Finally, $g_3(q^2)$, for electron or positron emission, has a negligible contribution to the decay rate due to the smallness of the factor $(m_e/M_B)^2$ which comes along with it.

Thus, only three form factors are relevant in the description of baryon semileptonic decays, namely, $f_1(q^2)$, $f_2(q^2)$ and $g_1(q^2)$. As for the $q^2$-dependence of the form factors, for $f_1(q^2)$ and $g_1(q^2)$ a linear expansion in $q^2$ is enough because higher powers amount to negligible contributions to the decay rate, less than a fraction of a percent, i.e., $f_1(q^2) = f_1(0) + (q^2/M_B^2) \lambda_1^f$ and $g_1(q^2) = g_1(0) + (q^2/M_B^2) \lambda_1^g$, where the slope parameters $\lambda_1^f$ and $\lambda_1^g$ are both of order unity \cite{gk}. In contrast, the $q^2$-dependence of $f_2$ can be ignored because it already contributes to order $q$ to the transition amplitude.

Second, we also should take into account in the analysis the issue of radiative corrections to the integrated observables. For practical purposes, we will include these corrections following the lines of Refs.~\cite{gk,flo04}.

Finally, we also implement the magnitudes of the CKM elements $V_{ud}$ and $V_{us}$ as recommended in Ref.~\cite{part} and for definiteness we set $\Delta = 0.231 \, \textrm{GeV}/c^2$, $f=93$ $\mathrm{MeV}/c^2$ and $\mu = 1 \, \mathrm{GeV}/c^2$.

Before we proceed with the numerical analysis, we should recall that among the available experimental data we can construct two different sets of observables. The first one is constituted by the decay rates and the $g_A/g_V$ ratios; the second one is constructed with the decay rates and the angular correlation and spin-asymmetry coefficients. Unless noted otherwise, we do not include simultaneously in the analysis the $g_A/g_V$ ratios and the angular and asymmetry coefficients, because the ratios are determined from the latter ones, i.e., these measurements are not independent.

The simplest possible fit we can perform is an SU(3) symmetric fit which involves only two parameters, namely, $a_1$ and $b_2$; this is equivalent to a fit using only $F$ and $D$ because at this level they are related as
\begin{eqnarray}
D = \frac12 a_1, \qquad F = \frac13 a_1 + \frac16 b_2.
\end{eqnarray}
By using the decay rates and the $g_A/g_V$ ratios, the best-fit values are $a_1=1.61 \pm 0.01$ and $b_2=-0.40 \pm 0.06$, or equivalently, $F=0.47 \pm 0.01$ and $D=0.81\pm 0.01$, with a $\chi^2=53.85$ for 12 degrees of freedom. Before drawing any conclusions about this rather high value of $\chi^2$, let us proceed to evaluate the effects of chiral loop corrections.

The next fit we can perform consists in neglecting the baryon mass splitting $\Delta$, which is equivalent to consider the degeneracy limit $\Delta/m_\Pi\to 0$. Actually, a fit under these assumptions was already performed in Ref.~\cite{fh06}. The primary goal of that analysis was not to be definitive about the determination of $g_A$, but rather to test the working assumptions. In the present analysis we are interested in quantifying the effects of a non-vanishing $\Delta$ on $g_A$. Thus, by using the decay rates and the $g_A/g_V$ ratios listed in Table \ref{tab:tab1}, we find $a_1 = 0.28 \pm 0.07$, $b_2 = -0.67 \pm 0.04$, $b_3 = 4.02 \pm 0.26$, and $c_3 = -13.95 \pm 2.92$, with a $\chi^2= 39.33$ for 10 degrees of freedom. Hereafter, the quoted errors of the best-fit parameters will be from the $\chi^2$ fit only, and will not include any theoretical uncertainties, unless stated otherwise. A close inspection of the output of the fit reveals that, except for $c_3$, the values of the parameters obtained are as expected from the $1/N_c$ expansion, namely, they are roughly of order 1. For $c_3$ the situation is radically different because it falls far away from any coherent expectation. Surprisingly, the effects of the loop corrections shift noticeably the values of $a_1$ and $b_2$ with respect to the SU(3) symmetric case previously discussed. This should not be a cause of concern. Actually, when loop corrections with both octet and decuplet baryons are taken into account, there appear two more coefficients $b_3$ and $c_3$, which are directly related to the couplings $\mathcal C$ and $\mathcal H$. For $N_c=3$ the relations are \cite{jen96}
\begin{eqnarray}
\begin{array}{l}
\displaystyle
D = \frac12 a_1 + \frac16 b_3, \\ [3mm]
\displaystyle
F = \frac13 a_1 + \frac16 b_2 + \frac19 b_3, \\ [3mm]
\displaystyle
\mathcal{C} = - a_1 - \frac12 c_3, \\ [3mm]
\displaystyle
\mathcal{H} = - \frac32 a_1 - \frac32 b_2 - \frac52 b_3.
\end{array} \label{eq:cid}
\end{eqnarray}
With the above values of the best-fit parameters, we get $F=0.43\pm 0.01$, $D=0.81 \pm 0.01$, $\mathcal{C}=6.70\pm 1.10$ and $\mathcal{H} = -9.47\pm 0.50$. Loop corrections thus introduce small shifts in $F$ and $D$ compared to the SU(3) symmetric fit, but the values of $\mathcal{C}$ and $\mathcal{H}$ are considerably different from expectations: $|\mathcal{C}|=1.6$ and $\mathcal{H}=-1.9\pm 0.7$, as found in Ref.~\cite{jm91b}.

\begingroup
\begin{table}
\caption{\label{tab:gasym}Predicted values of $g_A$ of some observed baryon semileptonic decays for vanishing $\Delta$. Contributions from the different SU(3) representations are listed. The decay rates and $g_A/g_V$ ratios are used in the fit.}
\begin{center}
\begin{tabular}{lrrrrrrrr}
\hline\hline
& & & \multicolumn{3}{c}{Fig.~\ref{fig:eins}(a,b,c)} & \multicolumn{3}{c}{Fig.~\ref{fig:eins}(d)} \\
Process & Total & Tree & $\mathbf{1}$ & $\mathbf{8}$ & $\mathbf{27}$ & $\mathbf{1}$ & $\mathbf{8}$ & $\mathbf{27}$ \\ \hline
$np$                 & $ 1.275$ & $ 1.238$ & $-0.403$ & $ 0.208$ & $ 0.006$ & $ 0.334$ & $-0.111$ & $ 0.002$ \\
$\Sigma^\pm \Lambda$ & $ 0.623$ & $ 0.661$ & $-0.219$ & $ 0.066$ & $-0.005$ & $ 0.179$ & $-0.059$ & $ 0.001$ \\
$\Lambda p$          & $-0.899$ & $-0.855$ & $ 0.274$ & $-0.058$ & $ 0.005$ & $-0.231$ & $-0.038$ & $ 0.005$ \\
$\Sigma^-n$          & $ 0.345$ & $ 0.381$ & $-0.134$ & $-0.024$ & $ 0.004$ & $ 0.103$ & $ 0.017$ & $-0.002$ \\
$\Xi^- \Lambda$      & $ 0.225$ & $ 0.194$ & $-0.055$ & $ 0.033$ & $-0.008$ & $ 0.053$ & $ 0.009$ & $-0.001$ \\
$\Xi^-\Sigma^0$      & $ 0.795$ & $ 0.875$ & $-0.285$ & $-0.073$ & $ 0.007$ & $ 0.236$ & $ 0.039$ & $-0.005$ \\
$\Xi^0\Sigma^+$      & $ 1.124$ & $ 1.238$ & $-0.403$ & $-0.104$ & $ 0.009$ & $ 0.334$ & $ 0.056$ & $-0.007$ \\
\hline\hline
\end{tabular}
\end{center}
\end{table}
\endgroup

As we can observe in Table \ref{tab:gasym}, the different SU(3) flavor contributions to $g_A$ follow the pattern dictated by the $1/N_c$ expansion. The singlet corrections --the most significant ones-- are roughly speaking $1/N_c$ suppressed with respect to the tree-level value. Subsequent suppressions of the octet and $\mathbf{27}$ contributions are also noticeable. Hence, in spite of the high value of $\chi^2$, the fit in the degenerate case yields predictions of $g_A$ which are in accordance with expectations. However, the price we need to pay relies in the rather high values of the parameters of the theory, which is not completely satisfactory.

As mentioned before, an analogous fit was performed in Ref.~\cite{fh06}. Our analysis here differs from the former one in two crucial aspects. First, in Ref.~\cite{fh06} we used the experimental information accessible at that time \cite{part2}. The values of $V_{ud}$ and $V_{us}$, however, have been updated along with the experimental information on the processes $n\to p$ and $\Xi^0\to \Sigma^+$ \cite{part}. These improvements introduce perceptible differences in our current analysis. Second, in Ref.~\cite{fh06} we performed a constrained fit in order to get $c_3$ from the baryon-meson coupling $|\mathcal{C}|=1.6$. Now we obtain $c_3$ from the data only, on the same footing as the other parameters $a_1$, $b_2$ and $b_3$. Hence we may say that our present numerical results supersede the ones of Ref.~\cite{fh06}.

We now proceed with the evaluation of the effects of a non-vanishing decuplet-octet mass difference $\Delta$. As in the previous fit, we use the experimental information on the decay rates and the $g_A/g_V$ ratios in order to determine the parameters $a_1$, $b_2$, $b_3$ and $c_3$. The best-fit values obtained are this time $a_1=-0.35\pm 0.02$, $b_2=-2.40\pm 0.16$, $b_3=6.53\pm 0.16$ and $c_3=5.86\pm 0.29$, with $\chi^2=17.80$ for 10 degrees of freedom. Although the values of $b_3$ and $c_3$ are slightly higher than expected, we can say that there is a noticeable improvement of the best-fit parameters in this case with respect to the previous one. Besides, $\chi^2$ reduces its value considerably to 1.78/dof, which indicates a much better fit. Also, $F=0.21\pm 0.02$, $D=0.91\pm 0.02$, $\mathcal{C}=-2.58 \pm 0.14$, and $\mathcal{H}=-12.2 \pm 0.16$. There are some rearrangements in the values of the parameters with respect to the previous case but still the output is not entirely satisfactory.

Let us now turn to the baryon axial vector couplings. The predicted values for $g_A$ are listed in Table \ref{tab:gaD}. We observe that there is an overall agreement in these predictions. The $1/N_c$ suppressions, dictated by the $1/N_c$ expansion, are evident in all the flavor contributions to $g_A$. While the singlet piece is the most significant one, the octet and $\mathbf{27}$ pieces exhibit suppressions relative to the tree-level value as expected. Still, we need to point out that the entries of process $\Xi^-\to \Lambda$ show worrisome deviations from the expected values. This behavior has been systematically observed in other analyses \cite{fjm98,fh06,flo04}.

\begingroup
\begin{table}
\caption{\label{tab:gaD}Predicted values of $g_A$ of some observed baryon semileptonic decays for non-vanishing $\Delta$. Contributions from the different SU(3) representations are listed. The decay rates and $g_A/g_V$ ratios are used in the fit.}
\begin{center}
\begin{tabular}{lrrrrrrrrrrrrrr}
\hline\hline
& & & \multicolumn{3}{c}{Fig.~\ref{fig:eins}(a,b,c), $\mathcal{O}(\Delta^0)$} & \multicolumn{3}{c}{Fig.~\ref{fig:eins}(a,b,c), $\mathcal{O}(\Delta)$}
& \multicolumn{3}{c}{Fig.~\ref{fig:eins}(a,b,c), $\mathcal{O}(\Delta^2)$} & \multicolumn{3}{c}{Fig.~\ref{fig:eins}(d)} \\
Process & Total & Tree & $\mathbf{1}$ & $\mathbf{8}$ & $\mathbf{27}$ & $\mathbf{1}$ & $\mathbf{8}$ & $\mathbf{27}$ & $\mathbf{1}$ & $\mathbf{8}$ & $\mathbf{27}$ & $\mathbf{1}$ & $\mathbf{8}$ & $\mathbf{27}$ \\ \hline
$np$ & $ 1.275$ & $ 1.121$ & $-0.550$ & $ 0.372$ & $ 0.003$ & $ 0.361$ & $-0.170$ & $-0.002$ & $-0.041$ & $-0.022$ & $ 0.000$ & $ 0.303$ & $-0.101$ & $ 0.002$ \\
$\Sigma^\pm\Lambda$ & $ 0.629$ & $ 0.745$ & $-0.364$ & $ 0.142$ & $-0.002$ & $ 0.038$ & $-0.040$ & $ 0.001$ & $-0.021$ & $-0.007$ & $ 0.000$ & $ 0.201$ & $-0.067$ & $ 0.001$ \\
$\Lambda p$ & $-0.879$ & $-0.628$ & $ 0.310$ & $-0.121$ & $ 0.003$ & $-0.404$ & $ 0.120$ & $-0.002$ & $ 0.030$ & $ 0.008$ & $ 0.000$ & $-0.170$ & $-0.028$ & $ 0.004$ \\
$\Sigma^- n$ & $ 0.340$ & $ 0.704$ & $-0.341$ & $-0.015$ & $ 0.007$ & $-0.268$ & $ 0.044$ & $-0.001$ & $-0.010$ & $ 0.002$ & $ 0.000$ & $ 0.190$ & $ 0.032$ & $-0.004$ \\
$\Xi^-\Lambda$ & $ 0.361$ & $-0.117$ & $ 0.054$ & $ 0.159$ & $-0.014$ & $ 0.366$ & $-0.041$ & $ 0.003$ & $-0.009$ & $-0.005$ & $ 0.001$ & $-0.032$ & $-0.005$ & $ 0.001$ \\
$\Xi^-\Sigma^0$ & $ 0.820$ & $ 0.793$ & $-0.389$ & $-0.132$ & $ 0.012$ & $ 0.255$ & $ 0.060$ & $-0.002$ & $-0.029$ & $ 0.008$ & $-0.001$ & $ 0.214$ & $ 0.036$ & $-0.005$ \\
$\Xi^0\Sigma^+$ & $ 1.160$ & $ 1.121$ & $-0.550$ & $-0.186$ & $ 0.017$ & $ 0.361$ & $ 0.085$ & $-0.003$ & $-0.041$ & $ 0.011$ & $-0.001$ & $ 0.303$ & $ 0.050$ & $-0.007$ \\
\hline\hline
\end{tabular}
\end{center}
\end{table}
\endgroup

Next, we provide in Table \ref{tab:ds1pre} the observables obtained with the best-fit parameters in order to compare them with the experimental values displayed in Table \ref{tab:tab1}. The most important deviations between theory and experiment arise from the decay rates of the processes $\Sigma^-\to \Lambda$, $\Lambda\to p$ and $\Xi^0\to \Sigma^+$ whose contributions to the total $\chi^2$ amount to $\chi^2_{\Sigma^-\Lambda}=5.31$, $\chi^2_{\Lambda p}=2.37$ and $\chi^2_{\Xi^0 \Sigma^+}=1.78$, respectively, and the $g_A/g_V$ ratio of $n\to p$, which contributes with $\chi^2_{np}=3.87$ to $\chi^2$.

\begingroup
\begin{table}
\caption{\label{tab:ds1pre} Values of predicted observables for eight observed baryon semileptonic decays. The units of $R$ are $10^{-3}\, \textrm{s}^{-1}$ for neutron decay and $10^6 \, \textrm{s}^{-1}$ for the others. The decay rates and $g_A/g_V$ ratios are used in the fit.}
\begin{center}
\begin{tabular}{lrrrrrrrr}
 \hline \hline
&
$n \to p e^- \overline \nu_e$ &
$\Sigma^+ \to \Lambda e^+ \nu_e$ &
$\Sigma^- \to \Lambda e^- \overline \nu_e$ &
$\Lambda \to p e^- \overline \nu_e$ &
$\Sigma^- \to n e^-\overline \nu_e$ &
$\Xi^- \to \Lambda e^- \overline \nu_e$ &
$\Xi^-\to \Sigma^0 e^- \overline \nu_e$ &
$\Xi^0\to \Sigma^+ e^- \overline \nu_e$ \\
\hline
$R$
& $1.135$ & $0.257$ & $0.428$ & $3.250$ & $6.741$ & $3.309$ & $0.454$ & $0.820$ \\
$\alpha_{e\nu}$
& $-0.080$ & $-0.406$ & $-0.414$ & $-0.026$ & $0.340$ & $0.500$ &  & \\
$\alpha_e$
& $-0.089$ & & & $0.018$ & $-0.630$ &  &  & \\
$\alpha_\nu$
& $0.987$ & & & $0.977$ & $-0.352$ &  &  &  \\
$\alpha_B$
& & & & $-0.590$ & $0.665$ &  & &  \\
$A$
& & & $0.046$ &  &  & $0.641$ &  & \\
$B$
& & & $0.888$ & & & & & \\
$g_A/g_V$
& $1.271$ & & & $0.717$ & $-0.340$ & $0.293$ & $1.162$ & $1.162$ \\
\hline \hline
\end{tabular}
\end{center}
\end{table}
\endgroup

Now we can redo the analysis by using the other set of experimental information discussed above, namely, the one constituted by the decay rates and the angular correlation and spin-asymmetry coefficients. At this stage we have at our disposal 8 decay rates and 17 coefficients. We have two more pieces of information available, the $g_A/g_V$ ratios of the processes $\Xi^-\to\Sigma^0$ and $\Xi^0\to \Sigma^+$; we will use them because their coefficients have not been measured yet.

As in the previous case, we proceed to perform the comparison between theory and experiment in the limit of vanishing $\Delta$. The fit produces $a_1 = 0.30 \pm 0.06$, $b_2 = -0.65 \pm 0.03$, $b_3 = 3.92 \pm 0.24$, and $c_3 = -13.79 \pm 2.17$, with a $\chi^2= 62.62$ for 23 degrees of freedom. We observe that the values of the best-fit parameters do not change substantially with respect to their analogues when the decay rates and $g_A/g_V$ ratios are used. Still, the differences, although small, are perceptible. The predicted $g_A$ are listed in Table \ref{tab:drcoef1} for the sake of completeness. The different flavor contributions behave in the same way as in Table \ref{tab:gasym}.

\begingroup
\begin{table}
\caption{\label{tab:drcoef1}Predicted values of $g_A$ of some observed baryon semileptonic decays for vanishing $\Delta$. Contributions from the different SU(3) representations are listed. The decay rates and angular correlation and spin-asymmetry coefficients are used in the fit.}
\begin{center}
\begin{tabular}{lrrrrrrrr}
\hline\hline
& & & \multicolumn{3}{c}{Fig.~\ref{fig:eins}(a,b,c)} & \multicolumn{3}{c}{Fig.~\ref{fig:eins}(d)} \\
Process & Total & Tree & $\mathbf{1}$ & $\mathbf{8}$ & $\mathbf{27}$ & $\mathbf{1}$ & $\mathbf{8}$ & $\mathbf{27}$ \\ \hline
$np$                 & $ 1.275$ & $ 1.236$ & $-0.404$ & $ 0.211$ & $ 0.007$ & $ 0.334$ & $-0.111$ & $ 0.002$ \\
$\Sigma^\pm \Lambda$ & $ 0.625$ & $ 0.659$ & $-0.221$ & $ 0.072$ & $-0.005$ & $ 0.178$ & $-0.059$ & $ 0.001$ \\
$\Lambda p$          & $-0.897$ & $-0.855$ & $ 0.274$ & $-0.057$ & $ 0.005$ & $-0.231$ & $-0.038$ & $ 0.005$ \\
$\Sigma^-n$          & $ 0.335$ & $ 0.378$ & $-0.137$ & $-0.028$ & $ 0.004$ & $ 0.102$ & $ 0.017$ & $-0.002$ \\
$\Xi^- \Lambda$      & $ 0.233$ & $ 0.196$ & $-0.053$ & $ 0.038$ & $-0.009$ & $ 0.053$ & $ 0.009$ & $-0.001$ \\
$\Xi^-\Sigma^0$      & $ 0.791$ & $ 0.874$ & $-0.286$ & $-0.075$ & $ 0.007$ & $ 0.236$ & $ 0.039$ & $-0.005$ \\
$\Xi^0\Sigma^+$      & $ 1.119$ & $ 1.236$ & $-0.404$ & $-0.106$ & $ 0.010$ & $ 0.334$ & $ 0.056$ & $-0.007$ \\
\hline\hline
\end{tabular}
\end{center}
\end{table}
\endgroup

When a non-vanishing $\Delta$ is considered, the fit yields $a_1 =-0.36 \pm 0.02$, $b_2 =-2.50 \pm 0.15$, $b_3 = 6.64 \pm 0.15$, and $c_3 = 5.81 \pm 0.25$, with a $\chi^2= 36.10$ for 23 degrees of freedom. The predicted $g_A$ are given in Table \ref{tab:drcoef2}. There are small but perceptible differences between the entries of this Table \ref{tab:drcoef2} and those of Table \ref{tab:gaD}.

\begingroup
\begin{table}
\caption{\label{tab:drcoef2}Predicted values of $g_A$ of some observed baryon semileptonic decays for non-vanishing $\Delta$. Contributions from the different SU(3) representations are listed. The decay rates and angular correlation and spin-asymmetry coefficients are used in the fit.}
\begin{center}
\begin{tabular}{lrrrrrrrrrrrrrr}
\hline\hline
& & & \multicolumn{3}{c}{Fig.~\ref{fig:eins}(a,b,c), $\mathcal{O}(\Delta^0)$} & \multicolumn{3}{c}{Fig.~\ref{fig:eins}(a,b,c), $\mathcal{O}(\Delta)$}
& \multicolumn{3}{c}{Fig.~\ref{fig:eins}(a,b,c), $\mathcal{O}(\Delta^2)$} & \multicolumn{3}{c}{Fig.~\ref{fig:eins}(d)} \\
Process & Total & Tree & $\mathbf{1}$ & $\mathbf{8}$ & $\mathbf{27}$ & $\mathbf{1}$ & $\mathbf{8}$ & $\mathbf{27}$ & $\mathbf{1}$ & $\mathbf{8}$ & $\mathbf{27}$ & $\mathbf{1}$ & $\mathbf{8}$ & $\mathbf{27}$ \\ \hline
$np$ & $ 1.275$ & $ 1.125$ & $-0.580$ & $ 0.389$ & $ 0.003$ & $ 0.388$ & $-0.182$ & $-0.003$ & $-0.046$ & $-0.024$ & $ 0.000$ & $ 0.304$ & $-0.101$ & $ 0.002$ \\
$\Sigma^\pm\Lambda$ & $ 0.630$ & $ 0.755$ & $-0.382$ & $ 0.152$ & $-0.002$ & $ 0.041$ & $-0.042$ & $ 0.001$ & $-0.023$ & $-0.007$ & $ 0.000$ & $ 0.204$ & $-0.068$ & $ 0.001$ \\
$\Lambda p$ & $-0.874$ & $-0.623$ & $ 0.328$ & $-0.125$ & $ 0.003$ & $-0.434$ & $ 0.129$ & $-0.002$ & $ 0.033$ & $ 0.009$ & $ 0.000$ & $-0.168$ & $-0.028$ & $ 0.004$ \\
$\Sigma^- n$ & $ 0.332$ & $ 0.724$ & $-0.355$ & $-0.018$ & $ 0.007$ & $-0.289$ & $ 0.048$ & $-0.001$ & $-0.011$ & $ 0.003$ & $ 0.000$ & $ 0.196$ & $ 0.033$ & $-0.004$ \\
$\Xi^-\Lambda$ & $ 0.373$ & $-0.132$ & $ 0.053$ & $ 0.168$ & $-0.015$ & $ 0.394$ & $-0.044$ & $ 0.003$ & $-0.009$ & $-0.006$ & $ 0.001$ & $-0.036$ & $-0.006$ & $ 0.001$ \\
$\Xi^-\Sigma^0$ & $ 0.819$ & $ 0.795$ & $-0.410$ & $-0.138$ & $ 0.013$ & $ 0.274$ & $ 0.064$ & $-0.003$ & $-0.032$ & $ 0.009$ & $-0.001$ & $ 0.215$ & $ 0.036$ & $-0.005$ \\
$\Xi^0\Sigma^+$ & $ 1.158$ & $ 1.125$ & $-0.580$ & $-0.195$ & $ 0.019$ & $ 0.388$ & $ 0.091$ & $-0.004$ & $-0.046$ & $ 0.012$ & $-0.001$ & $ 0.304$ & $ 0.051$ & $-0.007$ \\
\hline\hline
\end{tabular}
\end{center}
\end{table}
\endgroup

We conclude this stage of the comparison by providing in Table \ref{tab:ds2pre} the observables obtained with the best-fit parameters in this last case in order to compare them with the experimental values given in Table \ref{tab:tab1}. The highest
deviations between theory and experiment come from $R$ from $\Sigma^- \to \Lambda$ $(\chi^2_{\Sigma^-\Lambda}=5.71)$, $\alpha_\nu$ from $\Lambda\to p$ $(\chi^2_{\Lambda p}=6.68)$ and $\Sigma^-\to n$ $(\chi^2_{\Sigma^- n}=4.11)$ and from $\alpha_e$ in the processes $n\to p$ $(\chi^2_{np}=2.10)$ and $\Lambda\to p$ $(\chi^2_{\Lambda p}=2.54)$, which together amount half the total $\chi^2$.

In summary, although the inclusion of one-loop corrections, both for vanishing and non-vanishing decuplet-octet mass difference, in the numerical analysis displays some interesting trends, the output of the fits, particularly translated into the determination of the axial-vector couplings $D$, $F$, $\mathcal{C}$, and $\mathcal{H}$ are not entirely satisfactory. We then learn that perturbative SU(3) symmetry breaking should play an important role in the numerical analysis, as we discuss in the next section.

\begingroup
\begin{table}
\caption{\label{tab:ds2pre} Values of predicted observables for eight observed baryon semileptonic decays. The units of $R$ are $10^{-3}\, \textrm{s}^{-1}$ for neutron decay and $10^6 \, \textrm{s}^{-1}$ for the others. The decay rates and angular correlation and spin-asymmetry coefficients are used in the fit.
}
\begin{center}
\begin{tabular}{lrrrrrrrr}
 \hline \hline
&
$n \to p e^- \overline \nu_e$ &
$\Sigma^+ \to \Lambda e^+ \nu_e$ &
$\Sigma^- \to \Lambda e^- \overline \nu_e$ &
$\Lambda \to p e^- \overline \nu_e$ &
$\Sigma^- \to n e^-\overline \nu_e$ &
$\Xi^- \to \Lambda e^- \overline \nu_e$ &
$\Xi^-\to \Sigma^0 e^- \overline \nu_e$ &
$\Xi^0\to \Sigma^+ e^- \overline \nu_e$ \\
\hline
$R$
& $1.135$ & $0.257$ & $0.430$ & $3.230$ & $6.661$ & $3.355$ & $0.453$ & $0.818$ \\
$\alpha_{e\nu}$
& $-0.080$ & $-0.406$ & $-0.414$ & $-0.022$ & $0.352$ & $0.485$ &  & \\
$\alpha_e$
& $-0.089$ & & & $0.020$ & $-0.618$ &  &  &  \\
$\alpha_\nu$
& $0.987$ & & & $0.976$ & $-0.354$ &  &  &  \\
$\alpha_B$
& & & & $-0.590$ & $0.658$ &  &  &  \\
$A$
& & & $0.046$ &  &  & $0.653$ &  & \\
$B$
& & & $0.888$ & & & & & \\
$g_A/g_V$
& $1.275$ & & & $0.714$ & $-0.332$ & $0.304$ & $1.158$ & $1.158$ \\
\hline \hline
\end{tabular}
\end{center}
\end{table}
\endgroup

\subsection{Fits to the data on the $\beta$ decay of the octet baryons and the strong decays of the decuplet baryons: Inclusion of both chiral and perturbative breaking corrections}

The off-diagonal matrix elements of $A^{kc}+\delta A^{kc}$ involving decuplet and octet baryons can be obtained through the strong transitions of decuplet baryons to octet baryons and pions. The available experimental information on the kinematically allowed strong decays $\Delta \to N\pi$, $\Sigma^*\to\Lambda\pi$, $\Sigma^*\to\Sigma\pi$, $\Xi^*\to\Xi\pi$ can be found in Ref.~\cite{part} in the form of widths.

The formalism to obtain the widths of the strong decays of the baryon decuplet within chiral perturbation theory was originally introduced by Peccei \cite{pecc} and further implemented in the analysis of Ref.~\cite{ddjm96}. In this formalism, the width of a decuplet baryon $B^\prime$ decaying into an octet baryon $B$ and a pion is given by
\begin{equation}
\Gamma_{B^\prime} = \frac{g^2 C(B,B^\prime)^2 (E_B+M_B) |\mathbf{q}|^3}{24\pi f^2 M_{B^\prime}}, \label{eq:decw}
\end{equation}
where $g$ is the axial vector coupling for this decay, $C(B,B^\prime)$ is a Clebsch-Gordan coefficient which equals to $\{1,1/\sqrt{2},1/\sqrt{3},1/\sqrt{2}\}$ for $\{\Delta \to N\pi, \Sigma^*\to\Lambda\pi, \Sigma^*\to\Sigma\pi, \Xi^*\to\Xi\pi\}$, $M_{B^\prime}$, $M_B$ are the masses of the decuplet and octet baryons, respectively, $f$ is the pion decay constant and $E_B$ and $\mathbf{q}$ are the octet baryon energy and the pion three-momentum in the rest frame of $B^\prime$, respectively. With the help of Eq.~(\ref{eq:decw}), the axial vector couplings $g$ can be determined for each decay and are listed in Table \ref{tab:decwidth}. Notice that the Clebsch-Gordan coefficients have been chosen in such a way that the $g$ couplings are all equal in the limit of exact SU(3) symmetry \cite{ddjm96}.

\begingroup
\squeezetable
\begin{table}
\caption{\label{tab:decwidth} Value of the axial vector coupling $g$ extracted from the widths of the strong decays of the decuplet baryons \cite{part} by using Eq.~(\ref{eq:decw}).
}
\begin{center}
\begin{tabular}{
l
r@{.}l@{\,$\pm$\,}r@{.}l r@{.}l@{\,$\pm$\,}r@{.}l
r@{.}l@{\,$\pm$\,}r@{.}l r@{.}l@{\,$\pm$\,}r@{.}l
r@{.}l@{\,$\pm$\,}r@{.}l r@{.}l@{\,$\pm$\,}r@{.}l
r@{.}l@{\,$\pm$\,}r@{.}l r@{.}l@{\,$\pm$\,}r@{.}l
} \hline \hline
Decay &
\multicolumn{4}{c}{$\Delta \to N\pi$} & \multicolumn{4}{c}{$\Sigma^*\to\Lambda\pi$} & \multicolumn{4}{c}{$\Sigma^*\to\Sigma\pi$} & \multicolumn{4}{c}{$\Xi^*\to\Xi\pi$} \\
\hline
$g$ & $-2$ & 04 & 0 & 01 & $-1$ & 69 & 0 & 02 & $-1$ & 59 & 0 & 10 & $-1$ & 46 & 0 & 04 \\ \hline \hline
\end{tabular}
\end{center}
\end{table}
\endgroup

At this stage we have four extra pieces of experimental information at our disposal. In order to perform a comparison between theory and experiment in a more pragmatic manner, we will perform a global fit by using the experimental information on $g_A$ and $g$ only, this time including both chiral and perturbative symmetry breaking corrections into our evaluation. The fits discussed in the previous section taught us that perturbative symmetry breaking, being of order $\mathcal{O}(\epsilon N_c)$, might have an important effect in the determination of the axial-vector couplings. Practically, they are of the same order as the singlet contribution in the loop corrections and should be retained.

Perturbative flavor SU(3) symmetry breaking involves several extra free parameters in the analysis. In order to overcome this difficulty, we will adopt the following strategy. We will incorporate perturbative symmetry breaking corrections up to next-to-leading order, i.e., we only retain order $\mathcal{O}(N_c^0)$ corrections to $A^{kc}$ in the $1/N_c$ expansion (\ref{eq:akcsb}), otherwise we loose predictive power. We also include SU(3) breaking in the strangeness zero sector only. Under these assumptions, $A^{kc}+\delta_{\mathrm{SB}}^{kc}$ takes on the simplified form
\begin{eqnarray}
A^{kc}+\delta_{\mathrm{SB}}^{kc} & = & a_1 G^{kc} + b_2 \frac{1}{N_c} \mathcal{D}_2^{kc} + b_3 \frac{1}{N_c^2} \mathcal{D}_3^{kc} + c_3 \frac{1}{N_c^2} \mathcal{O}_3^{kc} + W_{\Delta S}\Bigg[ d_1 d^{c8e} G^{ke} + d_2 \frac{1}{N_c} d^{c8e} \mathcal{D}_2^{ke} \nonumber \\
&  & \mbox{}  + d_3 \frac{1}{N_c} \left( \{G^{kc},T^8\}-\{G^{k8},T^c\} \right) +
d_4 \frac{1}{N_c} \left( \{G^{kc},T^8\}+\{G^{k8},T^c\} \right) \Bigg], \label{eq:akcsbfit}
\end{eqnarray}
where $W_{\Delta S}=1$ for $\Delta S=1$ processes ($c=4\pm i 5$) and vanishes for $\Delta S=0$ processes ($c=1\pm i2$). Let us notice that the $d_i$ coefficients have been redefined according to the discussion of the previous section.

In order to perform the fit in the limit $\Delta \to 0$ in a consistent fashion, we should set $b_3=0$ in Eq.~(\ref{eq:akcsbfit}) and remove from $\delta A_{\mathrm{1L}}^{kc}$ all the terms of relative order $1/N_c^3$, namely, those terms proportional to $b_2^3$, $a_1b_2b_3$, and $a_1b_2c_3$, and set also $b_3=0$ in there. Keeping the $c_3$ term in Eq.~(\ref{eq:akcsbfit}) will avoid mixing up symmetry breaking effects with $1/N_c$ corrections in the symmetric couplings $D$, $F$, $\mathcal{C}$, and $\mathcal{H}$ \cite{ddjm96}. As for the nonvanishing $\Delta$ case, we should keep Eq.~(\ref{eq:akcsbfit}) the way it stands due to the next-to-next-to-leading order contributions that come along with $\Delta$.

Proceeding with the fit in the limit of vanishing $\Delta$, using the experimental data on $g_A$ (or, alternatively, $g_A/g_V$), the fit yields $a_1= 1.00 \pm 0.02$, $b_2=0.73\pm 0.06$, $b_3=0.0$, and $c_3=0.82\pm 0.04$, and $d_1=-0.67\pm 0.04$, $d_2=6.66 \pm 0.41$, $d_3= 0.09 \pm 0.03$, and $d_4= -0.01 \pm 0.06$ with $\chi^2=14.96$ for three degrees of freedom. The highest contributions to $\chi^2$ come from $g_A$ of the process $\Xi^0\Sigma^+$ ($\chi^2_{\Xi^0\Sigma^+}=4.88)$ and $g$ from the processes $\Sigma^*\Sigma$ and $\Xi^*\Xi$ ($\chi^2_{\Sigma^*\Sigma}= 3.15$ and $\chi^2_{\Xi^*\Xi}=3.33$). With these best-fit parameters, the SU(3) symmetric couplings become $D=0.50 \pm 0.01$, $F=0.45 \pm 0.01$, $\mathcal{C}=-1.41 \pm 0.01$, and $\mathcal{H} =-2.59 \pm 0.07$. Also, the various symmetry breaking contributions to $g_A$ and $g$ are listed in Table \ref{tab:avcds0} for the sake of completeness.

\begingroup
\begin{table}
\caption{\label{tab:avcds0}Predicted axial vector couplings for vanishing $\Delta$. The experimental information on $g_A$ and $g$ are used in the fit. Note that SU(3) flavor symmetry breaking is taken into account in two ways: explicitly through perturbative symmetry breaking (SB) and implicitly through the integrals occurring in the one-loop corrections.}
\begin{center}
\begin{tabular}{lrrrrrrrrr}
\hline\hline
& & & & \multicolumn{3}{c}{Fig.~\ref{fig:eins}(a,b,c)} & \multicolumn{3}{c}{Fig.~\ref{fig:eins}(d)} \\
Process & Total & Tree & SB & $\mathbf{1}$ & $\mathbf{8}$ & $\mathbf{27}$ & $\mathbf{1}$ & $\mathbf{8}$ & $\mathbf{27}$ \\ \hline
$np$                 & $ 1.270$ & $ 0.953$ & $ 0.000$ & $ 0.064$ & $ 0.082$ & $-0.003$ & $ 0.257$ & $-0.086$ & $ 0.002$ \\
$\Sigma^\pm \Lambda$ & $ 0.309$ & $ 0.407$ & $-0.158$ & $-0.018$ & $ 0.004$ & $-0.001$ & $ 0.110$ & $-0.037$ & $ 0.001$ \\
$\Lambda p$          & $-0.903$ & $-0.760$ & $ 0.254$ & $-0.097$ & $-0.063$ & $-0.003$ & $-0.205$ & $-0.034$ & $ 0.005$ \\
$\Sigma^-n$          & $ 0.349$ & $ 0.045$ & $ 0.371$ & $-0.108$ & $ 0.025$ & $ 0.003$ & $ 0.012$ & $ 0.002$ & $ 0.000$ \\
$\Xi^- \Lambda$      & $ 0.301$ & $ 0.353$ & $-0.328$ & $ 0.114$ & $ 0.051$ & $ 0.002$ & $ 0.095$ & $ 0.016$ & $-0.002$ \\
$\Xi^-\Sigma^0$      & $ 0.778$ & $ 0.674$ & $-0.122$ & $ 0.046$ & $-0.029$ & $ 0.001$ & $ 0.182$ & $ 0.030$ & $-0.004$ \\
$\Xi^0\Sigma^+$      & $ 1.100$ & $ 0.953$ & $-0.172$ & $ 0.064$ & $-0.041$ & $ 0.001$ & $ 0.257$ & $ 0.043$ & $-0.006$ \\
$\Delta N$           & $-2.039$ & $-1.409$ & $ 0.000$ & $-0.412$ & $ 0.031$ & $ 0.008$ & $-0.381$ & $ 0.127$ & $-0.003$ \\
$\Sigma^*\Lambda$    & $-1.680$ & $-1.409$ & $ 0.388$ & $-0.412$ & $ 0.012$ & $-0.002$ & $-0.381$ & $ 0.127$ & $-0.003$ \\
$\Sigma^*\Sigma$     & $-1.413$ & $-1.409$ & $ 0.505$ & $-0.412$ & $ 0.159$ & $ 0.000$ & $-0.381$ & $ 0.127$ & $-0.003$ \\
$\Xi^*\Xi$           & $-1.533$ & $-1.409$ & $ 0.491$ & $-0.412$ & $ 0.067$ & $-0.013$ & $-0.381$ & $ 0.127$ & $-0.003$ \\
\hline\hline
\end{tabular}
\end{center}
\end{table}
\endgroup
We find a fairly good agreement between the predicted and the observed couplings $g_A$ and $g$. Also, the pattern dictated by the various breaking pieces are in accordance with expectations. Of particular interest are the values of SU(3) symmetry breaking listed in the third column from left to right in that table.

When we finally redo the fit in the limit of nonvanishing $\Delta$, we find $a_1= 0.64\pm 0.22$, $b_2=0.21\pm 25$, $b_3= 1.35\pm 0.06$, and $c_3=1.90\pm 0.41$, and $d_1= -0.44\pm 0.12$, $d_2= 6.48\pm 0.37 $, $d_3= 0.04\pm 0.03$, and $d_4= 0.08 \pm 0.07$ with $\chi^2=2.28$ for two degrees of freedom. The highest contribution to $\chi^2$ come from $g_A$ of the process $\Xi^-\Lambda$ ($\chi^2_{\Xi^-\Lambda}=1.58)$. Similarly, with the best-fit parameters in this case we find $D= 0.54\pm 0.03$, $F= 0.40\pm 0.03$, $\mathcal{C}= -1.59\pm 0.05 $, and $\mathcal{H} = -4.64 \pm 1.30$. Also, the several breaking contributions to $g_A$ and $g$ are listed in Table \ref{tab:avcds1} for the sake of completeness. Let us remark that in Tables \ref{tab:avcds0} and \ref{tab:avcds1} the total correction to the (flavor-symmetric) tree-level value includes SU(3) flavor symmetry breaking in two ways: explicitly through perturbative symmetry breaking (SB) and implicitly through the integrals occurring in the one-loop corrections.

\begingroup
\begin{table}
\caption{\label{tab:avcds1}Predicted axial vector couplings for non-vanishing $\Delta$. The
experimental information on $g_A$ and $g$ are used in the fit. Note that SU(3) flavor symmetry breaking is taken into account in two ways: explicitly through perturbative symmetry breaking (SB) and implicitly through the integrals occurring in the one-loop corrections.}
\begin{center}
\begin{tabular}{lrrrrrrrrrrrrrrr}
\hline\hline
& & & & \multicolumn{3}{c}{Fig.~\ref{fig:eins}(a,b,c), $\mathcal{O}(\Delta^0)$} & \multicolumn{3}{c}{Fig.~\ref{fig:eins}(a,b,c), $\mathcal{O}(\Delta)$}
& \multicolumn{3}{c}{Fig.~\ref{fig:eins}(a,b,c), $\mathcal{O}(\Delta^2)$} & \multicolumn{3}{c}{Fig.~\ref{fig:eins}(d)} \\
Process & Total & Tree & SB & $\mathbf{1}$ & $\mathbf{8}$ & $\mathbf{27}$ & $\mathbf{1}$ & $\mathbf{8}$ & $\mathbf{27}$ & $\mathbf{1}$ & $\mathbf{8}$ & $\mathbf{27}$ & $\mathbf{1}$ & $\mathbf{8}$ & $\mathbf{27}$ \\ \hline
$np$                 & $ 1.270$ & $ 0.939$ & $ 0.000$ & $-0.115$ & $ 0.110$ & $ 0.001$ & $ 0.330$ & $-0.161$ & $-0.003$ & $-0.003$ & $ 0.000$ & $ 0.000$ & $ 0.254$ & $-0.084$ & $ 0.002$ \\
$\Sigma^\pm \Lambda$ & $ 0.389$ & $ 0.443$ & $-0.104$ & $-0.043$ & $ 0.010$ & $-0.001$ & $ 0.048$ & $-0.042$ & $ 0.002$ & $-0.008$ & $ 0.004$ & $ 0.000$ & $ 0.120$ & $-0.040$ & $ 0.001$ \\
$\Lambda p$          & $-0.881$ & $-0.707$ & $ 0.286$ & $ 0.098$ & $-0.078$ & $ 0.000$ & $-0.356$ & $ 0.100$ & $-0.002$ & $-0.004$ & $ 0.000$ & $ 0.000$ & $-0.191$ & $-0.032$ & $ 0.004$ \\
$\Sigma^-n$          & $ 0.337$ & $ 0.145$ & $ 0.333$ & $ 0.010$ & $-0.003$ & $ 0.002$ & $-0.212$ & $ 0.038$ & $-0.002$ & $-0.016$ & $-0.004$ & $ 0.000$ & $ 0.039$ & $ 0.007$ & $-0.001$ \\
$\Xi^- \Lambda$      & $ 0.230$ & $ 0.265$ & $-0.375$ & $-0.055$ & $ 0.028$ & $-0.001$ & $ 0.308$ & $-0.039$ & $ 0.003$ & $ 0.012$ & $ 0.003$ & $ 0.000$ & $ 0.072$ & $ 0.012$ & $-0.002$ \\
$\Xi^-\Sigma^0$      & $ 0.871$ & $ 0.664$ & $-0.166$ & $-0.081$ & $-0.039$ & $ 0.002$ & $ 0.234$ & $ 0.057$ & $-0.002$ & $-0.002$ & $ 0.000$ & $ 0.000$ & $ 0.179$ & $ 0.030$ & $-0.004$ \\
$\Xi^0\Sigma^+$      & $ 1.232$ & $ 0.939$ & $-0.234$ & $-0.115$ & $-0.055$ & $ 0.003$ & $ 0.330$ & $ 0.081$ & $-0.003$ & $-0.003$ & $ 0.000$ & $ 0.000$ & $ 0.254$ & $ 0.042$ & $-0.006$ \\
$\Delta N$           & $-2.040$ & $-1.587$ & $ 0.000$ & $-0.226$ & $-0.050$ & $ 0.008$ & $ 0.183$ & $-0.121$ & $-0.011$ & $ 0.042$ & $ 0.013$ & $ 0.000$ & $-0.429$ & $ 0.143$ & $-0.003$ \\
$\Sigma^*\Lambda$    & $-1.693$ & $-1.587$ & $ 0.255$ & $-0.226$ & $-0.044$ & $-0.005$ & $ 0.183$ & $-0.036$ & $ 0.006$ & $ 0.042$ & $ 0.008$ & $ 0.000$ & $-0.429$ & $ 0.143$ & $-0.003$ \\
$\Sigma^*\Sigma$     & $-1.530$ & $-1.587$ & $ 0.208$ & $-0.226$ & $ 0.210$ & $ 0.013$ & $ 0.183$ & $-0.089$ & $ 0.005$ & $ 0.042$ & $ 0.000$ & $ 0.000$ & $-0.429$ & $ 0.143$ & $-0.003$ \\
$\Xi^*\Xi$           & $-1.460$ & $-1.587$ & $ 0.296$ & $-0.226$ & $ 0.090$ & $-0.017$ & $ 0.183$ & $ 0.023$ & $ 0.028$ & $ 0.042$ & $-0.001$ & $ 0.000$ & $-0.429$ & $ 0.143$ & $-0.003$ \\
\hline\hline
\end{tabular}
\end{center}
\end{table}
\endgroup

All along, the fits performed in this work point in one direction. While the values of the parameters $a_1$ and $b_2$ appear to be quite stable, the parameters $\mathcal{C}$ and $\mathcal{H}$ are only fairly well determined. Still, the fits performed here, as compared to the fits discussed in the previous subsection, have improved substantially. We thus conclude that it was important to also take into account perturbative SU(3) symmetry breaking effects into our analysis.

To close this section, we would like to emphasize that in the literature there are some analysis available to compare with, although it is difficult to assess the success of the many calculations of SU(3) symmetry breaking corrections to the axial vector form factors. Predictions that vary substantially from one another are obtained. For instance, in the paper by Dai and collaborators \cite{ddjm96} the issue of $1/N_c$ and perturbative SU(3) breaking corrections are discussed using information on both octet and decuplet baryons. Also, in the paper by Zhu, Sacco, and Ramsey-Musolf \cite{zsr02} chiral corrections are discussed, including SU(3) breaking perturbatively; they apply their results to octet baryons only. We limit ourselves to claim that, on general grounds, there is a good agreement between these calculations and ours.

\section{Conclusions}
\label{conclusions}

The present study was devoted to the evaluation of the baryon axial vector current up to one-loop order within large-$N_c$ baryon chiral perturbation theory, taking into account the mass difference between decuplet and octet baryons as well as perturbative flavor SU(3) symmetry breaking. In this framework, the one-loop correction to the baryon axial vector current amounts to an infinite series, each term representing a rather complicated structure of commutators and/or anticommutators, involving the baryon axial vector current $A^k$ and the baryon mass operator $\mathcal{M}$. We have taken into account the first three contributions in this expansion, i.e., the degeneracy limit ($AAA$), the leading ($AAA\mathcal{M}$) and the next-to-leading ($AAA\mathcal{M}\mathcal{M}$) order correction, respectively. We have explicitly evaluated these expressions -- individually for the flavor singlet, flavor octet and flavor $\mathbf{27}$ contributions -- up to order $1/N_c^2$ relative to the tree-level value. Large-$N_c$ cancellations occur between various Feynman diagrams at various levels which are a consequence of the SU(6) spin-flavor symmetry. We have also incorporated perturbative flavor SU(3) symmetry breaking at leading order in $\epsilon = m_s-\hat{m}$ into out analysis, which resulted in taking into account four more operators in the axial current, as compared to the SU(3) symmetry limit.

The order of the calculation envisaged in the present work allowed us to perform various fits. More precisely, fitting our analytical expressions against experimental data on baryon semileptonic decays, we are able to extract the basic parameters $a_1$, $b_2$, $b_3$ and $c_3$ of the $1/N_c$ baryon chiral Lagrangian as well as the axial vector couplings $g_A$ for octet baryons. In a first approach we have neglected the mass difference $\Delta= M_T - M_B$ between decuplet and octet baryons. This analysis thus follows the lines of the fit carried out in Ref.~\cite{fh06}. In a more refined approach, we then incorporate the effects of a non-vanishing mass difference.

In the first part of the analysis referring to the degeneracy limit $\Delta \to 0$, the comparison between the experimental data and the theoretical expressions through a least-squares fit yields a rather poor $\chi^2/\mathrm{dof}=3.95/\mathrm{dof}$. In the second part, where the $\Delta$-effects are taken into account, the fit yields $\chi^2=1.78/\mathrm{dof}$, which can be considered as a better fit. Although in both cases the predicted observables, i.e. the decay rates as well as angular correlation and spin-asymmetry coefficients in baryon semileptonic decays are in good agreement with their experimental counterparts, the latter fit is preferred over the former one. This is because the best-fit parameters $a_1$, $b_2$, $b_3$ and $c_3$, introduced in the definition of the axial-vector current Eq.~(\ref{eq:akc}), are in accordance with the pattern to be expected from the $1/N_c$ expansion, i.e., they are roughly of order 1.

While the above fits have been obtained by using the decay rates and the $g_A/g_V$ ratios, to corroborate our analysis, we have performed the analogous fits but this time using as an input the other set of experimental data available, i.e. the decays rates and the angular correlation and spin-asymmetry coefficients. The results are perfectly consistent with those obtained with the first set of experimental data, both in the degeneracy limit and in the case of a nonvanishing decuplet-octet mass difference. Still, although the fits improve when a nonvanishing $\Delta$ is considered, they are not completely satisfactory.

We thus perform yet another set of fits, which apart from baryon semileptonic decays also involves the four kinematically allowed strong decays of decuplet baryons. Now the fits improve substantially and we find a fairly good agreement between the predicted and observed axial couplings and the parameters $D, F, {\cal C}$ and ${\cal H}$ of the chiral Lagrangian.

In conclusion, it was essential to systematically consider the mass difference between decuplet and octet baryons as well as perturbative flavor SU(3) symmetry breaking in our analysis and in our subsequent fits. It allowed us to establish that the large-$N_c$ baryon chiral perturbation theory predictions regarding the renormalization of the baryon axial vector current are in very good agreement both with the expectations from the $1/N_c$ expansion and with the experimental results.

\acknowledgments

This work has been partially supported by Consejo Nacional de Ciencia y Tecnolog{\'\i}a (Mexico).

\appendix

\section{\label{app:loopint}Loop integrals}

The loop graphs involved in Fig.~\ref{fig:eins}(a,b,c) can be expressed in terms of a single loop integral, namely
\begin{equation}
\delta^{ij} \, F(m,\Delta,\mu) = \frac{i}{f^2} \int \frac{d^4k}{(2\pi)^4} \frac{(\mathbf{k}^i)(-\mathbf{k}^j)}
{(k^2-m^2)(k\cdot v-\Delta+i\epsilon)}, \label{eq:Fdef}
\end{equation}
where $\mu$ is the scale parameter of dimensional regularization. The solution of this integral takes the form \cite{fhjm00}\footnote{For $m < \left|\Delta\right|$ the function $F(m,\Delta,\mu)$ (and its derivatives) develops an imaginary part which is required in order to meet the condition $\lim_{\Delta\to 0^-} F(m,\Delta,\mu)=\lim_{\Delta\to 0^+} F(m,\Delta,\mu)$. This imaginary part has been omitted in previous works \cite{fhjm00,fh06}. It is understood, of course, that it is the real part of $F(m,\Delta,\mu)$ that is to correspond to the actual numerical evaluation.}
\begin{eqnarray}
24\pi^2f^2 \, F(m,\Delta,\mu) & = & \Delta \left[\Delta^2 - \frac32 m^2 \right] \ln \frac{m^2}{\mu^2} - \frac83 \Delta^3 - \frac72 \Delta m^2 \nonumber \\
& & \mbox{} + \left\{ \begin{array}{ll} \displaystyle 2(m^2-\Delta^2)^{3/2} \left[ \frac{\pi}{2} - \textrm{arctan}
\left( \frac{\Delta}{\sqrt{m^2-\Delta^2}} \right) \right], & m \ge \left|\Delta\right|, \\[6mm]
\displaystyle - (\Delta^2-m^2)^{3/2} \left[-2i\pi + \ln \left( { \frac{\Delta-\sqrt{\Delta^2-m^2}}{\Delta+\sqrt{\Delta^2-m^2}}} \right) \right], & m < \left|\Delta\right|.
\end{array} \right.
\end{eqnarray}

The derivatives of $F(m,\Delta,\mu)$ required here read
\begin{eqnarray}
24\pi^2f^2 \, F^{(1)} (m,\Delta,\mu) & = & 3\left[\Delta^2- \frac12 m^2\right]\ln \frac{m^2}{\mu^2} - 6\Delta^2 - \frac{11}{2} m^2
\nonumber \\
& & \mbox{} - \left\{ \begin{array}{ll} \displaystyle 6\Delta\sqrt{m^2-\Delta^2} \left[\frac{\pi}{2} - \textrm{arctan}
\left(\frac{\Delta}{\sqrt{m^2-\Delta^2}} \right) \right], & m \geq |\Delta|, \\[6mm]
\displaystyle 3\Delta\sqrt{\Delta^2-m^2}\left[-2i\pi+ \ln \left( \frac{\Delta-\sqrt{\Delta^2-m^2}}{\Delta+\sqrt{\Delta^2-m^2}} \right) \right], & m < |\Delta|, \end{array} \right.
\end{eqnarray}
\begin{eqnarray}
24\pi^2f^2 \, F^{(2)} (m,\Delta,\mu) & = & 6\Delta \left[\ln\frac{m^2}{\mu^2}-1\right] \nonumber \\
& & \mbox{} - \left\{ \begin{array}{ll} \displaystyle \frac{6(m^2-2\Delta^2)}{\sqrt{m^2-\Delta^2}} \left[\frac{\pi}{2} - \textrm{arctan} \left( \frac{\Delta} {\sqrt{m^2-\Delta^2}} \right) \right], & m \geq |\Delta|, \\[6mm]
\displaystyle \frac{3(2\Delta^2-m^2)}{\sqrt{\Delta^2-m^2}} \left[-2i\pi + \ln \left(\frac{\Delta-\sqrt{\Delta^2-m^2}}{\Delta+\sqrt{\Delta^2-m^2}} \right) \right], & m < |\Delta|, \end{array} \right.
\end{eqnarray}
and
\begin{eqnarray}
24\pi^2f^2 \, F^{(3)} (m,\Delta,\mu) & = & 6\ln\frac{m^2}{\mu^2} - \frac{6\Delta^2}{m^2-\Delta^2} \nonumber \\
& & \mbox{} + \left\{ \begin{array}{ll} \displaystyle \frac{6\Delta(3m^2-2\Delta^2)}{(m^2-\Delta^2)^{3/2}} \left[\frac{\pi}{2} - \textrm{arctan} \left( \frac{\Delta}{\sqrt{m^2-\Delta^2}} \right) \right], & m \geq |\Delta|, \\[6mm]
\displaystyle \frac{3\Delta(3m^2-2\Delta^2)}{(\Delta^2-m^2)^{3/2}} \left[-2i\pi + \ln \left( \frac{\Delta-\sqrt{\Delta^2-m^2}}{\Delta+\sqrt{\Delta^2-m^2}} \right) \right], & m < |\Delta|. \end{array} \right.
\end{eqnarray}

\section{Reduction of baryon operators\label{app:reduc}}

The evaluation of the commutator-anticommutator structure
\begin{eqnarray}
\left\{ A^{ja}, \left[A^{kc}, \left[ \mathcal{M}, A^{jb} \right] \right] \right\},\nonumber
\end{eqnarray}
which represents the leading contribution to the renormalized baryon axial vector current for finite decuplet-octet mass difference, yields the following terms: \\

\textit{1.\ Flavor singlet contribution}

\begin{equation}
\label{GJ2GG-singlet}
\{G^{ia},[G^{kc},[J^2,G^{ia}]]\} = - \frac12(N_f-2)G^{kc} + \frac12(N_c+N_f)\mathcal{D}_2^{kc} - \frac12\mathcal{D}_3^{kc} - \mathcal{O}_3^{kc},
\end{equation}

\begin{eqnarray}
\label{GJ2GD2-singlet}
& & \{G^{ia},[\mathcal{D}_2^{kc},[J^2,G^{ia}]]\} + \{\mathcal{D}_2^{ia},[G^{kc},[J^2,G^{ia}]]\} = 2(N_c+N_f)G^{kc} + \frac12[N_c(N_c+2N_f)-9N_f-2] \mathcal{D}_2^{kc} \nonumber \\
& & \mbox{} + \frac12(N_c+N_f) \mathcal{D}_3^{kc} - 2\mathcal{D}_4^{kc},
\end{eqnarray}

\begin{eqnarray}
\{\mathcal{D}_2^{ia},[\mathcal{D}_2^{kc},[J^2,G^{ia}]]\} = (N_f+2) \mathcal{O}_3^{kc},
\end{eqnarray}

\begin{eqnarray}
& & \{G^{ia},[\mathcal{D}_3^{kc},[J^2,G^{ia}]]\} + \{\mathcal{D}_3^{ia},[G^{kc},[J^2,G^{ia}]]\} = 2[N_c(N_c+2N_f)+2N_f] G^{kc} + 10(N_c+N_f) \mathcal{D}_2^{kc} \nonumber \\
& & \mbox{} + \frac12[2N_c(N_c+2N_f)-17N_f-2] \mathcal{D}_3^{kc} - 2(N_f-2) \mathcal{O}_3^{kc} + (N_c+N_f) \mathcal{D}_4^{kc} - 3\mathcal{D}_5^{kc},
\end{eqnarray}

\begin{eqnarray}
& & \{G^{ia},[\mathcal{O}_3^{kc},[J^2,G^{ia}]]\} + \{G^{ia},[G^{kc},[J^2,\mathcal{O}_3^{ia}]]\} +
\{\mathcal{O}_3^{ia},[G^{kc},[J^2,G^{ia}]]\} = 3(N_c+N_f) \mathcal{D}_2^{kc} - \frac32 N_f \mathcal{D}_3^{kc} \nonumber \\
& & \mbox{} + \frac12[2N_c(N_c+2N_f)-13N_f+2] \mathcal{O}_3^{kc} + (N_c+N_f) \mathcal{D}_4^{kc} - \mathcal{D}_5^{kc} - 5\mathcal{O}_5^{kc}.
\end{eqnarray}

\textit{2.\ Flavor octet contribution}

\begin{eqnarray}
\label{GJ2GG-octet}
& & d^{ab8}\{G^{ia},[G^{kc},[J^2,G^{ib}]]\} = - \frac14(N_f-4)d^{c8e}G^{ke} + \frac{N_c(N_c+2N_f)-2N_f+4}{4N_f} \delta^{c8}J^k + \frac14(N_c+N_f) d^{c8e}\mathcal{D}_2^{ke} \nonumber \\
& & \mbox{} + \frac14(N_c+N_f) [J^2,[T^8,G^{kc}]] - \frac14 d^{c8e}\mathcal{D}_3^{ke} -
\frac12 d^{c8e}\mathcal{O}_3^{ke} - \frac12 \{G^{kc},\{J^r,G^{r8}\}\} + \frac{1}{N_f} \{G^{k8},\{J^r,G^{rc}\}\} \nonumber \\
& & \mbox{} + \frac18 \{J^k,\{T^c,T^8\}\} - \frac{1}{N_f} \{J^k,\{G^{rc},G^{r8}\}\} - \frac{1}{2 N_f} \delta^{c8}\{J^2,J^k\},
\end{eqnarray}

\begin{eqnarray}
\label{GJ2GD2-octet}
& & d^{ab8}\left(\{G^{ia},[\mathcal{D}_2^{kc},[J^2,G^{ib}]]\} + \{\mathcal{D}_2^{ia},[G^{kc},[J^2,G^{ib}]]\}
\right) = (N_c+N_f)d^{c8e}G^{ke} - \frac{7N_f+4}{4} d^{c8e}\mathcal{D}_2^{ke} + \{G^{kc},T^8\} \nonumber \\
& & \mbox{} - \frac{N_f}{2} \{T^c,G^{k8}\} - \frac{N_f^2+4}{4N_f} [J^2,[T^8,G^{kc}]] + \frac14 (N_c+N_f) d^{c8e}\mathcal{D}_3^{ke} - \frac{N_f-2}{2N_f}(N_c+N_f) \{G^{k8},\{J^r,G^{rc}\}\} \nonumber \\
& & \mbox{} + \frac14(N_c+N_f) \{J^k,\{T^c,T^8\}\} + \frac{N_f-2}{2N_f}(N_c+N_f) \{J^k,\{G^{rc},G^{r8}\}\} - \frac12 d^{c8e}\mathcal{D}_4^{ke} - \frac{N_f+1}{N_f} \{\mathcal{D}_2^{kc},\{J^r,G^{r8}\}\} \nonumber \\
& & \mbox{} + \frac12 \{\mathcal{D}_2^{k8},\{J^r,G^{rc}\}\} - \frac{N_f-2}{2N_f}\{J^2,\{G^{k8},T^c\}\},
\end{eqnarray}

\begin{eqnarray}
& & d^{ab8}\{\mathcal{D}_2^{ia},[\mathcal{D}_2^{kc},[J^2,G^{ib}]]\} = - \frac{N_c+N_f}{N_f} [J^2,[T^8,G^{kc}]]
+ \frac{N_f+2}{2} d^{c8e}\mathcal{O}_3^{ke} + \{G^{kc},\{J^r,G^{r8}\}\} \nonumber \\
& & \mbox{} - \{G^{k8},\{J^r,G^{rc}\}\} + \frac{(N_c+N_f)(N_f-2)}{2N_f} \{\mathcal{D}_2^{kc},\{J^r,G^{r8}\}\}
- \frac{(N_c+N_f)(N_f-2)}{2N_f} \{J^2,\{G^{k8},T^c\}\},
\end{eqnarray}

\begin{eqnarray}
& & d^{ab8} \left( \{G^{ia},[\mathcal{D}_3^{kc},[J^2,G^{ib}]]\} + \{\mathcal{D}_3^{ia},[G^{kc},[J^2,G^{ib}]]\} \right) = 2N_f d^{c8e}G^{ke}
+ \frac{5N_c(N_c+2N_f)}{N_f} \delta^{c8}J^k \nonumber \\
& & \mbox{} + 5(N_c+N_f)d^{c8e} \mathcal{D}_2^{ke} + 2(N_c+N_f) \{G^{kc},T^8\} - (N_c+N_f) [J^2,[T^8,G^{kc}]]
- \frac54 N_f d^{c8e}\mathcal{D}_3^{ke} \nonumber \\
& & \mbox{} + \frac{2(N_f+2)}{N_f} d^{c8e}\mathcal{O}_3^{ke} - \frac{N_f^2-2N_f+4}{N_f} \{G^{kc},\{J^r,G^{r8}\}\}
- \frac{3N_f^2-2N_f-4}{N_f} \{G^{k8},\{J^r,G^{rc}\}\} \nonumber \\
& & \mbox{} + \frac52 \{J^k,\{T^c,T^8\}\} - 2(N_f+3) \{J^k,\{G^{rc},G^{r8}\}\}
+ \frac{N_c(N_c+2N_f)-10N_f}{2N_f} \delta^{c8}\{J^2,J^k\} \nonumber \\
& & \mbox{} + \frac12(N_c+N_f) d^{c8e}\mathcal{D}_4^{ke} + 2(N_c+N_f) \{\mathcal{D}_2^{k8},\{J^r,G^{rc}\}\}
- \frac12 d^{c8e}\mathcal{D}_5^{ke} \nonumber \\
& & \mbox{} - \frac{2(N_f-2)}{N_f} \{J^2,\{G^{k8},\{J^r,G^{rc}\}\}\} + \frac14 \{J^2,\{J^k,\{T^c,T^8\}\}\}
- \frac{2}{N_f} \{J^2,\{J^k,\{G^{rc},G^{r8}\}\}\} \nonumber \\
& & \mbox{} - \frac{3N_f+2}{2N_f} \{J^k,\{\{J^r,G^{rc}\},\{J^m,G^{m8}\}\}\} - \frac{1}{N_f} \delta^{c8}\{J^2,\{J^2,J^k\}\},
\end{eqnarray}

\begin{eqnarray}
& & d^{ab8} \left( \{G^{ia},[\mathcal{O}_3^{kc},[J^2,G^{ib}]]\} + \{G^{ia},[G^{kc},[J^2,\mathcal{O}_3^{ib}]]\} +
\{\mathcal{O}_3^{ia},[G^{kc},[J^2,G^{ib}]]\} \right) = \frac{3N_c(N_c+2N_f)}{2N_f}\delta^{c8}J^k \nonumber \\
& & + \frac32(N_c+N_f)d^{c8e}\mathcal{D}_2^{ke} + \frac32(N_c+N_f)[J^2,[T^8,G^{kc}]] - \frac{(3N_f+4)(N_f-2)}{4N_f}d^{c8e}\mathcal{D}_3^{ke} - \frac{7N_f-4}{4}d^{c8e}\mathcal{O}_3^{ke} \nonumber \\
& & \mbox{} - \frac32 N_f\{G^{kc},\{J^r,G^{r8}\}\} + \frac32 N_f\{G^{k8},\{J^r,G^{rc}\}\} + \frac34\{J^k,\{T^c,T^8\}\} - \frac{N_f+4}{N_f}\{J^k,\{G^{rc},G^{r8}\}\} \nonumber \\
& & \mbox{} + \frac{N_cN_f(N_c+2N_f)-2N_f(3N_f-1)+8}{2N_f^2}\delta^{c8}\{J^2,J^k\} + \frac12 (N_c+N_f)d^{c8e}\mathcal{D}_4^{ke} - (N_c+N_f)\{\mathcal{D}_2^{k8},\{J^r,G^{rc}\}\} \nonumber \\
& & \mbox{} + (N_c+N_f)\{J^2,\{G^{kc},T^8\}\} + \frac34(N_c+N_f)\{J^2,[J^2,[T^8,G^{kc}]]\} - \frac12 d^{c8e}\mathcal{D}_5^{ke} - \frac32 d^{c8e}\mathcal{O}_5^{ke}
- \frac72\{J^2,\{G^{kc},\{J^r,G^{r8}\}\}\} \nonumber \\
& & \mbox{} + \frac{N_f+1}{N_f}\{J^2,\{G^{k8},\{J^r,G^{rc}\}\}\} + \frac14\{J^2,\{J^k,\{T^c,T^8\}\}\}
- \frac{2}{N_f}\{J^2,\{J^k,\{G^{rc},G^{r8}\}\}\} \nonumber \\
& & \mbox{} + \frac{3N_f+2}{4N_f}\{J^k,\{\{J^r,G^{rc}\},\{J^m,G^{m8}\}\}\} - \frac{1}{N_f} \delta^{c8}\{J^2,\{J^2,J^k\}\}.
\end{eqnarray}

\textit{3.\ Flavor $\mathbf{27}$ contribution}

\begin{eqnarray}
\label{GJ2GG-27}
& & \{G^{i8},[G^{kc},[J^2,G^{i8}]]\} = - \frac{1}{N_f} \delta^{c8}\mathcal{O}_3^{k8} + \frac12 d^{c8e}\{J^k,\{G^{re},G^{r8}\}\} - \frac12 d^{c8e}\{G^{k8},\{J^r,G^{re}\}\} \nonumber \\
& & \mbox{} - \frac14 \epsilon^{kim}f^{c8e}\{T^e,\{J^i,G^{m8}\}\} ,
\end{eqnarray}

\begin{eqnarray}
\label{GJ2GD2-27}
& & \{G^{i8}, [\mathcal{D}_2^{kc}, [J^2,G^{i8}]] \} + \{G^{i8}, [G^{kc},[J^2, \mathcal{D}_2^{i8}]] \}
+ \{\mathcal{D}_2^{i8}, [G^{kc}, [J^2,G^{i8}]] \} = - \frac{15}{4}f^{c8e}f^{8eg}\mathcal{D}_2^{kg} + \frac{i}{2}f^{c8e}[G^{ke},\{J^r,G^{r8}\}] \nonumber \\
& & - if^{c8e}[G^{k8},\{J^r,G^{re}\}] - \frac12 f^{c8e}f^{8eg}\mathcal{D}_4^{kg} + \{\mathcal{D}_2^{kc},\{G^{r8},G^{r8}\}\} + \{\mathcal{D}_2^{k8},\{G^{rc},G^{r8}\}\}
- \frac12 \{\{J^r,G^{rc}\},\{G^{k8},T^8\}\} \nonumber \\
& & \mbox{} - \frac12 \{\{J^r,G^{r8}\},\{G^{k8},T^c\}\} + \frac{i}{2}f^{c8e}\{J^k,[\{J^i,G^{ie}\},\{J^r,G^{r8}\}]\},
\end{eqnarray}

\begin{eqnarray}
& & \{\mathcal{D}_2^{i8},[\mathcal{D}_2^{kc},[J^2,G^{i8}]]\} = - \frac14 f^{c8e}f^{8eg}\mathcal{D}_3^{kg}
+ \frac12 f^{c8e}f^{8eg}\mathcal{O}_3^{kg} + \frac12 \epsilon^{kim}f^{c8e}\{T^e,\{J^i,G^{m8}\}\} \nonumber \\
& & \mbox{} - \frac12 \epsilon^{kim}f^{c8e}\{T^8,\{J^i,G^{me}\}\} + \frac12 \{\mathcal{D}_2^{kc},\{T^8,\{J^r,G^{r8}\}\}\}
- \frac12 \{J^2,\{G^{k8},\{T^c,T^8\}\}\},
\end{eqnarray}

\begin{eqnarray}
& & \{G^{i8},[\mathcal{D}_3^{kc},[J^2,G^{i8}]]\} + \{\mathcal{D}_3^{i8},[G^{kc},[J^2,G^{i8}]]\} = 3 f^{c8e} f^{8eg} G^{kg}
- \frac12 d^{c8e} d^{8eg} G^{kg} - \frac{1}{2N_f} d^{c88} J^k - 2 d^{c8e} d^{8eg} \mathcal{D}_3^{kg} \nonumber \\
& & \mbox{} + \frac{4}{N_f} \delta^{c8} \mathcal{D}_3^{k8}
- \frac{4}{N_f} \delta^{88} \mathcal{D}_3^{kc} - d^{c8e} d^{8eg} \mathcal{O}_3^{kg} + f^{c8e} f^{8eg} \mathcal{O}_3^{kg} + \frac{2}{N_f} \delta^{c8} \mathcal{O}_3^{k8} - \frac{2}{N_f} \delta^{88} \mathcal{O}_3^{kc} + 4 \{G^{kc},\{G^{r8},G^{r8}\}\} \nonumber \\
& & \mbox{} - 4 \{G^{k8},\{G^{rc},G^{r8}\}\} + 6 d^{c8e} \{J^k,\{G^{re},G^{r8}\}\} - 2 d^{88e} \{J^k,\{G^{rc},G^{re}\}\} + \frac12 d^{c8e} \{G^{ke},\{J^r,G^{r8}\}\} \nonumber \\
& & \mbox{} + 2 d^{c8e} \{G^{k8},\{J^r,G^{re}\}\} - d^{88e} \{G^{kc},\{J^r,G^{re}\}\} - d^{88e} \{G^{ke},\{J^r,G^{rc}\}\} - \frac{4}{N_f} d^{c88} \{J^2,J^k\} \nonumber \\
& & \mbox{} + \epsilon^{kim} f^{c8e} \{T^e,\{J^i,G^{m8}\}\}
- 2 \{\{J^r,G^{rc}\},\{G^{k8},\{J^i,G^{i8}\}\}\} + 2 \{J^k,\{\{J^i,G^{ic}\},\{G^{r8},G^{r8}\}\}\} \nonumber \\
& & \mbox{} - \frac12 d^{c8e} \{\mathcal{D}_3^{k8},\{J^r,G^{re}\}\}
+ d^{c8e} \{J^2,\{J^k,\{G^{re},G^{r8}\}\}\},
\end{eqnarray}

\begin{eqnarray}
& & \{G^{i8},[\mathcal{O}_3^{kc},[J^2,G^{i8}]]\} + \{G^{i8},[G^{kc},[J^2,\mathcal{O}_3^{i8}]]\} +
\{\mathcal{O}_3^{i8},[G^{kc},[J^2,G^{i8}]]\} = - \frac12 d^{c8e} d^{8eg} G^{kg} \nonumber \\
& & \mbox{} - \frac{1}{2N_f} d^{c88} J^k - d^{c8e} d^{8eg} \mathcal{D}_3^{kg} - \frac12 d^{c8e} d^{8eg} \mathcal{O}_3^{kg}
- \frac32 f^{c8e} f^{8eg} \mathcal{O}_3^{kg} - \frac{7}{N_f} \delta^{c8} \mathcal{O}_3^{k8} - \frac{1}{N_f} \delta^{88} \mathcal{O}_3^{kc}
\nonumber \\
& & \mbox{} + 2 d^{c8e} \{J^k,\{G^{re},G^{r8}\}\} + \frac72 d^{c8e} \{G^{ke},\{J^r,G^{r8}\}\} - 3 d^{c8e} \{G^{k8},\{J^r,G^{re}\}\}
- \frac12 d^{88e} \{G^{kc},\{J^r,G^{re}\}\} \nonumber \\
& & \mbox{} + \frac12 d^{88e} \{G^{ke},\{J^r,G^{rc}\}\} - \frac{2}{N_f} d^{c88} \{J^2,J^k\} - \epsilon^{kim} f^{c8e} \{T^e,\{J^i,G^{m8}\}\}
- \frac{3}{N_f} \delta^{c8} \mathcal{O}_5^{k8} \nonumber \\
& & \mbox{} - \{G^{kc},\{\{J^i,G^{i8}\},\{J^r,G^{r8}\}\}\} + \{\{J^r,G^{rc}\},\{G^{k8},\{J^i,G^{i8}\}\}\}
- \{J^k,\{\{J^i,G^{ic}\},\{G^{r8},G^{r8}\}\}\} \nonumber \\
& & \mbox{} + 2 \{J^2,\{G^{kc},\{G^{r8},G^{r8}\}\}\} + \frac14 d^{c8e} \{\mathcal{D}_3^{k8},\{J^r,G^{re}\}\}
+ d^{c8e} \{J^2,\{J^k,\{G^{re},G^{r8}\}\}\} \nonumber \\
& & \mbox{} - \frac34 d^{c8e} \{J^2,\{G^{k8},\{J^r,G^{re}\}\}\} - \frac34 \epsilon^{kim} f^{c8e} \{J^2,\{T^e,\{J^i,G^{m8}\}\}\}.
\end{eqnarray}

The next-to-leading order contribution to the renormalized baryon axial vector current for finite decuplet-octet mass difference involves the two operator structures
\begin{equation*}
\left[A^{ja}, \left[\left[\mathcal{M}, \left[ \mathcal{M},A^{jb}\right]\right],A^{kc}\right] \right] ,
\quad \mbox{and} \quad \left[\left[\mathcal{M},A^{ja}\right], \left[\left[\mathcal{M},A^{jb}\right],A^{kc}\right]\right],
\end{equation*}
with two mass insertions. The explicit evaluation of the various contributions reads: \\

\textit{1.\ Flavor singlet contribution}

\begin{equation}
[G^{ia},[[J^2,[J^2,G^{ia}]],G^{kc}]] = - \frac32 (N_c+N_f) \mathcal{D}_2^{kc} + \frac12 (N_f+1) \mathcal{D}_3^{kc} + N_f \mathcal{O}_3^{kc},
\end{equation}

\begin{eqnarray}
& & [G^{ia},[[J^2,[J^2,G^{ia}]],\mathcal{D}_2^{kc}]] + [\mathcal{D}_2^{ia},[[J^2,[J^2,G^{ia}]],G^{kc}]] =
- \frac32 \left[ N_c(N_c+2N_f)-6 N_f \right] \mathcal{D}_2^{kc} \nonumber \\
& & \mbox{} - \frac{7}{2} (N_c+N_f) \mathcal{D}_3^{kc} - 2 (N_c+N_f) \mathcal{O}_3^{kc} + (3N_f+8) \mathcal{D}_4^{kc},
\end{eqnarray}

\begin{eqnarray}
& & [[J^2,G^{ia}],[[J^2,G^{ia}],G^{kc}]] = - \left[N_c(N_c+2 N_f)-N_f \right] G^{kc} + \frac52 (N_c+N_f) \mathcal{D}_2^{kc}
- \frac12 (N_f+1) \mathcal{D}_3^{kc} - (N_f-1) \mathcal{O}_3^{kc}, \nonumber \\
\end{eqnarray}

\begin{equation}
[[J^2,G^{ia}],[[J^2,G^{ia}],\mathcal{D}_2^{kc}]] = \frac32 \left[N_c(N_c+2N_f)-4 N_f\right] \mathcal{D}_2^{kc}
+ \frac52 (N_c+N_f) \mathcal{D}_3^{kc} - 3 (N_f+2) \mathcal{D}_4^{kc}.
\end{equation}

\textit{2.\ Flavor octet contribution}

\begin{eqnarray}
& & d^{ab8} [G^{ia},[[J^2,[J^2,G^{ib}]],G^{kc}]] = - \frac{3N_c(N_c+2N_f)}{4N_f} \delta^{c8} J^k
- \frac34 (N_c+N_f) d^{c8e} \mathcal{D}_2^{ke} - \frac14 (N_c+N_f) [J^2,[T^8,G^{kc}]] \nonumber \\
& & \mbox{} + \frac{N_f^2+N_f-4}{4N_f} d^{c8e} \mathcal{D}_3^{ke} + \frac{N_f^2-2}{2N_f} d^{c8e}
\mathcal{O}_3^{ke} - \frac38 \{J^k,\{T^c,T^8\}\} + \frac{N_f+4}{2N_f} \{J^k,\{G^{rc},G^{r8}\}\} + \frac{1}{N_f} \{G^{kc},\{J^r,G^{r8}\}\}
\nonumber \\
& & \mbox{} - \frac{1}{N_f} \{G^{k8},\{J^r,G^{rc}\}\} + \frac{2 N_f^2+N_f-4}{2N_f^2} \delta^{c8} \{J^2,J^k\} ,
\end{eqnarray}

\begin{eqnarray}
& & d^{ab8} \left( [G^{ia},[[J^2,[J^2,G^{ib}]],\mathcal{D}_2^{kc}]] + [\mathcal{D}_2^{ia},[[J^2,[J^2,G^{ib}]],G^{kc}]] \right) =
\frac92 N_f d^{c8e} \mathcal{D}_2^{ke} + \frac12 (N_f-2) [J^2,[T^8,G^{kc}]] \nonumber \\
& & \mbox{} - \frac{(N_c+N_f)(5N_f+4)}{4N_f} d^{c8e} \mathcal{D}_3^{ke} - \frac{(N_c+N_f)(N_f+2)}{2N_f} d^{c8e} \mathcal{O}_3^{ke}
- \frac{(N_c+N_f)(N_f-2)}{2N_f} \{G^{kc},\{J^r,G^{r8}\}\} \nonumber \\
& & \mbox{} + \frac{(N_c+N_f)(N_f-2)}{2N_f} \{G^{k8},\{J^r,G^{rc}\}\} - \frac34 (N_c+N_f) \{J^k,\{T^c,T^8\}\}
- \frac{(N_c+N_f)(N_f-2)}{N_f} \{J^k,\{G^{rc},G^{r8}\}\} \nonumber \\
& & \mbox{} + \frac{(N_c+N_f)(N_f-2)}{N_f^2} \delta^{c8}\{J^2,J^k\} + \frac12 (N_f+7)d^{c8e} \mathcal{D}_4^{ke}
+ \frac{N_f^2+7N_f+4}{2N_f} \{\mathcal{D}_2^{kc},\{J^r,G^{r8}\}\} - \frac{5}{2} \{\mathcal{D}_2^{k8},\{J^r,G^{rc}\}\} \nonumber \\
& & \mbox{} - \{J^2,\{G^{kc},T^8\}\} + \frac{N_f^2+2N_f-4}{2N_f} \{J^2,\{G^{k8},T^c\}\} + \frac{N_f+2}{2N_f} \{J^2,[J^2,[T^8,G^{kc}]]\},
\end{eqnarray}

\begin{eqnarray}
& & d^{ab8} [[J^2,G^{ia}],[[J^2,G^{ib}],G^{kc}]] = \frac{N_f}{2} d^{c8e} G^{ke} + \frac{5N_c(N_c+2N_f)}{4N_f} \delta^{c8}J^k
+ \frac54 (N_c+N_f) d^{c8e} \mathcal{D}_2^{ke} - (N_c+N_f) \{G^{kc},T^8\} \nonumber \\
& & \mbox{} + \frac12 (N_c+N_f) [J^2,[T^8,G^{kc}]] - \frac14 (N_f+1) d^{c8e} \mathcal{D}_3^{ke}
- \frac{N_f}{2} d^{c8e} \mathcal{O}_3^{ke} + \{G^{kc},\{J^r,G^{r8}\}\} + \frac{N_f}{2} \{G^{k8},\{J^r,G^{rc}\}\} \nonumber \\
& & \mbox{} + \frac58 \{J^k,\{T^c,T^8\}\} - \frac12 (N_f+3) \{J^k,\{G^{rc},G^{r8}\}\} - \frac{2N_f+1}{2N_f} \delta^{c8} \{J^2,J^k\},
\end{eqnarray}

\begin{eqnarray}
& & d^{ab8} [[J^2,G^{ia}],[[J^2,G^{ib}],\mathcal{D}_2^{kc}]] = - 3 N_f d^{c8e} \mathcal{D}_2^{ke}
+ \frac54 (N_c+N_f) d^{c8e} \mathcal{D}_3^{ke} + \frac34 (N_c+N_f) \{J^k,\{T^c,T^8\}\} \nonumber \\
& & \mbox{} - \frac12 (N_f+5) d^{c8e} \mathcal{D}_4^{ke}
- \frac{2N_f+7}{2} \{\mathcal{D}_2^{kc},\{J^r,G^{r8}\}\} + \frac52 \{\mathcal{D}_2^{k8},\{J^r,G^{rc}\}\}.
\end{eqnarray}

\textit{3.\ Flavor $\mathbf{27}$ contribution}

\begin{eqnarray}
& & [G^{i8},[[J^2,[J^2,G^{i8}]],G^{kc}]] = \frac12 d^{c8e} d^{8eg} \mathcal{D}_3^{kg} + \frac12 f^{c8e} f^{8eg} \mathcal{O}_3^{kg}
+ \frac12 d^{c8e} d^{8eg} \mathcal{O}_3^{kg} + \frac{2}{N_f} \delta^{c8} \mathcal{O}_3^{k8} \nonumber \\
& & \mbox{} - d^{c8e} \{J^k,\{G^{re},G^{r8}\}\} - \frac12 d^{c8e}\{G^{ke},\{J^r,G^{r8}\}\} + \frac12 d^{c8e}\{G^{k8},\{J^r,G^{re}\}\}
+ \frac{1}{N_f} d^{c88}\{J^2,J^k\},
\end{eqnarray}

\begin{eqnarray}
& & [G^{i8},[[J^2,[J^2,G^{i8}]],\mathcal{D}_2^{kc}]] + [\mathcal{D}_2^{i8},[[J^2,[J^2,G^{i8}]],G^{kc}]] = \frac{15}{2} f^{c8e} f^{8eg}
\mathcal{D}_2^{kg} + \frac{i}{2} f^{c8e} [G^{k8},\{J^r,G^{re}\}] \nonumber \\
& & \mbox{} + 4 f^{c8e} f^{8eg}\mathcal{D}_4^{kg} + \frac{2}{N_f} \delta^{c8} \mathcal{D}_4^{k8}
+ \frac{2}{N_f} \delta^{88} \mathcal{D}_4^{kc} - 2 \{\mathcal{D}_2^{kc},\{G^{r8},G^{r8}\}\} - 2 \{\mathcal{D}_2^{k8},\{G^{rc},G^{r8}\}\}
\nonumber \\
& & \mbox{} + \frac12 d^{c8e} \{J^2,\{G^{ke},T^8\}\} + \frac12 d^{88e} \{J^2,\{G^{ke},T^c\}\}
+ \frac12 d^{c8e} \{\mathcal{D}_2^{k8},\{J^r,G^{re}\}\} + \frac12 d^{88e} \{\mathcal{D}_2^{kc},\{J^r,G^{re}\}\} \nonumber \\
& & \mbox{} + \frac12 \{\{J^r,G^{rc}\},\{G^{k8},T^8\}\} - \frac12 \{\{J^r,G^{r8}\},\{G^{kc},T^8\}\}
- if^{c8e} \{\{J^r,G^{re}\},[J^2,G^{k8}]\} \nonumber \\
& & \mbox{} + \frac{i}{2} f^{c8e} \{\{J^r,G^{r8}\},[J^2,G^{ke}]\} - 3if^{c8e} \{J^k,[\{J^i,G^{ie}\},\{J^r,G^{r8}\}]\},
\end{eqnarray}

\begin{eqnarray}
& & [[J^2,G^{i8}],[[J^2,G^{i8}],G^{kc}]] = \frac34 f^{c8e} f^{8eg} G^{kg} - \frac12 d^{c8e}d^{8eg}\mathcal{D}_3^{kg}
+ \frac{1}{N_f} \delta^{c8}\mathcal{D}_3^{k8} - \frac12 f^{c8e}f^{8eg}\mathcal{O}_3^{kg} \nonumber \\
& & \mbox{} - \frac12 d^{c8e} d^{8eg}\mathcal{O}_3^{kg} - \frac{2}{N_f} \delta^{c8}\mathcal{O}_3^{k8} - \{G^{kc},\{G^{r8},G^{r8}\}\}
- \{G^{k8},\{G^{rc},G^{r8}\}\} + \frac32 d^{c8e} \{J^k,\{G^{re},G^{r8}\}\} \nonumber \\
& & \mbox{} - \frac12 d^{88e} \{J^k,\{G^{rc},G^{re}\}\} + d^{c8e} \{G^{ke},\{J^r,G^{r8}\}\} - \frac12 d^{c8e} \{G^{k8},\{J^r,G^{re}\}\}
+ \frac12 d^{88e} \{G^{ke},\{J^r,G^{rc}\}\} \nonumber \\
& & \mbox{} - \frac{1}{N_f} d^{c88} \{J^2,J^k\} - \frac14 \epsilon^{kim} f^{c8e} \{T^e,\{J^i,G^{m8}\}\},
\end{eqnarray}

\begin{eqnarray}
& & [[J^2,G^{i8}],[[J^2,G^{i8}],\mathcal{D}_2^{kc}]] = -6 f^{c8e} f^{8eg}\mathcal{D}_2^{kg} - \frac72 f^{c8e} f^{8eg} \mathcal{D}_4^{kg}
- \frac{2}{N_f} \delta^{88} \mathcal{D}_4^{kc} + 2 \{\mathcal{D}_2^{kc},\{G^{r8},G^{r8}\}\} \nonumber \\
& & \mbox{} - d^{88e} \{\mathcal{D}_2^{kc},\{J^r,G^{re}\}\} + if^{c8e} \{J^2,[G^{ke},\{J^r,G^{r8}\}]\}
- if^{c8e} \{J^2,[G^{k8},\{J^r,G^{re}\}]\} \nonumber \\
& & \mbox{} - if^{c8e} \{\{J^r,G^{re}\},[J^2,G^{k8}]\} + if^{c8e} \{\{J^r,G^{r8}\},[J^2,G^{ke}]\}
+ \frac32 if^{c8e} \{J^k,[\{J^i,G^{ie}\},\{J^r,G^{r8}\}]\}.
\end{eqnarray}

\section{Flavor $\mathbf{8}$ and $\mathbf{27}$ contributions to $g_A$\label{sec:827}}

The octet contributions to $\delta A^{kc}$, Eq.~(\ref{eq:dasplit}), can be written as
\begin{equation}
\delta A_{\mathbf{8}}^{kc} = \sum_{i=1}^{30} o_i O_i^{kc}, \label{eq:da8}
\end{equation}
where the operators $O_i^{kc}$ that occur at this order are
\begin{eqnarray}
\begin{array}{lll}
O_{1}^{kc} = d^{c8e} G^{ke}, &
O_{2}^{kc} = \delta^{c8} J^k, &
O_{3}^{kc} = d^{c8e} \mathcal{D}_2^{ke}, \\
O_{4}^{kc} = \{G^{kc},T^8\}, &
O_{5}^{kc} = \{G^{k8},T^c\}, &
O_{6}^{kc} = d^{c8e} \mathcal{D}_3^{ke}, \\
O_{7}^{kc} = d^{c8e} \mathcal{O}_3^{ke}, &
O_{8}^{kc} = \{G^{kc},\{J^r,G^{r8}\}\}, &
O_{9}^{kc} = \{G^{k8},\{J^r,G^{rc}\}\}, \\
O_{10}^{kc} = \{J^k,\{T^c,T^8\}\}, &
O_{11}^{kc} = \{J^k,\{G^{rc},G^{r8}\}\}, &
O_{12}^{kc} = \delta^{c8} \{J^2,J^k\}, \\
O_{13}^{kc} = d^{c8e} \mathcal{D}_4^{ke}, &
O_{14}^{kc} = \{\mathcal{D}_2^{kc},\{J^r,G^{r8}\}\}, &
O_{15}^{kc} = \{\mathcal{D}_2^{k8},\{J^r,G^{rc}\}\}, \\
O_{16}^{kc} = \{J^2,\{G^{kc},T^8\}\}, &
O_{17}^{kc} = \{J^2,\{G^{k8},T^c\}\}, &
O_{18}^{kc} = \{J^2,[ G^{kc},\{J^r,G^{r8}\}]\}, \\
O_{19}^{kc} = \{J^2,[G^{k8},\{J^r,G^{rc}\}]\}, &
O_{20}^{kc} = \{[J^2,G^{kc}],\{J^r,G^{r8}\}\}, &
O_{21}^{kc} = \{[J^2,G^{k8}],\{J^r,G^{rc}\}\}, \\
O_{22}^{kc} = \{J^k,[ \{J^m,G^{mc}\},\{J^r,G^{r8}\}]\}, &
O_{23}^{kc} = d^{c8e} \mathcal{D}_5^{ke}, &
O_{24}^{kc} = d^{c8e} \mathcal{O}_5^{ke}, \\
O_{25}^{kc} = \{J^2,\{G^{kc},\{J^r,G^{r8}\}\}\}, &
O_{26}^{kc} = \{J^2,\{G^{k8},\{J^r,G^{rc}\}\}\}, &
O_{27}^{kc} = \{J^2,\{J^k,\{T^c,T^8\}\}\}, \\
O_{28}^{kc} = \{J^2,\{J^k,\{G^{rc},G^{r8}\}\}\}, &
O_{29}^{kc} = \{J^k,\{\{J^r,G^{rc}\},\{J^m,G^{m8}\}\}\}, &
O_{30}^{kc} = \delta^{c8} \{J^2,\{J^2,J^k\}\},
\end{array}
\end{eqnarray}
and the corresponding coefficients read
\begin{eqnarray}
o_{1} & = & \left[ \frac{11}{48}a_1^3 - \frac{N_c+3}{3N_c}a_1^2b_2 - \frac{9}{4N_c^2}a_1b_2^2 - \frac{5}{2N_c^2}a_1^2b_3 + \frac{3}{4N_c^2}a_1^2c_3 - \frac{3(N_c+3)}{N_c^3}a_1b_2b_3 \right] F_{\mathbf{8}}^{(1)} \nonumber \\
& & \mbox{} + \left[ - \frac18 a_1^3 - \frac{N_c+3}{2N_c} a_1^2b_2 - \frac{3}{N_c^2} a_1^2b_3 \right] \frac{\Delta}{N_c} F_{\mathbf{8}}^{(2)} - \frac18 a_1^3 \frac{\Delta^2}{N_c^2} F_{\mathbf{8}}^{(3)},
\end{eqnarray}
\begin{eqnarray}
o_{2} & = & \left[ \frac{5}{36} a_1^3 + \frac{N_c+3}{18N_c} a_1^2b_2 - \frac{3N_c^2+18N_c-8}{12N_c^2} a_1^2b_3
- \frac{N_c+6}{24N_c} a_1^2c_3 \right] F_{\mathbf{8}}^{(1)} \nonumber \\
& & \mbox{} + \left[ - \frac{N_c^2+6N_c-2}{24} a_1^3 - \frac{5N_c+30}{6N_c} a_1^2b_3 - \frac{N_c+6}{4N_c} a_1^2c_3 \right] \frac{\Delta}{N_c} F_{\mathbf{8}}^{(2)} - \frac{11N_c(N_c+6)}{144} a_1^3 \frac{\Delta^2}{N_c^2} F_{\mathbf{8}}^{(3)},
\end{eqnarray}
\begin{eqnarray}
o_{3} & = & \left[ \frac{23}{16N_c} a_1^2b_2 - \frac{3(N_c+3)}{4N_c^2} a_1^2b_3 - \frac{N_c+3}{8N_c^2} a_1^2c_3 - \frac{3}{4N_c^3}b_2^3 - \frac{3}{2N_c^3} a_1b_2b_3 + \frac{9}{N_c^3} a_1b_2c_3 \right] F_{\mathbf{8}}^{(1)} \nonumber \\
& & \mbox{} +\left[ - \frac{N_c+3}{8} a_1^3 + \frac{25}{8N_c} a_1^2b_2 - \frac{5(N_c+3)}{2N_c^2} a_1^2b_3
- \frac{3(N_c+3)}{4N_c^2} a_1^2c_3 \right] \frac{\Delta}{N_c} F_{\mathbf{8}}^{(2)} \nonumber \\
& & \mbox{} + \left[ - \frac{11 (N_c+3)}{48} a_1^3 + \frac{3}{N_c} a_1^2b_2 \right] \frac{\Delta^2}{N_c^2} F_{\mathbf{8}}^{(3)},
\end{eqnarray}
\begin{eqnarray}
o_{4} & = & \left[ - \frac{1}{3N_c} a_1^2b_2 - \frac{N_c+3}{12N_c^2} a_1b_2^2 - \frac{N_c+3}{2N_c^2} a_1^2b_3- \frac{N_c+3}{2N_c^2} a_1^2c_3 - \frac{3}{N_c^3} a_1b_2b_3 \right] F_{\mathbf{8}}^{(1)} \nonumber \\
& & \mbox{} + \left[ - \frac{1}{2N_c} a_1^2b_2 - \frac{N_c+3}{N_c^2} a_1^2b_3 \right] \frac{\Delta}{N_c} F_{\mathbf{8}}^{(2)} + \frac{N_c+3}{12} a_1^3 \frac{\Delta^2}{N_c^2} F_{\mathbf{8}}^{(3)},
\end{eqnarray}
\begin{equation}
o_{5} = \left[ \frac{11}{12N_c} a_1^2b_2 + \frac{N_c+3}{6N_c^2} a_1b_2^2 + \frac{4}{N_c^3} a_1b_2b_3 \right] F_{\mathbf{8}}^{(1)} + \frac{3}{4N_c} a_1^2b_2 \frac{\Delta}{N_c} F_{\mathbf{8}}^{(2)},
\end{equation}
\begin{eqnarray}
o_{6} & = & \left[ \frac{9}{16N_c^2} a_1b_2^2 + \frac{65}{48N_c^2} a_1^2b_3 + \frac{1}{6N_c^2} a_1^2c_3
+ \frac{3(N_c+3)}{4N_c^3} a_1b_2b_3 - \frac{23(N_c+3)}{24N_c^3} a_1b_2c_3\right] F_{\mathbf{8}}^{(1)} \nonumber \\
& & \mbox{} + \left[ \frac18 a_1^3 - \frac{N_c+3}{8N_c} a_1^2b_2 + \frac{15}{8N_c^2} a_1^2b_3 + \frac{13}{24N_c^2} a_1^2c_3 \right] \frac{\Delta}{N_c} F_{\mathbf{8}}^{(2)} + \left[ \frac{7}{36} a_1^3 - \frac{53(N_c+3)}{144N_c} a_1^2b_2 \right] \frac{\Delta^2}{N_c^2} F_{\mathbf{8}}^{(3)},
\end{eqnarray}
\begin{eqnarray}
o_{7} & = & \left[ \frac{1}{8N_c^2} a_1b_2^2 - \frac{3}{4N_c^2} a_1^2b_3 + \frac{71}{48N_c^2} a_1^2c_3 - \frac{N_c+3}{6N_c^3} a_1b_2b_3 - \frac{N_c+3}{3N_c^3} a_1b_2c_3 \right] F_{\mathbf{8}}^{(1)} \nonumber \\
& & \mbox{} + \left[ \frac14 a_1^3 - \frac{5}{4N_c^2} a_1b_2^2 - \frac{5}{3N_c^2} a_1^2b_3 + \frac{17}{8N_c^2} a_1^2c_3 \right] \frac{\Delta}{N_c} F_{\mathbf{8}}^{(2)} + \left[ \frac{23}{72} a_1^3 - \frac{5(N_c+3)}{36N_c} a_1^2b_2 \right] \frac{\Delta^2}{N_c^2} F_{\mathbf{8}}^{(3)},
\end{eqnarray}
\begin{eqnarray}
o_{8} & = & \left[ \frac{1}{4N_c^2} a_1b_2^2 - \frac{1}{6N_c^2} a_1^2b_3 + \frac{5}{2N_c^2} a_1^2c_3
+ \frac{N_c+3}{3N_c^3} a_1b_2b_3 \right] F_{\mathbf{8}}^{(1)} \nonumber \\
& & \mbox{} +\left[ \frac14 a_1^3 - \frac{1}{2N_c^2} a_1b_2^2 + \frac{7}{6N_c^2} a_1^2b_3 + \frac{9}{4N_c^2} a_1^2c_3 \right] \frac{\Delta}{N_c} F_{\mathbf{8}}^{(2)} + \left[ - \frac{1}{36} a_1^3 - \frac{N_c+3}{36N_c} a_1^2b_2 \right] \frac{\Delta^2}{N_c^2} F_{\mathbf{8}}^{(3)},
\end{eqnarray}
\begin{eqnarray}
o_{9} & = & \left[ \frac{1}{4N_c^2} a_1b_2^2 + \frac{7}{3N_c^2} a_1^2b_3 - \frac{5}{4N_c^2} a_1^2c_3 + \frac{N_c+3}{3N_c^3}a_1b_2b_3 \right] F_{\mathbf{8}}^{(1)} \nonumber \\
& & \mbox{} + \left[ - \frac16 a_1^3 + \frac{N_c+3}{12N_c} a_1^2b_2 + \frac{1}{2N_c^2} a_1b_2^2
+ \frac{17}{6N_c^2} a_1^2b_3 - \frac{9}{4N_c^2} a_1^2c_3 \right] \frac{\Delta}{N_c} F_{\mathbf{8}}^{(2)} + \left[
- \frac{13}{72} a_1^3 + \frac{N_c+3}{36N_c} a_1^2b_2 \right] \frac{\Delta^2}{N_c^2} F_{\mathbf{8}}^{(3)},
\end{eqnarray}
\begin{eqnarray}
o_{10} & = & \left[ \frac{1}{12N_c^2} a_1b_2^2 - \frac{N_c+3}{24N_c^3}b_2^3 - \frac{3}{8N_c^2} a_1^2b_3 - \frac{1}{16N_c^2} a_1^2c_3 + \frac{N_c+3}{4N_c^3} a_1b_2b_3 - \frac{3(N_c+3)}{8N_c^3} a_1b_2c_3 \right] F_{\mathbf{8}}^{(1)} \nonumber \\
& & \mbox{} + \left[ - \frac{1}{16} a_1^3 - \frac{N_c+3}{8N_c} a_1^2b_2 - \frac{5}{4N_c^2} a_1^2b_3 - \frac{3}{8N_c^2} a_1^2c_3 \right] \frac{\Delta}{N_c} F_{\mathbf{8}}^{(2)} + \left[ - \frac{11}{96} a_1^3
- \frac{3(N_c+3)}{16N_c} a_1^2b_2\right] \frac{\Delta^2}{N_c^2} F_{\mathbf{8}}^{(3)},
\end{eqnarray}
\begin{eqnarray}
o_{11} & = & \left[ \frac{2}{N_c^2} a_1^2b_3 - \frac{1}{3N_c^2} a_1^2c_3
- \frac{N_c+3}{3N_c^3} a_1b_2c_3 \right] F_{\mathbf{8}}^{(1)} + \left[ \frac16 a_1^3 - \frac{N_c+3}{12N_c} a_1^2b_2 + \frac{6}{N_c^2} a_1^2b_3 + \frac{7}{6N_c^2} a_1^2c_3 \right] \frac{\Delta}{N_c} F_{\mathbf{8}}^{(2)} \nonumber \\
& & \mbox{} + \left[ \frac49 a_1^3 - \frac{N_c+3}{18N_c} a_1^2b_2 \right] \frac{\Delta^2}{N_c^2} F_{\mathbf{8}}^{(3)},
\end{eqnarray}
\begin{eqnarray}
o_{12} & = & \left[ \frac{7}{9N_c^2} a_1^2b_3 + \frac{13}{36N_c^2} a_1^2c_3
+ \frac{N_c+3}{9N_c^3} a_1b_2c_3 \right] F_{\mathbf{8}}^{(1)} \nonumber \\
& & \mbox{} + \left[ \frac{1}{12} a_1^3 - \frac{N_c^2+6N_c-30}{12N_c^2} a_1^2b_3 - \frac{3N_c^2+18N_c-40}{36N_c^2} a_1^2c_3 \right] \frac{\Delta}{N_c} F_{\mathbf{8}}^{(2)} + \left[ \frac{55}{216} a_1^3 + \frac{N_c+3}{54N_c} a_1^2b_2 \right] \frac{\Delta^2}{N_c^2} F_{\mathbf{8}}^{(3)},
\end{eqnarray}
\begin{eqnarray}
o_{13} & = & \left[ \frac{3}{8N_c^3}b_2^3 + \frac{3}{4N_c^3} a_1b_2b_3 + \frac{3}{N_c^3} a_1b_2c_3 \right] F_{\mathbf{8}}^{(1)} + \left[ \frac{1}{4N_c} a_1^2b_2 - \frac{N_c+3}{4N_c^2} a_1^2b_3
- \frac{N_c+3}{4N_c^2} a_1^2c_3 \right] \frac{\Delta}{N_c} F_{\mathbf{8}}^{(2)} \nonumber \\
& & \mbox{} + \frac{7}{6N_c} a_1^2b_2 \frac{\Delta^2}{N_c^2} F_{\mathbf{8}}^{(3)},
\end{eqnarray}
\begin{equation}
o_{14} = \left[ \frac{1}{2N_c^3}b_2^3 - \frac{7}{2N_c^3} a_1b_2b_3 + \frac{10}{3N_c^3} a_1b_2c_3 \right] F_{\mathbf{8}}^{(1)} + \left[ \frac{2}{3N_c} a_1^2b_2 - \frac{N_c+3}{12N_c^2} a_1b_2^2 \right] \frac{\Delta}{N_c} F_{\mathbf{8}}^{(2)} + \frac{107}{72N_c} a_1^2b_2 \frac{\Delta^2}{N_c^2} F_{\mathbf{8}}^{(3)},
\end{equation}
\begin{equation}
o_{15} = \left[ \frac{2}{N_c^3} a_1b_2b_3 - \frac{23}{12N_c^3} a_1b_2c_3 \right] F_{\mathbf{8}}^{(1)} + \left[ - \frac{1}{4N_c} a_1^2b_2 - \frac{N_c+3}{N_c^2} a_1^2b_3 + \frac{N_c+3}{2N_c^2} a_1^2c_3 \right] \frac{\Delta}{N_c} F_{\mathbf{8}}^{(2)} - \frac{5}{8N_c} a_1^2b_2 \frac{\Delta^2}{N_c^2} F_{\mathbf{8}}^{(3)},
\end{equation}
\begin{equation}
o_{16} = \left[ \frac{1}{6N_c^3} a_1b_2b_3 - \frac{1}{3N_c^3} a_1b_2c_3 \right] F_{\mathbf{8}}^{(1)} - \frac{N_c+3}{2N_c^2} a_1^2c_3 \frac{\Delta}{N_c} F_{\mathbf{8}}^{(2)} - \frac{1}{6N_c} a_1^2b_2 \frac{\Delta^2}{N_c^2} F_{\mathbf{8}}^{(3)},
\end{equation}
\begin{equation}
o_{17} = \left[ \frac{5}{3N_c^3} a_1b_2b_3 + \frac{11}{12N_c^3} a_1b_2c_3 \right] F_{\mathbf{8}}^{(1)} + \left[ \frac{1}{12N_c} a_1^2b_2 + \frac{N_c+3}{12N_c^2} a_1b_2^2 \right] \frac{\Delta}{N_c} F_{\mathbf{8}}^{(2)} + \frac{11}{36N_c} a_1^2b_2 \frac{\Delta^2}{N_c^2} F_{\mathbf{8}}^{(3)},
\end{equation}
\begin{eqnarray}
o_{18} & = & \left[ - \frac{1}{16N_c^3} a_1b_2b_3 - \frac{15}{64N_c^3} a_1b_2c_3 \right] F_{\mathbf{8}}^{(1)} +\left[ - \frac{1}{32N_c} a_1^2b_2 - \frac{3}{32N_c^2} a_1b_2^2 + \frac{9N_c-91}{384N_c^2} a_1^2b_3
+ \frac{42N_c-65}{3072N_c^2} a_1^2c_3 \right] \frac{\Delta}{N_c} F_{\mathbf{8}}^{(2)} \nonumber \\
& & \mbox{} + \left[ \frac{1}{2304} a_1^3 - \frac{1}{24N_c} a_1^2b_2 \right] \frac{\Delta^2}{N_c^2} F_{\mathbf{8}}^{(3)},
\end{eqnarray}
\begin{eqnarray}
o_{19} & = & \left[ \frac{1}{16N_c^3} a_1b_2b_3 + \frac{15}{64N_c^3} a_1b_2c_3 \right] F_{\mathbf{8}}^{(1)} + \left[ \frac{1}{32N_c} a_1^2b_2 + \frac{3}{32N_c^2} a_1b_2^2 - \frac{9N_c-91}{384N_c^2} a_1^2b_3 - \frac{42N_c-65}{3072N_c^2} a_1^2c_3 \right] \frac{\Delta}{N_c} F_{\mathbf{8}}^{(2)} \nonumber \\
& & \mbox{} + \left[ - \frac{1}{2304} a_1^3 + \frac{1}{24N_c} a_1^2b_2 \right] \frac{\Delta^2}{N_c^2} F_{\mathbf{8}}^{(3)},
\end{eqnarray}
\begin{eqnarray}
o_{20} & = & \left[ - \frac{1}{16N_c^3} a_1b_2b_3 - \frac{15}{64N_c^3} a_1b_2c_3 \right] F_{\mathbf{8}}^{(1)} + \left[ - \frac{1}{32N_c} a_1^2b_2 - \frac{3}{32N_c^2} a_1b_2^2 + \frac{9N_c-91}{384N_c^2} a_1^2b_3 + \frac{42N_c-65}{3072N_c^2} a_1^2c_3 \right] \frac{\Delta}{N_c} F_{\mathbf{8}}^{(2)} \nonumber \\
& & \mbox{} + \left[ \frac{1}{2304} a_1^3 - \frac{1}{24N_c} a_1^2b_2 \right] \frac{\Delta^2}{N_c^2} F_{\mathbf{8}}^{(3)},
\end{eqnarray}
\begin{eqnarray}
o_{21} & = & \left[ \frac{1}{16N_c^3} a_1b_2b_3 + \frac{15}{64N_c^3} a_1b_2c_3 \right] F_{\mathbf{8}}^{(1)} + \left[ \frac{1}{32N_c} a_1^2b_2 + \frac{3}{32N_c^2} a_1b_2^2
- \frac{9N_c-91}{384N_c^2} a_1^2b_3
- \frac{42N_c-65}{3072N_c^2} a_1^2c_3 \right] \frac{\Delta}{N_c} F_{\mathbf{8}}^{(2)} \nonumber \\
& & \mbox{} + \left[ - \frac{1}{2304} a_1^3 + \frac{1}{24N_c} a_1^2b_2 \right] \frac{\Delta^2}{N_c^2} F_{\mathbf{8}}^{(3)},
\end{eqnarray}
\begin{eqnarray}
o_{22} & = & \left[ \frac{1}{16N_c^3} a_1b_2b_3 + \frac{15}{64N_c^3} a_1b_2c_3 \right] F_{\mathbf{8}}^{(1)} +
\left[ \frac{1}{32N_c} a_1^2b_2 + \frac{3}{32N_c^2} a_1b_2^2 - \frac{9N_c-91}{384N_c^2} a_1^2b_3 - \frac{42N_c-65}{3072N_c^2} a_1^2c_3 \right] \frac{\Delta}{N_c} F_{\mathbf{8}}^{(2)} \nonumber \\
& & \mbox{} + \left[ - \frac{1}{2304} a_1^3 + \frac{1}{24N_c} a_1^2b_2 \right] \frac{\Delta^2}{N_c^2} F_{\mathbf{8}}^{(3)},
\end{eqnarray}
\begin{equation}
o_{23} = \left[ \frac{1}{4N_c^2} a_1^2b_3 + \frac{1}{4N_c^2} a_1^2c_3 \right] \frac{\Delta}{N_c} F_{\mathbf{8}}^{(2)},
\end{equation}
\begin{equation}
o_{24} = \frac{3}{4 N_c^2} a_1^2c_3 \frac{\Delta}{N_c} F_{\mathbf{8}}^{(2)},
\end{equation}
\begin{equation}
o_{25} = \frac{7}{4N_c^2} a_1^2c_3 \frac{\Delta}{N_c} F_{\mathbf{8}}^{(2)},
\end{equation}
\begin{equation}
o_{26} = \left[ \frac{1}{3N_c^2} a_1^2b_3 - \frac{2}{3N_c^2} a_1^2c_3 \right] \frac{\Delta}{N_c} F_{\mathbf{8}}^{(2)},
\end{equation}
\begin{equation}
o_{27} = \left[ - \frac{1}{8N_c^2} a_1^2b_3 - \frac{1}{8N_c^2} a_1^2c_3 \right] \frac{\Delta}{N_c} F_{\mathbf{8}}^{(2)},
\end{equation}
\begin{equation}
o_{28} = \left[ \frac{1}{3N_c^2} a_1^2b_3 + \frac{1}{3N_c^2} a_1^2c_3 \right] \frac{\Delta}{N_c} F_{\mathbf{8}}^{(2)},
\end{equation}
\begin{equation}
o_{29} = \left[ \frac{11}{12N_c^2} a_1^2b_3 - \frac{11}{24N_c^2} a_1^2c_3 \right] \frac{\Delta}{N_c} F_{\mathbf{8}}^{(2)},
\end{equation}
\begin{equation}
o_{30} = \left[ \frac{1}{6N_c^2} a_1^2b_3 + \frac{1}{6N_c^2} a_1^2 c_3 \right] \frac{\Delta}{N_c} F_{\mathbf{8}}^{(2)}.
\end{equation}

Finally, for the $\mathbf{27}$ representation we obtain
\begin{equation}
\delta A_{\mathbf{27}}^{kc} = \sum_{i=1}^{61} t_i T_i^{kc}, \label{eq:da27}
\end{equation}
where the operator basis is
\begin{eqnarray}
\begin{array}{lll}
T_{1}^{kc} = f^{c8e} f^{8eg} G^{kg}, &
T_{2}^{kc} = d^{c8e} d^{8eg} G^{kg}, &
T_{3}^{kc} = \delta^{c8} G^{k8}, \\
T_{4}^{kc} = d^{c88} J^k, &
T_{5}^{kc} = f^{c8e} f^{8eg} \mathcal{D}_2^{kg}, &
T_{6}^{kc} = \delta^{c8} \mathcal{D}_2^{k8}, \\
T_{7}^{kc} = \delta^{88} \mathcal{D}_2^{kc}, &
T_{8}^{kc} = d^{c8e} \{G^{ke},T^8\}, &
T_{9}^{kc} = d^{88e} \{G^{ke},T^c\}, \\
T_{10}^{kc} = if^{c8e} [G^{ke},\{J^r,G^{r8}\}], &
T_{11}^{kc} = if^{c8e} [G^{k8},\{J^r,G^{re}\}], &
T_{12}^{kc} = f^{c8e} f^{8eg} \mathcal{D}_3^{kg}, \\
T_{13}^{kc} = d^{c8e} d^{8eg} \mathcal{D}_3^{kg}, &
T_{14}^{kc} = \delta^{c8} \mathcal{D}_3^{k8}, &
T_{15}^{kc} = \delta^{88} \mathcal{D}_3^{kc}, \\
T_{16}^{kc} = f^{c8e} f^{8eg} \mathcal{O}_3^{kg}, &
T_{17}^{kc} = d^{c8e} d^{8eg} \mathcal{O}_3^{kg}, &
T_{18}^{kc} = \delta^{c8} \mathcal{O}_3^{k8}, \\
T_{19}^{kc} = \delta^{88} \mathcal{O}_3^{kc}, &
T_{20}^{kc} = \{G^{kc},\{T^8,T^8\}\}, &
T_{21}^{kc} = \{G^{k8},\{T^c,T^8\}\}, \\
T_{22}^{kc} = \{G^{kc},\{G^{r8},G^{r8}\}\}, &
T_{23}^{kc} = \{G^{k8},\{G^{rc},G^{r8}\}\}, &
T_{24}^{kc} = d^{c8e} \{J^k,\{G^{re},G^{r8}\}\}, \\
T_{25}^{kc} = d^{88e} \{J^k,\{G^{rc},G^{re}\}\}, &
T_{26}^{kc} = d^{c8e} \{G^{ke},\{J^r,G^{r8}\}\}, &
T_{27}^{kc} = d^{c8e} \{G^{k8},\{J^r,G^{re}\}\}, \\
T_{28}^{kc} = d^{88e} \{G^{kc},\{J^r,G^{re}\}\}, &
T_{29}^{kc} = d^{88e} \{G^{ke},\{J^r,G^{rc}\}\}, &
T_{30}^{kc} = d^{c88} \{J^2,J^k\}, \\
T_{31}^{kc} = \epsilon^{kim} f^{c8e} \{T^e,\{J^i,G^{m8}\}\}, &
T_{32}^{kc} = \epsilon^{kim} f^{c8e} \{T^8,\{J^i,G^{me}\}\}, &
T_{33}^{kc} = f^{c8e} f^{8eg} \mathcal{D}_4^{kg}, \\
T_{34}^{kc} = \delta^{c8} \mathcal{D}_4^{k8}, &
T_{35}^{kc} = \delta^{88} \mathcal{D}_4^{kc}, &
T_{36}^{kc} = \{\mathcal{D}_2^{kc},\{T^8,T^8\}\}, \\
T_{37}^{kc} = \{\mathcal{D}_2^{kc},\{G^{r8},G^{r8}\}\}, &
T_{38}^{kc} = \{\mathcal{D}_2^{k8},\{G^{rc},G^{r8}\}\}, &
T_{39}^{kc} = d^{c8e} \{J^2,\{G^{ke},T^8\}\}, \\
T_{40}^{kc} = d^{88e} \{J^2,\{G^{ke},T^c\}\}, &
T_{41}^{kc} = d^{c8e} \{\mathcal{D}_2^{k8},\{J^r,G^{re}\}\}, &
T_{42}^{kc} = d^{88e} \{\mathcal{D}_2^{kc},\{J^r,G^{re}\}\}, \\
T_{43}^{kc} = \{\{J^r,G^{rc}\},\{G^{k8},T^8\}\}, &
T_{44}^{kc} = \{\{J^r,G^{r8}\},\{G^{kc},T^8\}\}, &
T_{45}^{kc} = \{\{J^r,G^{r8}\},\{G^{k8},T^c\}\}, \\
T_{46}^{kc} = if^{c8e} \{J^2,[G^{ke},\{J^r,G^{r8}\}]\}, &
T_{47}^{kc} = if^{c8e} \{J^2,[G^{k8},\{J^r,G^{re}\}]\}, &
T_{48}^{kc} = if^{c8e} \{J^k,[\{J^i,G^{ie}\},\{J^r,G^{r8}\}]\}, \\
T_{49}^{kc} = if^{c8e} \{\{J^r,G^{re}\},[J^2,G^{k8}]\}, &
T_{50}^{kc} = if^{c8e} \{\{J^r,G^{r8}\},[J^2,G^{ke}]\}, &
T_{51}^{kc} = \delta^{c8} \mathcal{O}_5^{k8}, \\
T_{52}^{kc} = \{G^{kc},\{\{J^i,G^{i8}\},\{J^r,G^{r8}\}\}\}, &
T_{53}^{kc} = \{\mathcal{D}_2^{kc},\{T^8,\{J^r,G^{r8}\}\}\}, &
T_{54}^{kc} = \{\{J^r,G^{rc}\},\{G^{k8},\{J^i,G^{i8}\}\}\}, \\
T_{55}^{kc} = \{J^k,\{\{J^i,G^{ic}\},\{G^{r8},G^{r8}\}\}\}, &
T_{56}^{kc} = \{J^2,\{G^{k8},\{T^c,T^8\}\}\}, &
T_{57}^{kc} = \{J^2,\{G^{kc},\{G^{r8},G^{r8}\}\}\}, \\
T_{58}^{kc} = d^{c8e} \{\mathcal{D}_3^{k8},\{J^r,G^{re}\}\}, &
T_{59}^{kc} = d^{c8e} \{J^2,\{J^k,\{G^{re},G^{r8}\}\}\}, &
T_{60}^{kc} = d^{c8e} \{J^2,\{G^{k8},\{J^r,G^{re}\}\}\}, \\
T_{61}^{kc} = \epsilon^{kim} f^{c8e} \{J^2,\{T^e,\{J^i,G^{m8}\}\}\}, & &
\end{array}
\end{eqnarray}
and the corresponding coefficients are
\begin{equation}
t_{1} = \left[ \frac18 a_1^3 - \frac{1}{N_c^2} a_1b_2^2 - \frac{3}{4N_c^2} a_1^2b_3 + \frac{3}{8N_c^2} a_1^2c_3 \right] F_{\mathbf{27}}^{(1)} - \frac{3}{2N_c^2} a_1^2b_3 \frac{\Delta}{N_c} F_{\mathbf{27}}^{(2)} - \frac{1}{16} a_1^3 \frac{\Delta^2}{N_c^2} F_{\mathbf{27}}^{(3)},
\end{equation}
\begin{equation}
t_{2} = \frac14 a_1^3 F_{\mathbf{27}}^{(1)} + \left[ \frac{1}{4N_c^2} a_1^2b_3 + \frac{1}{4N_c^2} a_1^2c_3 \right] \frac{\Delta}{N_c} F_{\mathbf{27}}^{(2)},
\end{equation}
\begin{equation}
t_{3} = \frac16 a_1^3 F_{\mathbf{27}}^{(1)},
\end{equation}
\begin{equation}
t_{4} = \frac{1}{12} a_1^3 F_{\mathbf{27}}^{(1)} + \left[ \frac{1}{12N_c^2} a_1^2b_3 + \frac{1}{12N_c^2} a_1^2c_3 \right] \frac{\Delta}{N_c} F_{\mathbf{27}}^{(2)},
\end{equation}
\begin{equation}
t_{5} = \left[ \frac{7}{8N_c} a_1^2b_2 - \frac{1}{2N_c^3}b_2^3 + \frac{9}{2N_c^3} a_1b_2c_3 \right] F_{\mathbf{27}}^{(1)} + \frac{15}{8N_c} a_1^2b_2 \frac{\Delta}{N_c} F_{\mathbf{27}}^{(2)} + \frac{7}{4N_c} a_1^2b_2 \frac{\Delta^2}{N_c^2} F_{\mathbf{27}}^{(3)},
\end{equation}
\begin{equation}
t_{6} = \frac{1}{3N_c} a_1^2b_2 F_{\mathbf{27}}^{(1)},
\end{equation}
\begin{equation}
t_{7} = \frac{1}{6N_c} a_1^2b_2 F_{\mathbf{27}}^{(1)},
\end{equation}
\begin{equation}
t_{8} = \frac{1}{2N_c} a_1^2b_2 F_{\mathbf{27}}^{(1)},
\end{equation}
\begin{equation}
t_{9} = \frac{1}{4N_c} a_1^2b_2 F_{\mathbf{27}}^{(1)},
\end{equation}
\begin{equation}
t_{10} = - \frac{2}{N_c^3} a_1b_2b_3 F_{\mathbf{27}}^{(1)} - \frac{1}{4N_c} a_1^2b_2 \frac{\Delta}{N_c} F_{\mathbf{27}}^{(2)},
\end{equation}
\begin{equation}
t_{11} = \left[ \frac{1}{2N_c} a_1^2b_2 + \frac{2}{N_c^3} a_1b_2b_3 \right] F_{\mathbf{27}}^{(1)} + \frac{1}{2N_c} a_1^2b_2 \frac{\Delta}{N_c} F_{\mathbf{27}}^{(2)} + \frac{1}{12N_c} a_1^2b_2 \frac{\Delta^2}{N_c^2} F_{\mathbf{27}}^{(3)},
\end{equation}
\begin{equation}
t_{12} = \left[ \frac{3}{8N_c^2} a_1b_2^2 + \frac{3}{8N_c^2} a_1^2b_3 \right] F_{\mathbf{27}}^{(1)} + \frac{1}{8N_c^2} a_1b_2^2 \frac{\Delta}{N_c} F_{\mathbf{27}}^{(2)},
\end{equation}
\begin{equation}
t_{13} = \left[ \frac{1}{4N_c^2} a_1^2b_3 + \frac{1}{4N_c^2} a_1^2c_3 \right] F_{\mathbf{27}}^{(1)} + \left[ \frac{1}{N_c^2} a_1^2b_3 + \frac{1}{2N_c^2} a_1^2c_3 \right] \frac{\Delta}{N_c} F_{\mathbf{27}}^{(2)} + \frac18 a_1^3 \frac{\Delta^2}{N_c^2} F_{\mathbf{27}}^{(3)},
\end{equation}
\begin{equation}
t_{14} = \left[ \frac{1}{6N_c^2} a_1^2b_3 + \frac{1}{6N_c^2} a_1^2c_3 \right] F_{\mathbf{27}}^{(1)} - \frac{2}{3N_c^2} a_1^2b_3 \frac{\Delta}{N_c} F_{\mathbf{27}}^{(2)} - \frac{1}{36} a_1^3 \frac{\Delta^2}{N_c^2} F_{\mathbf{27}}^{(3)},
\end{equation}
\begin{equation}
t_{15} = \frac{1}{3N_c^2} a_1^2b_3 F_{\mathbf{27}}^{(1)} + \frac{2}{3N_c^2} a_1^2b_3 \frac{\Delta}{N_c} F_{\mathbf{27}}^{(2)},
\end{equation}
\begin{equation}
t_{16} = \left[ \frac{1}{4N_c^2} a_1b_2^2 + \frac{3}{8N_c^2} a_1^2c_3 \right] F_{\mathbf{27}}^{(1)} + \left[ - \frac{1}{4N_c^2} a_1b_2^2 - \frac{1}{2N_c^2} a_1^2b_3 + \frac{3}{4N_c^2} a_1^2c_3 \right] \frac{\Delta}{N_c} F_{\mathbf{27}}^{(2)} + \frac18 a_1^3 \frac{\Delta^2}{N_c^2} F_{\mathbf{27}}^{(3)},
\end{equation}
\begin{equation}
t_{17} = \left[ \frac{1}{2N_c^2} a_1^2b_3 + \frac{1}{2N_c^2} a_1^2c_3 \right] F_{\mathbf{27}}^{(1)} + \left[ \frac{1}{2N_c^2} a_1^2b_3 + \frac{1}{4N_c^2} a_1^2c_3 \right] \frac{\Delta}{N_c} F_{\mathbf{27}}^{(2)} + \frac18 a_1^3 \frac{\Delta^2}{N_c^2} F_{\mathbf{27}}^{(3)},
\end{equation}
\begin{equation}
t_{18} = \frac{5}{6N_c^2} a_1^2c_3 F_{\mathbf{27}}^{(1)} + \left[ \frac16 a_1^3 - \frac{1}{3N_c^2} a_1^2b_3 + \frac{7}{6N_c^2} a_1^2c_3 \right] \frac{\Delta}{N_c} F_{\mathbf{27}}^{(2)} + \frac16 a_1^3 \frac{\Delta^2}{N_c^2} F_{\mathbf{27}}^{(3)},
\end{equation}
\begin{equation}
t_{19} = \frac{1}{3N_c^2} a_1^2c_3 F_{\mathbf{27}}^{(1)} + \left[ \frac{1}{3N_c^2} a_1^2b_3 + \frac{1}{6N_c^2} a_1^2c_3 \right] \frac{\Delta}{N_c} F_{\mathbf{27}}^{(2)},
\end{equation}
\begin{equation}
t_{20} = \frac{1}{4N_c^2} a_1b_2^2 F_{\mathbf{27}}^{(1)},
\end{equation}
\begin{equation}
t_{21} = \frac{1}{2N_c^2} a_1b_2^2 F_{\mathbf{27}}^{(1)},
\end{equation}
\begin{equation}
t_{22} = \left[ - \frac{1}{N_c^2} a_1^2b_3 - \frac{1}{2N_c^2} a_1^2c_3 \right] F_{\mathbf{27}}^{(1)} - \frac{2}{N_c^2} a_1^2b_3 \frac{\Delta}{N_c} F_{\mathbf{27}}^{(2)} + \frac{1}{12} a_1^3 \frac{\Delta^2}{N_c^2} F_{\mathbf{27}}^{(3)},
\end{equation}
\begin{equation}
t_{23} = \left[ \frac{1}{N_c^2} a_1^2b_3 - \frac{1}{2N_c^2} a_1^2c_3 \right] F_{\mathbf{27}}^{(1)} + \frac{2}{N_c^2} a_1^2b_3 \frac{\Delta}{N_c} F_{\mathbf{27}}^{(2)} + \frac{1}{12} a_1^3 \frac{\Delta^2}{N_c^2} F_{\mathbf{27}}^{(3)},
\end{equation}
\begin{equation}
t_{24} = \left[ - \frac{3}{2N_c^2} a_1^2b_3 - \frac{1}{4N_c^2} a_1^2c_3 \right] F_{\mathbf{27}}^{(1)} + \left[ - \frac14 a_1^3 - \frac{3}{N_c^2} a_1^2b_3 - \frac{1}{N_c^2} a_1^2c_3 \right] \frac{\Delta}{N_c} F_{\mathbf{27}}^{(2)} - \frac{7}{24} a_1^3 \frac{\Delta^2}{N_c^2} F_{\mathbf{27}}^{(3)},
\end{equation}
\begin{equation}
t_{25} = \left[ \frac{1}{2N_c^2} a_1^2b_3 - \frac{1}{4N_c^2} a_1^2c_3 \right] F_{\mathbf{27}}^{(1)} + \frac{1}{N_c^2} a_1^2b_3 \frac{\Delta}{N_c} F_{\mathbf{27}}^{(2)} + \frac{1}{24} a_1^3 \frac{\Delta^2}{N_c^2} F_{\mathbf{27}}^{(3)},
\end{equation}
\begin{equation}
t_{26} = \left[ \frac{2}{N_c^2} a_1^2b_3 - \frac{1}{2N_c^2} a_1^2c_3 \right] F_{\mathbf{27}}^{(1)} + \left[ - \frac{1}{4N_c^2} a_1^2b_3 - \frac{7}{4N_c^2} a_1^2c_3 \right] \frac{\Delta}{N_c} F_{\mathbf{27}}^{(2)} - \frac16 a_1^3 \frac{\Delta^2}{N_c^2} F_{\mathbf{27}}^{(3)},
\end{equation}
\begin{equation}
t_{27} = \left[ - \frac{1}{2N_c^2} a_1^2b_3 + \frac{3}{4N_c^2} a_1^2c_3 \right] F_{\mathbf{27}}^{(1)} + \left[ \frac14 a_1^3 - \frac{1}{N_c^2} a_1^2b_3 + \frac{3}{2N_c^2} a_1^2c_3 \right] \frac{\Delta}{N_c} F_{\mathbf{27}}^{(2)} + \frac18 a_1^3 \frac{\Delta^2}{N_c^2} F_{\mathbf{27}}^{(3)},
\end{equation}
\begin{equation}
t_{28} = \frac{1}{2N_c^2} a_1^2c_3 F_{\mathbf{27}}^{(1)} + \left[ \frac{1}{2N_c^2} a_1^2b_3 + \frac{1}{4N_c^2} a_1^2c_3 \right] \frac{\Delta}{N_c} F_{\mathbf{27}}^{(2)},
\end{equation}
\begin{equation}
t_{29} = \left[ \frac{1}{2N_c^2} a_1^2b_3 - \frac{1}{4N_c^2} a_1^2c_3 \right] F_{\mathbf{27}}^{(1)} + \left[ \frac{1}{2N_c^2} a_1^2b_3 - \frac{1}{4N_c^2} a_1^2c_3 \right] \frac{\Delta}{N_c} F_{\mathbf{27}}^{(2)} - \frac{1}{24} a_1^3 \frac{\Delta^2}{N_c^2} F_{\mathbf{27}}^{(3)},
\end{equation}
\begin{equation}
t_{30} = \left[ \frac{1}{6N_c^2} a_1^2b_3 + \frac{1}{6N_c^2} a_1^2c_3 \right] F_{\mathbf{27}}^{(1)} + \left[
\frac{2}{3N_c^2} a_1^2b_3 + \frac{1}{3N_c^2} a_1^2c_3 \right] \frac{\Delta}{N_c} F_{\mathbf{27}}^{(2)} + \frac{1}{12} a_1^3 \frac{\Delta^2}{N_c^2} F_{\mathbf{27}}^{(3)},
\end{equation}
\begin{equation}
t_{31} = \left[ - \frac{1}{4N_c^2} a_1b_2^2 - \frac{1}{4N_c^2} a_1^2b_3 + \frac{3}{8N_c^2} a_1^2c_3 \right] F_{\mathbf{27}}^{(1)} + \left[ \frac18 a_1^3 - \frac{1}{4N_c^2} a_1b_2^2 - \frac{1}{2N_c^2} a_1^2b_3 + \frac{1}{2N_c^2} a_1^2c_3\right] \frac{\Delta}{N_c} F_{\mathbf{27}}^{(2)} + \frac{1}{48} a_1^3 \frac{\Delta^2}{N_c^2} F_{\mathbf{27}}^{(3)},
\end{equation}
\begin{equation}
t_{32} = \frac{1}{4N_c^2} a_1b_2^2 \frac{\Delta}{N_c} F_{\mathbf{27}}^{(2)},
\end{equation}
\begin{equation}
t_{33} = \left[ \frac{1}{4N_c^3} b_2^3 + \frac{9}{4N_c^3} a_1b_2c_3 \right] F_{\mathbf{27}}^{(1)} + \frac{1}{4N_c} a_1^2b_2 \frac{\Delta}{N_c} F_{\mathbf{27}}^{(2)} + \frac{23}{24N_c} a_1^2b_2 \frac{\Delta^2}{N_c^2} F_{\mathbf{27}}^{(3)},
\end{equation}
\begin{equation}
t_{34} = \frac{2}{3N_c^3} a_1b_2c_3 F_{\mathbf{27}}^{(1)} + \frac{1}{9N_c} a_1^2b_2 \frac{\Delta^2}{N_c^2} F_{\mathbf{27}}^{(3)},
\end{equation}
\begin{equation}
t_{35} = \frac{1}{3N_c^3} a_1b_2c_3 F_{\mathbf{27}}^{(1)} + \frac{1}{6N_c} a_1^2b_2 \frac{\Delta^2}{N_c^2} F_{\mathbf{27}}^{(3)},
\end{equation}
\begin{equation}
t_{36} = \frac{1}{4N_c^3} b_2^3 F_{\mathbf{27}}^{(1)},
\end{equation}
\begin{equation}
t_{37} = - \frac{1}{N_c^3} a_1b_2c_3 F_{\mathbf{27}}^{(1)} - \frac{1}{2N_c} a_1^2b_2 \frac{\Delta}{N_c} F_{\mathbf{27}}^{(2)} - \frac{1}{2N_c} a_1^2b_2 \frac{\Delta^2}{N_c^2} F_{\mathbf{27}}^{(3)},
\end{equation}
\begin{equation}
t_{38} = - \frac{2}{N_c^3} a_1b_2c_3 F_{\mathbf{27}}^{(1)} - \frac{1}{2N_c} a_1^2b_2 \frac{\Delta}{N_c} F_{\mathbf{27}}^{(2)} - \frac{1}{3N_c} a_1^2b_2 \frac{\Delta^2}{N_c^2} F_{\mathbf{27}}^{(3)},
\end{equation}
\begin{equation}
t_{39} = \left[ \frac{1}{N_c^3} a_1b_2b_3 + \frac{1}{2N_c^3} a_1b_2c_3 \right] F_{\mathbf{27}}^{(1)} + \frac{1}{12N_c} a_1^2b_2 \frac{\Delta^2}{N_c^2} F_{\mathbf{27}}^{(3)},
\end{equation}
\begin{equation}
t_{40} = \left[ \frac{1}{2N_c^3} a_1b_2b_3 + \frac{1}{4N_c^3} a_1b_2c_3 \right] F_{\mathbf{27}}^{(1)} + \frac{1}{12N_c} a_1^2b_2 \frac{\Delta^2}{N_c^2} F_{\mathbf{27}}^{(3)},
\end{equation}
\begin{equation}
t_{41} = \left[ - \frac{1}{N_c^3} a_1b_2b_3 + \frac{1}{2N_c^3} a_1b_2c_3 \right] F_{\mathbf{27}}^{(1)} + \frac{1}{12N_c} a_1^2b_2 \frac{\Delta^2}{N_c^2} F_{\mathbf{27}}^{(3)},
\end{equation}
\begin{equation}
t_{42} = \left[ - \frac{1}{2N_c^3} a_1b_2b_3 + \frac{1}{4N_c^3} a_1b_2c_3 \right] F_{\mathbf{27}}^{(1)} + \frac{1}{6N_c} a_1^2b_2 \frac{\Delta^2}{N_c^2} F_{\mathbf{27}}^{(3)},
\end{equation}
\begin{equation}
t_{43} = \frac{1}{N_c^3} a_1b_2b_3 F_{\mathbf{27}}^{(1)} + \frac{1}{4N_c} a_1^2b_2 \frac{\Delta}{N_c} F_{\mathbf{27}}^{(2)} + \frac{1}{12N_c} a_1^2b_2 \frac{\Delta^2}{N_c^2} F_{\mathbf{27}}^{(3)},
\end{equation}
\begin{equation}
t_{44} = \frac{1}{N_c^3} a_1b_2b_3 F_{\mathbf{27}}^{(1)} - \frac{1}{12N_c} a_1^2b_2 \frac{\Delta^2}{N_c^2} F_{\mathbf{27}}^{(3)},
\end{equation}
\begin{equation}
t_{45} = \frac{1}{N_c^3} a_1b_2b_3 F_{\mathbf{27}}^{(1)} + \frac{1}{4N_c} a_1^2b_2 \frac{\Delta}{N_c} F_{\mathbf{27}}^{(2)},
\end{equation}
\begin{equation}
t_{46} = \left[ \frac{1}{N_c^3} a_1b_2b_3- \frac{1}{2N_c^3} a_1b_2c_3 \right] F_{\mathbf{27}}^{(1)} - \frac{1}{12N_c} a_1^2b_2 \frac{\Delta^2}{N_c^2} F_{\mathbf{27}}^{(3)},
\end{equation}
\begin{equation}
t_{47} = \frac{1}{N_c^3} a_1b_2c_3 F_{\mathbf{27}}^{(1)} + \frac{1}{12N_c} a_1^2b_2 \frac{\Delta^2}{N_c^2} F_{\mathbf{27}}^{(3)},
\end{equation}
\begin{equation}
t_{48} = \left[ \frac{1}{N_c^3} a_1b_2b_3 - \frac{1}{N_c^3} a_1b_2c_3 \right] F_{\mathbf{27}}^{(1)} - \frac{1}{4N_c} a_1^2b_2 \frac{\Delta}{N_c} F_{\mathbf{27}}^{(2)} - \frac{5}{8N_c}a_1^2b_2 \frac{\Delta^2}{N_c^2} F_{\mathbf{27}}^{(3)},
\end{equation}
\begin{equation}
t_{49} = \left[ - \frac{1}{N_c^3} a_1b_2b_3 + \frac{1}{2N_c^3} a_1b_2c_3 \right] F_{\mathbf{27}}^{(1)} - \frac{1}{12N_c} a_1^2b_2 \frac{\Delta^2}{N_c^2} F_{\mathbf{27}}^{(3)},
\end{equation}
\begin{equation}
t_{50} = - \frac{1}{2N_c^3} a_1b_2c_3 F_{\mathbf{27}}^{(1)},
\end{equation}
\begin{equation}
t_{51} = \frac{1}{2N_c^2} a_1^2c_3 \frac{\Delta}{N_c} F_{\mathbf{27}}^{(2)},
\end{equation}
\begin{equation}
t_{52} = \frac{1}{2N_c^2} a_1^2c_3 \frac{\Delta}{N_c} F_{\mathbf{27}}^{(2)},
\end{equation}
\begin{equation}
t_{53} = - \frac{1}{4N_c^2} a_1b_2^2 \frac{\Delta}{N_c} F_{\mathbf{27}}^{(2)},
\end{equation}
\begin{equation}
t_{54} = \left[ \frac{1}{N_c^2} a_1^2b_3 - \frac{1}{2N_c^2} a_1^2c_3 \right] \frac{\Delta}{N_c} F_{\mathbf{27}}^{(2)},
\end{equation}
\begin{equation}
t_{55} = \left[ - \frac{1}{N_c^2} a_1^2b_3 + \frac{1}{2N_c^2} a_1^2c_3 \right] \frac{\Delta}{N_c} F_{\mathbf{27}}^{(2)},
\end{equation}
\begin{equation}
t_{56} = \frac{1}{4N_c^2} a_1b_2^2 \frac{\Delta}{N_c} F_{\mathbf{27}}^{(2)},
\end{equation}
\begin{equation}
t_{57} = - \frac{1}{N_c^2} a_1^2c_3 \frac{\Delta}{N_c} F_{\mathbf{27}}^{(2)},
\end{equation}
\begin{equation}
t_{58} = \left[ \frac{1}{4N_c^2} a_1^2b_3 - \frac{1}{8N_c^2} a_1^2c_3 \right] \frac{\Delta}{N_c} F_{\mathbf{27}}^{(2)},
\end{equation}
\begin{equation}
t_{59} = \left[ - \frac{1}{2N_c^2} a_1^2b_3 - \frac{1}{2N_c^2} a_1^2c_3 \right] \frac{\Delta}{N_c} F_{\mathbf{27}}^{(2)},
\end{equation}
\begin{equation}
t_{60} = \frac{3}{8N_c^2} a_1^2c_3 \frac{\Delta}{N_c} F_{\mathbf{27}}^{(2)},
\end{equation}
\begin{equation}
t_{61} = \frac{3}{8N_c^2} a_1^2c_3 \frac{\Delta}{N_c} F_{\mathbf{27}}^{(2)}.
\end{equation}
Notice that in Eq.~(\ref{eq:da27}) the singlet and octet pieces must be subtracted off in order to have a truly $\mathbf{27}$ contribution.

\section{Matrix elements of baryon operators\label{secc:mtx}}

In this appendix we list the values of the matrix elements of baryon operators between SU(6) symmetric states that make up the axial vector couplings $g_A$ and $g$. In Ref.~\cite{fh06} a somewhat different operator basis was used; in the present analysis there appear some other operators whose matrix elements have not been evaluated yet. We thus consider it convenient to provide all the matrix elements required here once and for all. Of course, only nontrivial matrix elements are given: matrix elements which either vanish or do not contribute to any observed processes concerned here are not listed. Examples of the first and second case are $\langle B_1|\delta^{c8}J^k|B_2\rangle$ and $\langle B_1|[J^2,[T^8,G^{kc}]]|B_2 \rangle$, respectively. Besides, matrix elements of higher-order operators obtained by anticommuting with $J^2$ are also trivial and will be omitted hereafter.

For the tree-level and singlet contributions, we identify the operators listed in Table \ref{tab:mtxS}.

\begingroup
\begin{table}[h]
\caption{\label{tab:mtxS}Matrix elements of baryon operators: Tree-level and singlet contributions.}
\begin{center}
\begin{tabular}{lccccccccccc}
\hline\hline
$B_1B_2$ & $np$ & $\Sigma^\pm \Lambda$ & $\Lambda p$ & $\Sigma^- n$ & $\Xi^-\Lambda$ & $\Xi^-\Sigma^0$ & $\Xi^0\Sigma^+$ & $\Delta N$ & $\Sigma^*\Lambda$ & $\Sigma^*\Sigma$ & $\Xi^*\Xi$\\
\hline
$\langle G^{kc} \rangle$ & $\frac{5}{6}$ & $\frac{1}{\sqrt{6}}$ & $-\frac12\sqrt{\frac{3}{2}}$ & $\frac{1}{6}$ & $\frac{1}{2 \sqrt{6}}$ & $\frac{5}{6 \sqrt{2}}$ & $\frac{5}{6}
$ & $-1$ & $-1$ & $-1$ & $-1$ \\
$\langle \mathcal{D}_{2}^{kc} \rangle$ & $\frac{1}{2}$ & $0$ & $-\frac12\sqrt{\frac{3}{2}}$ & $-\frac{1}{2}$ & $\frac12\sqrt{\frac{3}{2}}$ & $\frac{1}{2 \sqrt{2}}$ & $\frac{1}{2}$ & $0$ & $0$ & $0$ & $0$ \\
$\langle \mathcal{D}_{3}^{kc} \rangle$ & $\frac{5}{2}$ & $\sqrt{\frac{3}{2}}$ & $-\frac32\sqrt{\frac{3}{2}}$ & $\frac{1}{2}$ & $\frac12\sqrt{\frac{3}{2}}$ & $\frac{5}{2 \sqrt{2}}
$ & $\frac{5}{2}$ & $0$ & $0$ & $0$ & $0$ \\
$\langle \mathcal{O}_{3}^{kc} \rangle$ & $0$ & $0$ & $0$ & $0$ & $0$ & $0$ & $0$ & $-\frac{9}{2}$ & $-\frac{9}{2}$ & $-\frac{9}{2}$ & $-\frac{9}{2}$ \\
\hline\hline
\end{tabular}
\end{center}
\end{table}
\endgroup

For the octet contribution the matrix elements are listed in Table \ref{tab:mtxO}.

\begingroup
\begin{table}[h]
\caption{\label{tab:mtxO}Matrix elements of baryon operators: Octet contribution.}
\begin{center}
\begin{tabular}{lccccccccccc}
\hline\hline
$B_1B_2$ & $np$ & $\Sigma^\pm \Lambda$ & $\Lambda p$ & $\Sigma^- n$ & $\Xi^-\Lambda$ & $\Xi^-\Sigma^0$ & $\Xi^0\Sigma^+$ & $\Delta N$ & $\Sigma^*\Lambda$ & $\Sigma^*\Sigma$ & $\Xi^*\Xi$ \\
\hline
$\langle O_{1}^{kc} \rangle$ & $\frac{5}{6 \sqrt{3}}$ & $\frac{1}{3 \sqrt{2}}$ & $\frac{1}{4 \sqrt{2}}$ & $-\frac{1}{12 \sqrt{3}}$ & $-\frac{1}{12 \sqrt{2}}$ & $-\frac{5}{12\sqrt{6}}$ & $-\frac{5}{12 \sqrt{3}}$ & $-\frac{1}{\sqrt{3}}$ & $-\frac{1}{\sqrt{3}}$ & $-\frac{1}{\sqrt{3}}$ & $-\frac{1}{\sqrt{3}}$ \\
$\langle O_{3}^{kc} \rangle$ & $\frac{1}{2 \sqrt{3}}$ & $0$ & $\frac{1}{4 \sqrt{2}}$ & $\frac{1}{4 \sqrt{3}}$ & $-\frac{1}{4 \sqrt{2}}$ & $-\frac{1}{4 \sqrt{6}}$ & $-\frac{1}{4\sqrt{3}}$ & $0$ & $0$ & $0$ & $0$ \\
$\langle O_{4}^{kc} \rangle$ & $\frac{5}{2 \sqrt{3}}$ & $0$ & $-\frac{3}{4 \sqrt{2}}$ & $\frac{1}{4 \sqrt{3}}$ & $-\frac{1}{4 \sqrt{2}}$ & $-\frac{5}{4 \sqrt{6}}$ & $-\frac{5}{4\sqrt{3}}$ & $-\sqrt{3}$ & $0$ & $0$ & $\sqrt{3}$ \\
$\langle O_{5}^{kc} \rangle$ & $\frac{1}{2 \sqrt{3}}$ & $0$ & $\frac{1}{4 \sqrt{2}}$ & $-\frac{\sqrt{3}}{4}$ & $-\frac{5}{4 \sqrt{2}}$ & $-\frac{1}{4 \sqrt{6}}$ & $-\frac{1}{4\sqrt{3}}$ & $0$ & $0$ & $-2 \sqrt{3}$ & $-\sqrt{3}$ \\
$\langle O_{6}^{kc} \rangle$ & $\frac{5}{2 \sqrt{3}}$ & $\frac{1}{\sqrt{2}}$ & $\frac{3}{4 \sqrt{2}}$ & $-\frac{1}{4 \sqrt{3}}$ & $-\frac{1}{4 \sqrt{2}}$ & $-\frac{5}{4 \sqrt{6}}$ & $-\frac{5}{4 \sqrt{3}}$ & $0$ & $0$ & $0$ & $0$ \\
$\langle O_{7}^{kc} \rangle$ & $0$ & $0$ & $0$ & $0$ & $0$ & $0$ & $0$ & $-\frac{3 \sqrt{3}}{2}$ & $-\frac{3 \sqrt{3}}{2}$ & $-\frac{3 \sqrt{3}}{2}$ & $-\frac{3 \sqrt{3}}{2}$ \\
$\langle O_{8}^{kc} \rangle$ & $\frac{5}{4 \sqrt{3}}$ & $0$ & $\frac{3}{8 \sqrt{2}}$ & $\frac{\sqrt{3}}{8}$ & $-\frac{5}{8 \sqrt{2}}$ & $-\frac{5}{8 \sqrt{6}}$ & $-\frac{5}{8\sqrt{3}}$ & $-\frac{3 \sqrt{3}}{2}$ & $\frac{\sqrt{3}}{2}$ & $-\frac{\sqrt{3}}{2}$ & $2 \sqrt{3}$ \\
$\langle O_{9}^{kc} \rangle$ & $\frac{5}{4 \sqrt{3}}$ & $0$ & $\frac{3}{8 \sqrt{2}}$ & $\frac{\sqrt{3}}{8}$ & $-\frac{5}{8 \sqrt{2}}$ & $-\frac{5}{8 \sqrt{6}}$ & $-\frac{5}{8\sqrt{3}}$ & $0$ & $\frac{\sqrt{3}}{2}$ & $-\frac{7 \sqrt{3}}{2}$ & $-\sqrt{3}$ \\
$\langle O_{10}^{kc} \rangle$ & $\sqrt{3}$ & $0$ & $-\frac{3}{2 \sqrt{2}}$ & $-\frac{\sqrt{3}}{2}$ & $-\frac{3}{2 \sqrt{2}}$ & $-\frac12\sqrt{\frac{3}{2}}$ & $-\frac{\sqrt{3}}{2}$ & $0$ & $0$ & $0$ & $0$ \\
$\langle O_{11}^{kc} \rangle$ & $\frac{5}{4 \sqrt{3}}$ & $-\frac{1}{\sqrt{2}}$ & $\frac{3}{8 \sqrt{2}}$ & $\frac{11}{8 \sqrt{3}}$ & $-\frac{13}{8 \sqrt{2}}$ & $-\frac{5}{8 \sqrt{6}}$ & $-\frac{5}{8 \sqrt{3}}$ & $0$ & $0$ & $0$ & $0$ \\
$\langle O_{14}^{kc} \rangle$ & $\frac{\sqrt{3}}{4}$ & $0$ & $\frac{3}{8 \sqrt{2}}$ & $-\frac{3 \sqrt{3}}{8}$ & $-\frac{15}{8 \sqrt{2}}$ & $-\frac18\sqrt{\frac{3}{2}}$ & $-\frac{\sqrt{3}}{8}$ & $0$ & $0$ & $0$ & $0$ \\
$\langle O_{15}^{kc} \rangle$ & $\frac{5 \sqrt{3}}{4}$ & $0$ & $-\frac{9}{8 \sqrt{2}}$ & $\frac{\sqrt{3}}{8}$ & $-\frac{3}{8 \sqrt{2}}$ & $-\frac58 \sqrt{\frac{3}{2}}$ & $-\frac{5\sqrt{3}}{8}$ & $0$ & $0$ & $0$ & $0$ \\
$\langle O_{18}^{kc} \rangle$ & $0$ & $-\frac{3}{2 \sqrt{2}}$ & $-\frac{27}{16 \sqrt{2}}$ & $-\frac{\sqrt{3}}{16}$ & $\frac{3}{16 \sqrt{2}}$ & $\frac{25}{16} \sqrt{\frac{3}{2}}$ & $\frac{25 \sqrt{3}}{16}$ & $\frac{9 \sqrt{3}}{2}$ & $\frac{9 \sqrt{3}}{4}$ & $-\frac{9 \sqrt{3}}{4}$ & $-\frac{9 \sqrt{3}}{4}$ \\
$\langle O_{19}^{kc} \rangle$ & $0$ & $\frac{3}{2 \sqrt{2}}$ & $\frac{27}{16 \sqrt{2}}$ & $\frac{\sqrt{3}}{16}$ & $-\frac{3}{16 \sqrt{2}}$ & $-\frac{25}{16} \sqrt{\frac{3}{2}}$ & $-\frac{25 \sqrt{3}}{16}$ & $0$ & $\frac{9 \sqrt{3}}{4}$ & $\frac{27 \sqrt{3}}{4}$ & $\frac{27 \sqrt{3}}{4}$ \\
$\langle O_{20}^{kc} \rangle$ & $0$ & $0$ & $0$ & $0$ & $0$ & $0$ & $0$ & $-\frac{9 \sqrt{3}}{2}$ & $\frac{3 \sqrt{3}}{2}$ & $-\frac{3 \sqrt{3}}{2}$ & $6 \sqrt{3}$ \\
$\langle O_{21}^{kc} \rangle$ & $0$ & $0$ & $0$ & $0$ & $0$ & $0$ & $0$ & $0$ & $\frac{3 \sqrt{3}}{2}$ & $-\frac{21 \sqrt{3}}{2}$ & $-3 \sqrt{3}$ \\
$\langle O_{22}^{kc} \rangle$ & $0$ & $-\frac{3}{\sqrt{2}}$ & $-\frac{27}{8 \sqrt{2}}$ & $-\frac{\sqrt{3}}{8}$ & $\frac{3}{8 \sqrt{2}}$ & $\frac{25}{8} \sqrt{\frac{3}{2}}$ & $\frac{25\sqrt{3}}{8}$ & $0$ & $0$ & $0$ & $0$ \\
$\langle O_{29}^{kc} \rangle$ & $\frac{5 \sqrt{3}}{4}$ & $0$ & $\frac{9}{8 \sqrt{2}}$ & $\frac{3 \sqrt{3}}{8}$ & $-\frac{15}{8 \sqrt{2}}$ & $-\frac58 \sqrt{\frac{3}{2}}$ & $-\frac{5 \sqrt{3}}{8}$ & $0$ & $0$ & $0$ & $0$ \\
\hline\hline
\end{tabular}
\end{center}
\end{table}
\endgroup

Finally, for the $\mathbf{27}$ contribution we provide Table \ref{tab:mtxT}.

\begingroup
\begin{table}[h]
\caption{\label{tab:mtxT}Matrix elements of baryon operators: $\mathbf{27}$ contribution.}
\begin{center}
\begin{tabular}{lccccccccccc}
\hline\hline
$B_1B_2$ & $np$ & $\Sigma^\pm \Lambda$ & $\Lambda p$ & $\Sigma^- n$ & $\Xi^-\Lambda$ & $\Xi^-\Sigma^0$ & $\Xi^0\Sigma^+$ & $\Delta N$ & $\Sigma^*\Lambda$ & $\Sigma^*\Sigma$ & $\Xi^*\Xi$ \\
\hline
$\langle T_{1}^{kc} \rangle$ & $0$ & $0$ & $-\frac38 \sqrt{\frac{3}{2}}$ & $\frac{1}{8}$ & $\frac18\sqrt{\frac{3}{2}}$ & $\frac{5}{8 \sqrt{2}}$ & $\frac{5}{8}$ & $0$ & $0$ & $0$ & $0$ \\
$\langle T_{2}^{kc} \rangle$ & $\frac{5}{18}$ & $\frac{1}{3 \sqrt{6}}$ & $-\frac{1}{8 \sqrt{6}}$ & $\frac{1}{72}$ & $\frac{1}{24 \sqrt{6}}$ & $\frac{5}{72 \sqrt{2}}$ & $\frac{5}{72} $ & $-\frac{1}{3}$ & $-\frac{1}{3}$ & $-\frac{1}{3}$ & $-\frac{1}{3}$ \\
$\langle T_{5}^{kc} \rangle$ & $0$ & $0$ & $-\frac38 \sqrt{\frac{3}{2}}$ & $-\frac{3}{8}$ & $\frac38\sqrt{\frac{3}{2}}$ & $\frac{3}{8 \sqrt{2}}$ & $\frac{3}{8}$ & $0$ & $0$ & $0$ & $0$ \\
$\langle T_{7}^{kc} \rangle$ & $\frac{1}{2}$ & $0$ & $-\frac12\sqrt{\frac{3}{2}}$ & $-\frac{1}{2}$ & $\frac12\sqrt{\frac{3}{2}}$ & $\frac{1}{2 \sqrt{2}}$ & $\frac{1}{2}$ & $0$ & $0$ & $0$ & $0$ \\
$\langle T_{8}^{kc} \rangle$ & $\frac{5}{6}$ & $0$ & $\frac18\sqrt{\frac{3}{2}}$ & $-\frac{1}{24}$ & $\frac{1}{8 \sqrt{6}}$ & $\frac{5}{24 \sqrt{2}}$ & $\frac{5}{24}$ & $-1$ & $0$ & $0$ & $1$ \\
$\langle T_{9}^{kc} \rangle$ & $-\frac{1}{6}$ & $0$ & $-\frac{1}{4 \sqrt{6}}$ & $\frac{1}{4}$ & $\frac{5}{4 \sqrt{6}}$ & $\frac{1}{12 \sqrt{2}}$ & $\frac{1}{12}$ & $0$ & $0$ & $2$ & $1$ \\
$\langle T_{10}^{kc} \rangle$ & $0$ & $0$ & $-\frac{9}{16} \sqrt{\frac{3}{2}}$ & $-\frac{1}{16}$ & $\frac{1}{16}\sqrt{\frac{3}{2}}$ & $\frac{25}{16 \sqrt{2}}$ & $\frac{25}{16}$ & $0$ & $0$ & $0$ & $0$ \\
$\langle T_{11}^{kc} \rangle$ & $0$ & $0$ & $\frac{9}{16}\sqrt{\frac{3}{2}}$ & $\frac{1}{16}$ & $-\frac{1}{16}\sqrt{\frac{3}{2}}$ & $-\frac{25}{16 \sqrt{2}}$ & $-\frac{25}{16}$ & $0$ & $0$ & $0$ & $0$ \\
$\langle T_{12}^{kc} \rangle$ & $0$ & $0$ & $-\frac98\sqrt{\frac{3}{2}}$ & $\frac{3}{8}$ & $\frac38\sqrt{\frac{3}{2}}$ & $\frac{15}{8 \sqrt{2}}$ & $\frac{15}{8}$ & $0$ & $0$ & $0$ & $0$ \\
$\langle T_{13}^{kc} \rangle$ & $\frac{5}{6}$ & $\frac{1}{\sqrt{6}}$ & $-\frac18\sqrt{\frac{3}{2}}$ & $\frac{1}{24}$ & $\frac{1}{8 \sqrt{6}}$ & $\frac{5}{24 \sqrt{2}}$ & $\frac{5}{24}$ & $0$ & $0$ & $0$ & $0$ \\
$\langle T_{15}^{kc} \rangle$ & $\frac{5}{2}$ & $\sqrt{\frac{3}{2}}$ & $-\frac32 \sqrt{\frac{3}{2}}$ & $\frac{1}{2}$ & $\frac12\sqrt{\frac{3}{2}}$ & $\frac{5}{2 \sqrt{2}}$ & $\frac{5}{2}$ & $0$ & $0$ & $0$ & $0$ \\
$\langle T_{17}^{kc} \rangle$ & $0$ & $0$ & $0$ & $0$ & $0$ & $0$ & $0$ & $-\frac{3}{2}$ & $-\frac{3}{2}$ & $-\frac{3}{2}$ & $-\frac{3}{2}$ \\
$\langle T_{19}^{kc} \rangle$ & $0$ & $0$ & $0$ & $0$ & $0$ & $0$ & $0$ & $-\frac{9}{2}$ & $-\frac{9}{2}$ & $-\frac{9}{2}$ & $-\frac{9}{2}$ \\
$\langle T_{20}^{kc} \rangle$ & $\frac{5}{2}$ & $0$ & $-\frac34\sqrt{\frac{3}{2}}$ & $\frac{1}{4}$ & $\frac14\sqrt{\frac{3}{2}}$ & $\frac{5}{4 \sqrt{2}}$ & $\frac{5}{4}$ & $-3$ & $0$ & $0$ & $-3$ \\
$\langle T_{21}^{kc} \rangle$ & $\frac{1}{2}$ & $0$ & $\frac18\sqrt{\frac{3}{2}}$ & $-\frac{3}{8}$ & $\frac58 \sqrt{\frac{3}{2}}$ & $\frac{1}{8 \sqrt{2}}$ & $\frac{1}{8}$ & $0$ & $0$ & $0$ & $3$ \\
$\langle T_{22}^{kc} \rangle$ & $\frac{5}{24}$ & $\sqrt{\frac{2}{3}}$ & $-\frac{5}{16}\sqrt{\frac{3}{2}}$ & $\frac{13}{48}$ & $\frac{7}{16}\sqrt{\frac{3}{2}}$ & $\frac{145}{48\sqrt{2}}$ & $\frac{145}{48}$ & $-\frac{3}{4}$ & $-1$ & $-2$ & $-\frac{13}{4}$ \\
$\langle T_{23}^{kc} \rangle$ & $\frac{5}{24}$ & $0$ & $-\frac{1}{32}\sqrt{\frac{3}{2}}$ & $\frac{11}{32}$ & $\frac{65}{32 \sqrt{6}}$ & $\frac{5}{96 \sqrt{2}}$ & $\frac{5}{96}$ & $0$ & $-\frac{1}{2}$ & $-\frac{5}{2}$ & $-\frac{5}{4}$ \\
$\langle T_{24}^{kc} \rangle$ & $\frac{5}{12}$ & $-\frac{1}{\sqrt{6}}$ & $-\frac{1}{16}\sqrt{\frac{3}{2}}$ & $-\frac{11}{48}$ & $\frac{13}{16 \sqrt{6}}$ & $\frac{5}{48 \sqrt{2}}$ & $\frac{5}{48}$ & $0$ & $0$ & $0$ & $0$ \\
$\langle T_{25}^{kc} \rangle$ & $-\frac{5}{12}$ & $\frac{1}{\sqrt{6}}$ & $-\frac18\sqrt{\frac{3}{2}}$ & $-\frac{11}{24}$ & $\frac{13}{8 \sqrt{6}}$ & $\frac{5}{24 \sqrt{2}}$ & $\frac{5}{24}$ & $0$ & $0$ & $0$ & $0$ \\
$\langle T_{26}^{kc} \rangle$ & $\frac{5}{12}$ & $0$ & $-\frac{1}{16}\sqrt{\frac{3}{2}}$ & $-\frac{1}{16}$ & $\frac{5}{16 \sqrt{6}}$ & $\frac{5}{48 \sqrt{2}}$ & $\frac{5}{48}$ & $-\frac{3}{2}$ & $\frac{1}{2}$ & $-\frac{1}{2}$ & $2$ \\
$\langle T_{27}^{kc} \rangle$ & $\frac{5}{12}$ & $0$ & $-\frac{1}{16}\sqrt{\frac{3}{2}}$ & $-\frac{1}{16}$ & $\frac{5}{16 \sqrt{6}}$ & $\frac{5}{48 \sqrt{2}}$ & $\frac{5}{48}$ & $-\frac{3}{2}$ & $\frac{1}{2}$ & $-\frac{1}{2}$ & $2$ \\
$\langle T_{28}^{kc} \rangle$ & $-\frac{5}{12}$ & $0$ & $-\frac18\sqrt{\frac{3}{2}}$ & $-\frac{1}{8}$ & $\frac{5}{8 \sqrt{6}}$ & $\frac{5}{24 \sqrt{2}}$ & $\frac{5}{24}$ & $\frac{3}{2}$ & $-\frac{1}{2}$ & $\frac{1}{2}$ & $-2$ \\
$\langle T_{29}^{kc} \rangle$ & $-\frac{5}{12}$ & $0$ & $-\frac18\sqrt{\frac{3}{2}}$ & $-\frac{1}{8}$ & $\frac{5}{8 \sqrt{6}}$ & $\frac{5}{24 \sqrt{2}}$ & $\frac{5}{24}$ & $0$ & $-\frac{1}{2}$ & $\frac{7}{2}$ & $1$ \\
$\langle T_{36}^{kc} \rangle$ & $\frac{3}{2}$ & $0$ & $-\frac34\sqrt{\frac{3}{2}}$ & $-\frac{3}{4}$ & $\frac34 \sqrt{\frac{3}{2}}$ & $\frac{3}{4 \sqrt{2}}$ & $\frac{3}{4}$ & $0$ & $0$ & $0$ & $0$ \\
$\langle T_{37}^{kc} \rangle$ & $\frac{1}{8}$ & $0$ & $-\frac{5}{16}\sqrt{\frac{3}{2}}$ & $-\frac{13}{16}$ & $\frac{21}{16} \sqrt{\frac{3}{2}}$ & $\frac{29}{16 \sqrt{2}}$ & $\frac{29}{16}$ & $0$ & $0$ & $0$ & $0$ \\
$\langle T_{38}^{kc} \rangle$ & $\frac{5}{8}$ & $0$ & $\frac{3}{32}\sqrt{\frac{3}{2}}$ & $\frac{11}{32}$ & $\frac{13}{32} \sqrt{\frac{3}{2}}$ & $\frac{5}{32 \sqrt{2}}$ & $\frac{5}{32}$ & $0$ & $0$ & $0$ & $0$ \\
$\langle T_{41}^{kc} \rangle$ & $\frac{5}{4}$ & $0$ & $\frac{3}{16}\sqrt{\frac{3}{2}}$ & $-\frac{1}{16}$ & $\frac{1}{16}\sqrt{\frac{3}{2}}$ & $\frac{5}{16 \sqrt{2}}$ & $\frac{5}{16}$ & $0$ & $0$ & $0$ & $0$ \\
$\langle T_{42}^{kc} \rangle$ & $-\frac{1}{4}$ & $0$ & $-\frac18\sqrt{\frac{3}{2}}$ & $\frac{3}{8}$ & $\frac58 \sqrt{\frac{3}{2}}$ & $\frac{1}{8 \sqrt{2}}$ & $\frac{1}{8}$ & $0$ & $0$ & $0$ & $0$ \\
$\langle T_{43}^{kc} \rangle$ & $\frac{5}{4}$ & $0$ & $-\frac38\sqrt{\frac{3}{2}}$ & $\frac{1}{8}$ & $\frac38 \sqrt{\frac{3}{2}}$ & $\frac{15}{8 \sqrt{2}}$ & $\frac{15}{8}$ & $0$ & $0$ & $0$ & $3$ \\
$\langle T_{44}^{kc} \rangle$ & $\frac{5}{4}$ & $0$ & $\frac{3}{16}\sqrt{\frac{3}{2}}$ & $\frac{3}{16}$ & $\frac{5}{16} \sqrt{\frac{3}{2}}$ & $\frac{5}{16 \sqrt{2}}$ & $\frac{5}{16}$ & $-\frac{9}{2}$ & $0$ & $0$ & $-6$ \\
$\langle T_{45}^{kc} \rangle$ & $\frac{1}{4}$ & $0$ & $-\frac{1}{16}\sqrt{\frac{3}{2}}$ & $-\frac{9}{16}$ & $\frac{25}{16} \sqrt{\frac{3}{2}}$ & $\frac{1}{16 \sqrt{2}}$ & $\frac{1}{16}$ & $0$ & $0$ & $-3$ & $6$ \\
$\langle T_{48}^{kc} \rangle$ & $0$ & $0$ & $-\frac{27}{16}\sqrt{\frac{3}{2}}$ & $-\frac{3}{16}$ & $\frac{3}{16} \sqrt{\frac{3}{2}}$ & $\frac{75}{16 \sqrt{2}}$ & $\frac{75}{16}$ & $0$ & $0$ & $0$ & $0$ \\
$\langle T_{52}^{kc} \rangle$ & $\frac{5}{8}$ & $\sqrt{\frac{3}{2}}$ & $-\frac{15}{16}\sqrt{\frac{3}{2}}$ & $\frac{5}{16}$ & $\frac{13}{16}\sqrt{\frac{3}{2}}$ & $\frac{65}{16\sqrt{2}}$ & $\frac{65}{16}$ & $-\frac{39}{4}$ & $-\frac{3}{2}$ & $-\frac{3}{2}$ & $-\frac{51}{4}$ \\
$\langle T_{53}^{kc} \rangle$ & $\frac{3}{4}$ & $0$ & $-\frac38\sqrt{\frac{3}{2}}$ & $-\frac{3}{8}$ & $\frac98 \sqrt{\frac{3}{2}}$ & $\frac{9}{8 \sqrt{2}}$ & $\frac{9}{8}$ & $0$ & $0$ & $0$ & $0$ \\
$\langle T_{54}^{kc} \rangle$ & $\frac{5}{8}$ & $\sqrt{\frac{3}{2}}$ & $-\frac{15}{16}\sqrt{\frac{3}{2}}$ & $\frac{5}{16}$ & $\frac{13}{16}\sqrt{\frac{3}{2}}$ & $\frac{65}{16\sqrt{2}}$ & $\frac{65}{16}$ & $0$ & $\frac{3}{4}$ & $-\frac{21}{4}$ & $6$ \\
$\langle T_{55}^{kc} \rangle$ & $\frac{5}{8}$ & $\sqrt{6}$ & $-\frac{15}{16}\sqrt{\frac{3}{2}}$ & $\frac{13}{16}$ & $\frac{21}{16}\sqrt{\frac{3}{2}}$ & $\frac{145}{16 \sqrt{2}}$ & $\frac{145}{16}$ & $0$ & $0$ & $0$ & $0$ \\
$\langle T_{58}^{kc} \rangle$ & $\frac{5}{4}$ & $0$ & $-\frac{3}{16}\sqrt{\frac{3}{2}}$ & $-\frac{3}{16}$ & $\frac{5}{16} \sqrt{\frac{3}{2}}$ & $\frac{5}{16 \sqrt{2}}$ & $\frac{5}{16}$ & $0$ & $0$ & $0$ & $0$ \\
$\langle T_{60}^{kc} \rangle$ & $\frac{5}{8}$ & $0$ & $-\frac{3}{32}\sqrt{\frac{3}{2}}$ & $-\frac{3}{32}$ & $\frac{5}{32} \sqrt{\frac{3}{2}}$ & $\frac{5}{32 \sqrt{2}}$ & $\frac{5}{32}$ & $-\frac{27}{4}$ & $\frac{9}{4}$ & $-\frac{9}{4}$ & $9$ \\
\hline\hline
\end{tabular}
\end{center}
\end{table}
\endgroup

\end{document}